\documentclass[10pt]{article}
\usepackage{amsmath}
\usepackage{graphicx}
\usepackage{amsfonts}
\usepackage{amssymb}
\usepackage{epsf}
\usepackage{latexsym}

\def\a{\underline{\mbox{a}}}
\def\b{\underline{\mbox{b}}}
 
\def\d{\mbox{d}}

\def\nb{\mbox{b}}

\def\B{\mbox{\tiny B}}

\def\be{\begin{equation}}
\def\ee{\end{equation}}
\def\bea{\begin{eqnarray}}
\def\eea{\end{eqnarray}}
\def\foo{\footnote}
\def\pa{\partial}
\def\d{\textrm{d}}
\def\dotb{\mbox{\Large$\circ$}_{\underline{\mbox{\scriptsize b}}}   }
\def\Circ{\mbox{\Large$\circ$}}
\def\de{\mbox{\Large $\ast$}_{\underline{\mbox{\scriptsize b}}}     }
\def\di{\mbox{d}_{\underline{{\mbox{\scriptsize b}}}}}
\def\DotB{\mbox{\Large$\circ$}_{\underline{{\mbox{\scriptsize B}}}}}
\def\De{\mbox{\Large $\ast$}_{\underline{\mbox{\scriptsize B}}}}

\def\sdot{\mbox{$\circ$}_{\underline{\mbox{\tiny  b}}}}
\def\sDot{\mbox{$\circ$}_{\underline{\mbox{\tiny  B}}}}

\def\Ast{\mbox{\Large$\ast$}}
\def\sAst{\mbox{$\ast$}}
\def\Di{\textrm{d}_{\underline{{\mbox{\scriptsize B}}}}}

\def\cr{\mbox{\scriptsize{\bf $\mbox{ } \times \mbox{ }$}}}
\def\crs{\mbox{\scriptsize{\bf $\mbox{ } \times \mbox{ }$}}}
\def\hat{\widehat}
\def\ip{i^{\prime}}
\def\jp{j^{\prime}}
\def\kp{k^{\prime}}
\def\lp{l^{\prime}}
\def\ipp{i^{\prime\prime}}
\def\jpp{j^{\prime\prime}}
\def\kpp{k^{\prime\prime}}
\def\lpp{l^{\prime\prime}}
\def\b{\mbox{\underline{b}}}

\def\B{\mbox{\underline{B}}}
\def\H{\mbox{\underline{H}}}
\def\L{\mbox{\underline{L}}}
\def\P{\mbox{\underline{P}}}
\def\R{\mbox{\underline{R}}}

\def\nm{\mbox{{m}}}
\def\nn{\mbox{{j}}}
\def\nH{\mbox{{H}}}
\def\nL{\mbox{{L}}}
\def\nM{\mbox{{M}}}
\def\nP{\mbox{{P}}}
\def\nR{\mbox{{R}}}
\def\nX{\mbox{{X}}}

\def\mR{\mbox{R}}
\def\mL{\mbox{L}}
\def\mH{\mbox{H}}
\def\mP{\mbox{P}} 
\def\mS{\mbox{S}}

\def\sa{\mbox{\scriptsize a}}

\def\sd{\mbox{\scriptsize d}}
\def\sa{\mbox{\scriptsize a}}

\def\sd{\mbox{\scriptsize d}}
\def\sg{\mbox{\scriptsize g}} 
\def\sh{\mbox{\scriptsize h}} 
\def\sj{\mbox{\scriptsize j}} 

\def\sll{\mbox{\scriptsize l}}  
\def\sm{\mbox{\scriptsize m}}

\def\sp{\mbox{\scriptsize p}}
\def\sq{\mbox{\scriptsize q}}

\def\ss{\mbox{\scriptsize s}}
\def\su{\mbox{\scriptsize u}}

\def\sw{\mbox{\scriptsize w}}

\def\sB{\mbox{\scriptsize B}}

\def\sE{\mbox{\scriptsize E}}

\def\sG{\mbox{\scriptsize G}}
\def\sH{\mbox{\scriptsize H}}
\def\sI{\mbox{\scriptsize I}}
\def\sJ{\mbox{\scriptsize J}}

\def\sL{\mbox{\scriptsize L}} 
\def\sM{\mbox{\scriptsize M}}

\def\sR{\mbox{\scriptsize R}}
\def\sS{\mbox{\scriptsize S}}
\def\sT{\mbox{\scriptsize T}}

\def\sV{\mbox{\scriptsize V}}
\def\sW{\mbox{\scriptsize W}}
\def\sX{\mbox{\scriptsize X}} 
\def\sY{\mbox{\scriptsize Y}} 
 
\def\scr{\mbox{\scriptsize (cross)}}
\def\spert{\mbox{\scriptsize pert}}

\def\sAM{\mbox{\scriptsize AM}}
\def\sBO{\mbox{\scriptsize BO}}
\def\sBOB{\mbox{\scriptsize BOB}}
\def\sAM{\mbox{\scriptsize ZAM}}

\def\sG{{\cal G}} 

\def\sJ{\mbox{\scriptsize Jacobi}}

\def\su{\mbox{\scriptsize universe}}
 
\def\eph{\mbox{\scriptsize eph}}
 
\def\eph(B){\mbox{\scriptsize emergent(LMB)}} 
\def\seB{\mbox{\scriptsize emergent(LMB)}}
\def\teB{\mbox{\tiny emergent(LMB)}} 
\def\seBH{\mbox{\scriptsize emergent(LMB: H)}}
\def\seBW{\mbox{\scriptsize emergent(WKB-LMB)}}

\def\emL{\mbox{\scriptsize emergent(WKB-LMB: L)}}

\def\sem{\mbox{\scriptsize em}}
\def\sWKB{\mbox{\scriptsize emergent(WKB)}}
\def\WKBH{\mbox{\scriptsize emergent(WKB: H)}}
\def\WKBL{\mbox{\scriptsize emergent(WKB: L)}}
\def\sE{\mbox{\scriptsize internal(Euler)}}
\def\sY{\mbox{\scriptsize internal(York)}}

\def\si{\mbox{\scriptsize internal}}

\def\semi{\mbox{\scriptsize emergent(WKB-LMB)}}

\def\suB{\mbox{\scriptsize\underline{B}}}
\def\suH{\mbox{\scriptsize\underline{H}}}

\def\tfA{\mbox{\tiny{\sffamily A}}}

\def\th{\mbox{\tiny h}}

\def\tE{\mbox{\tiny E}}

\def\tH{\mbox{\tiny H}}

\def\tL{\mbox{\tiny L}}

\def\tR{\mbox{\tiny R}}

\def\tT{\mbox{\tiny T}}

\def\tE{\mbox{\tiny internal(Euler)}}
\def\tint{\mbox{\tiny internal}}

\def\uttL{\underline{\mbox{\tt M}}}
\def\ttL{{\mbox{\tt M}}}

\def\ttH{\mbox{\tt{H}}}
\def\ttM{\mbox{\tt{M}}}
\def\fj{\mbox{\sffamily j}}

\def\fl{\mbox{\sffamily l}}

\def\fE{\mbox{\sffamily E}}
\def\fF{\mbox{\sffamily F}}

\def\fI{\mbox{\sffamily I}}
\def\fJ{\mbox{\sffamily J}}

\def\fL{\mbox{\sffamily L}}

\def\fQ{\mbox{\sffamily Q}}

\def\fS{\mbox{\sffamily S}}
\def\fT{\mbox{\sffamily T}}
\def\fU{\mbox{\sffamily U}}
\def\fV{\mbox{\sffamily V}}
\def\fW{\mbox{\sffamily W}}

\def\sfj{\mbox{\sffamily{\scriptsize j}}}

\def\sfl{\mbox{\sffamily{\scriptsize l}}}

\def\sfp{\mbox{\sffamily{\scriptsize p}}}
\def\sfq{\mbox{\sffamily{\scriptsize q}}}

\def\sfA{\mbox{\sffamily{\scriptsize A}}}

\def\sfE{\mbox{\sffamily{\scriptsize E}}}
\def\sfF{\mbox{\sffamily{\scriptsize F}}}

\def\sfT{\mbox{\sffamily{\scriptsize T}}}
\def\sfU{\mbox{\sffamily{\scriptsize U}}}

\def\sfW{\mbox{\sffamily{\scriptsize W}}}

\def\ssWKB{\mbox{\scriptsize WKB}}

\def\aha{\sqrt{    \{   \fT_{\sH} + \fT_{\sL}   \}/
                   \{   \fE + \fU_{\sH} + \fU_{\sL} + \fJ_{\sH\sL}   \}      }   }
\def\a{\dot{\mbox{a}}}
\def\Sun{\mbox{\scriptsize Sun}}
\def\Earth{\mbox{\scriptsize Earth}}
\def\Jupiter{\mbox{\scriptsize Jupiter}}

\def\chin{\chi_{\mbox{\scriptsize j}}}

\begin{document}

\begin{titlepage}

\begin{center}

\vspace{.3in}

{\Large{\bf EMERGENT SEMICLASSICAL TIME IN QUANTUM GRAVITY.}} 

\vspace{.1in}

{\large{\bf I. MECHANICAL MODELS.}} 

\vspace{.3in}

{\large{\bf Edward Anderson$^*$}}

\vspace{.3in}

\noindent{\it Peterhouse, Cambridge, U.K., CB21RD;}

\noindent{\it DAMTP, Centre for Mathemetical Sciences, Wilberforce Road, Cambridge, U.K., CB30WA.}

\noindent Work stated at {\it Department of Physics, Avadh Bhatia Physics Laboratory, University of 
Alberta, Edmonton, Canada}

\end{center}

\vspace{.3in}

%\baselineskip=24pt 

%===========================================================ABSTRACT=======================================================================
\begin{abstract}

\noindent The emergent semiclassical time approach to resolving the problem of time in quantum gravity 
is considered in the arena of relational particle toy models.  
In situations with `heavy' and `light' degrees of freedom, two notions of emergent semiclassical WKB 
time emerge; these are furthermore equivalent to two notions of emergent classical 
`Leibniz--Mach--Barbour' time.  
I study the semiclassical approach, in a geometric phase formalism, extended to include linear 
constraints, and with particular care to make explicit the approximations and assumptions used, 
which are an important part of the semiclassical approach.     
I propose a new iterative scheme for the semiclassical approach in the cosmologically-motivated case 
with one heavy degree of freedom.  
I find that the usual semiclassical quantum cosmology emergence of time comes hand in hand 
with the emergence of other qualitatively significant terms, 
including back-reactions on the heavy subsystem and second time derivatives.  
I take my analysis further for relational particle models with linearly-coupled harmonic oscillator 
potentials, which, being exactly soluble by means outside the semiclassical approach to quantum 
cosmology, are additionally useful for testing the justifiability of some of the approximations and 
assumptions habitually made therein.  
Finally, I contrast emergent semiclassical time with its hidden dilational Euler time counterpart. 

\end{abstract}

%==========================================================================================================================================

\vspace{.3in}

\mbox{ }

%===============================================COMMENTS FOR PREPRINT ARCHIVE==============================================================

%\noindent{\bf Keywords:} 

%==========================================================================================================================================

\noindent{\bf PACS numbers 04.60-m, 04.60.Ds}

\vspace{4in}

\noindent$^*$ ea212@cam.ac.uk

\end{titlepage}

%=====================================================================================================
%=====================================================================================================
%==========================================================================================================================================
\section{Introduction}
%==========================================================================================================================================
%==========================================================================================================================================
%==========================================================================================================================================

The problem of time  
\cite{Kuchar92, Isham93, Wheeler, DeWitt, EarlyKuchar, Kuchar81, UW89, K91, B94II, EOT, Kuchar99, 
Kieferbook, OII, TLESS}  
in quantum gravity is a pervasive multi-faceted conceptual problem in attempting to put 
together relativistic gravitational theory and quantum theory.  
Such a theoretical framework is relevant if physics is to be whole, 
while its study is also important toward acquiring more solid foundations 
for the gradually-developing discipline of quantum cosmology (see e.g. \cite{HT, Wiltshire, Kiefer99}).   
Via e.g. inflation, quantum cosmology may then contribute to the understanding and prediction of 
cosmic microwave background fluctuations and the origin of galaxies \cite{Inf, HallHaw}.
Inflation is currently serving reasonably well \cite{WMAP} at providing an explanation for these; this  
remains an `observationally active area', with the Planck experiment \cite{Planck} about to be performed.

Some aspects of the problem of time manifest themselves if one attempts to canonically quantize 
general relativity (GR) in its traditional, 3-metric variables.  
In this Paper, I approach this via a variant of the Baierlein--Sharp--Wheeler action \cite{BSW, GRT, RWR, ABFKO, Phan} for 
GR\foo{
%%%%%%%%%%%%%%%%%%%%%%%%%%%%%%%%%%%%%%%%%%%%%%%%%%%%%%%%%%%%%%%%%%%%%%%%%%%%%%%%%%%%%%%%%%%%%%%%%%%%%%%
I use 
( ) for function dependence, 
[ ] for functional dependence,
( ; ] for a mixture of function dependence before the semi-colon and functional dependence after it, 
and  $\lfloor \mbox{ } \rfloor$ to enclose those functions on which a derivative acts.  
$h_{\alpha\beta}(x)$ is the spatial 3-metric, 
with determinant $h = h(h_{\alpha\beta}(x))$, 
covariant derivative $\nabla_{\beta}$, 
and Ricci scalar ${\cal R}(x; h_{\alpha\beta}(x)]$.
$\dot{\mbox{}}$ denotes $\frac{\pa}{\pa\lambda}$.  
$ _{\sG}||\mbox{ }||$ is the norm with respect to the array ${\cal G}$, with the ${\cal G}$ 
ordered to the left; this array is the DeWitt supermetric, 
${\cal G}_{\alpha\beta\gamma\delta} = 
\frac{1}{\sqrt{h}}
\left\{
h_{\alpha\gamma}h_{\beta\delta} - \frac{1}{2}h_{\alpha\beta}h_{\gamma\delta}
\right\}$.  
${\cal G}^{-1}$ is the inverse array
${\cal G}^{\alpha\beta\gamma\delta} = \sqrt{h}
\{  h^{\alpha\gamma}h^{\beta\delta}  -  h^{\alpha\beta}h^{\gamma\delta}   \}$. 
This ordering to the left is a relatively simple choice used e.g in \cite{Isham93}, although 
that does furthermore argue for the more complicated super-coordinate invariant Laplacian ordering.  
This is considered in my present semiclassical program in \cite{SemiclIII}.
$\fT_{\th} = \mbox{}_{\underline{{\sG}}^{-1}}|| \sDot h||^2$ is the gravitational kinetic term 
up to factors of $c$ and $16\pi G$.  
The $\sDot$ symbol is explained in Appendix A. 
The underline on the ${\cal G}^{-1}$ denotes de-densitization by division by $\sqrt{h}$.
$\delta_{\sh}$ is shorthand for the functional derivative with components 
$\frac{\delta}{\delta h_{\alpha\beta}(x)}$.  
$\Lambda$ is the cosmological constant. 
$\Psi$ is the wavefunction of the universe.  
%
%%%%%%%%%%%%%%%%%%%%%%%%%%%%%%%%%%%%%%%%%%%%%%%%%%%%%%%%%%%%%%%%%%%%%%%%%%%%%%%%%%%%%%%%%%%%%%%%%%%%%%%
} 
\be
\fS[h_{\alpha\beta}, \dot{h}_{\alpha\beta}, \dot{\mbox{B}}_{\alpha}] = \frac{c^3}{16\pi G}
\int \d\lambda\int \d^3x \sqrt{h}\sqrt{   \{\Lambda + {\cal R}\}   \fT_{\sh}  }
\label{SBSW}
\mbox{ } .  
\ee
This action is {\it temporally relational} \cite{RWR} in that the radicand therein is homogeneous quadratic 
so the integrand is overall homogeneous linear in the velocities and hence is invariant under 
reparametrization of the label-time $\lambda$.
Moreover, the corrections involving the auxiliary variable $\B$ 
are of the form which renders irrelevant the coordinatization of space 
(in which sense it renders the theory {\it spatially relational}).   
These relational features underpin how actions of this type lead to constraints.  
Preliminarily, the canonical momenta are
\be
\pi^{\alpha\beta} = \frac{c^3}{16\pi G}\sqrt{\frac{\Lambda + {\cal R}}{\fT_{\sh}}} 
{\cal G}^{\alpha\beta\gamma\delta}\DotB h_{\gamma\delta} \mbox{ } .
\ee
Reparametrization invariance then implies \cite{Dirac} that there is at least one primary constraint  
(i.e. relation between these momenta arising purely from the form of the Lagrangian).  
Moreover, the local square root form (i.e. take the square root and then integrate) 
implies, by a working \cite{RWR} that is in close analogy with Pythagoras' theorem, that there is 
precisely one such constraint per space point, 
\be
{\cal H} \equiv \frac{16\pi G}{c^2}\mbox{}_{\sG}||\pi||^2 -\frac{c^4}{16\pi G}\sqrt{h}\{{\cal R} + \Lambda\} = 0 \mbox{ } \mbox{ (Hamiltonian constraint) } \mbox{ } .
\ee
There is, additionally, as a secondary constraint from variation with respect to the 3-diffeomorphism 
correcting variable $\B$,  
\be
{\cal M}^{\alpha} \equiv -2\nabla_{\beta}\pi^{\alpha\beta} = 0 \mbox{ } \mbox{ (momentum constraint) } .
\ee
Finally, applying the Dirac procedure \cite{Dirac} to these gives rise to no more constraints. 
Indeed it places 
heavy consistency restrictions on trying to construct actions similar to the above \cite{RWR, San, Phan}.  
Moreover, a full enough set of fundamental matter fields to describe nature can be included 
in this formalism \cite{AB, Van, Lan}.

Then, assuming standard commutation relations are to be used, and working in the configuration representation 
(and only at a formal level, with regularization and operator ordering issues left essentially 
unaddressed by the convenient `configurations to the left of momenta' choice of ordering) 
one arrives at the quantum momentum constraint
\be
\hat{{\cal M}}^{\alpha}\Psi \equiv  -\frac{\hbar}{i} 2\nabla_{\beta} {\delta_{\sh}\mbox{}^{\alpha\beta}} \Psi = 0
\label{GRQmom}
\ee
and the {\it Wheeler-DeWitt equation} (WDE) \cite{Wheeler, DeWitt} 
\be
\hat{{\cal H}}\Psi \equiv 
- \frac{16\pi G}{c^2}\hbar^2 \mbox{}_{\sG}||\delta_{\sh}||^2\Psi - 
\frac{c^4}{16\pi G}\ \sqrt{h}\{ {\cal R} + \Lambda\}\Psi = 0 \mbox{ } .  
\label{WDE}
\ee
The former is conceptually clear enough: $\Psi$ is to depend on spatial 3-geometry rather than 
on its coordinatization.  
But the latter is a time-independet Schr\"{o}dinger equation (TISE) 
rather than the kind of time-containing equation
[e.g. time-dependent Schr\"{o}dinger equation (TDSE) or Klein--Gordon (KG) type equation] 
that one would expect ab initio in a quantum theory.  
Thus, in flagrant contradiction to what we appear to experience, 
canonical quantum GR predicts that the universe is timeless in that it is stationary.  
This `frozen formalism' `paradox' is one of the various aspects of the problem of time.  
Ways that have been suggested around this which are conceptually reasonable, 
albeit both technically and conceptually unsurmountable in detail, include: 
strategy A), that the universe is truly timeless 
\cite{TLESSLIT, UW89, Kuchar92, Isham93, EOT, B94I, B94II, EOT, H99, HalliRec, HT, Kieferbook, TLESS}.
Strategy B), that instead of the usual sort of quantum theory based just on configurations, 
one should consider a quantum theory of histories \cite{Histories, Kuchar92, Isham93}. 
Strategy C), that there is no time fundamentally, but that it does emerge, at least in some regimes 
\cite{DeWitt, Gerlach, LR79, Banks, HallHaw, BV89, KS91, 
Kuchar92, Isham93, Semicl, Parentani, Kiefer99, Kuchar99, Kieferbook}.
Strategy D), that time is actually hidden within the conventional picture's variables, and 
could be identified by transforming to new variables such that it is disentangled from the true 
dynamical variables of the theory \cite{York72, Yorktime, Kuchar81, Kuchar92, Isham93, IB75, crit}.

In this Paper, I consider some versions of C) and D), i.e. approaches involving the search for 
timefunctions
It is thus appropriate to qualify what are desirable properties for such timefunctions.  
At the classical level, these include: 
1) that their {\sl global validity} is preferable.  
2) Within whatever range of validity one has, {\sl monotonicity} (gradient $g \geq 0$) is required, 
while $g = 0$ (frozen) and $g = \infty$ points may be problematic features.  
3) It also makes sense for a time function to be {\sl operationally meaningful} 
(computible from observable quantities).

In particular, I consider a substrategy of C), the semiclassical approach to quantum gravity, in which 
time emerges only in the semiclassical regime. 
This approach is conceptually attractive in being a `resolution' of the `paradox' between canonical 
quantum general relativity having a frozen formalism and local, light physics appearing to be dynamical.  
This resolution is via the Wheeler--DeWitt equation being replaced by an emergent time-dependent 
Schr\"{o}dinger equation, at least approximately, from which point one can give the system a 
probabilistic interpretation using the obvious associated Schr\"{o}dinger inner product.  
This is furthermore attractive through the semiclassical approach being likely applicable and similar 
for a variety of different viable gravitational theories (while the internal time approach would depend 
more on the technical details of one's theory of gravitation).  
It also makes contact with a fairly widely understood body of techniques: the Born--Oppenheimer 
approximation and Berry phase.  
In the semiclassical approach, time is then an approximate semiclassical concept \cite{Isham93}, e.g. 
there isn't such a thing as time in some early universe settings in quantum cosmology.  
Because of this intimate tie between semiclassical explanations of the problem of time and the 
approximations required for this regime to be valid (a topic much developed in Papers I-II), indications  
of the circumstances under which this breakdown occurs arise from detailed examination of the 
approximations involved (see Paper II).

The semiclassical approach also has ties to histories theory and timeless records theory approaches.  
Within histories theory, decoherence produces records, and decoherence is habitually linked with the emergence of (semi)classicality (see e.g. \cite{H99}).

The semiclassical approach is furthermore the setting for Halliwell--Hawking type schemes in which CMB 
inhomogeneities and galaxy formation are to follow from quantum cosmology, which is of current interest 
due to the considerable and ongoing improvements in observational cosmology.
This Paper provides a toy model for, and a number of techniques extendible to, a more thorough treatment 
of this more realistic arena.

In this Paper, I study relational particle models (RPM's) 
\cite{BB82, BS, Kuchar92, B94I, EOT, RWR, B03, Lan, Than, Paris, OI, OII, 2d} 
(also see \cite{RPMAPPS, ABFKO} for applications of these and \cite{RPMCONCAT, Gergely, GM} for 
technically-related work) as toy models toward understanding some aspects of the problem of time in 
quantum GR.
My subsequent study of GR in this light is in Paper II \cite{PaperII} (and is in a minimally-coupled multi-scalar 
matter setting directly relevant to theoretical cosmology).    
I approach this from my variant \cite{OII} of the Barbour--Bertotti action \cite{BB82},\foo{I 
continue to use summation convention for space indices here, but sum explicitly over interparticle 
(cluster) separations.  
The $\nR_{\gamma i}$ are the $i$ relative Jacobi coordinates \cite{Marchal}
with corresponding (cluster) relative masses $\mu_{i}$.  
$\fV$ is the potential, which has the relational dependence 
$\fV = \fV(||\R_i||, (\R_i, \R_j) \mbox{ alone})$.       
$\fU \equiv - \fV$.  
$ _{\mu}||\mbox{ }||$ is the norm with respect to the array 
${\mu}^{ij\alpha\beta} = 2\mu_i\delta^{\alpha\beta}\delta^{ij}$ (the mass metric).   
$ _{{\mu}^{-1}}||\mbox{ }||$ is the norm with respect to the array 
${\mu}_{ij\alpha\beta} = \delta^{\alpha\beta}\delta^{ij}/2\mu_i$. 
The inner products corresponding to the above are denoted by 
$( \mbox{ } , \mbox{ } )$ with the same prefixes.  
$\fT = \mbox{}_{{\mu}}||\sdot \nR ||^2$ is four times the usual kinetic term $\widetilde{\fT}$, 
where the $\sdot$ symbol is explained in Appendix A.  
$\pa_{\tR}$ is shorthand for the partial derivative with components  
$\frac{\pa}{\pa \sR_{\alpha i}}$.
$\fE$ is the energy of the system.   
$\Phi$ is the wavefunction of the toy universe.  } 
\be
\fS[\R_i, \dot{\R}_i, \dot{\b}] = \int\d\lambda\sqrt{\{\fE + \fU\}\fT}
\label{BBA} \mbox{ } .
\ee
The momenta are 
\be
\nP^{i\alpha} = \sqrt{\{\fE + \fU\}/{\fT}} \mbox{ } 2\mu_i   \dotb \nR_{i\alpha}  = 
\sqrt{\{\fE + \fU\}/{\fT}} \mbox{ } \mu^{ij\alpha\beta}\dotb \nR_{j\beta}
\mbox{ } . 
\label{momvelrel}
\ee
Just as above, temporal relationalism and the square root form of the action lead 
to a single quadratic primary constraint 
\be
\ttH \equiv \mbox{}_{\mu^{-1}}||\P||^2 + \fV = \fE \mbox{ (energy constraint) },
\ee
while a secondary constraint arises from variation 
with respect to the rotation auxiliary, $\b$:   
\be
\underline{\ttM} \equiv \sum_i \R_i \cr \P^i = 0  \mbox{ (zero angular momentum (ZAM) constraint) }.
\ee
Then, quantizing in the relative postition representation, we have 
\be
\hat{\tt{M}}^{\alpha}\Phi \equiv
\frac{\hbar}{i}{\epsilon^{\alpha\beta}}_{\gamma}
\sum_{i = 1}^n
\nR_{\beta i}{\pa_{\sR}\mbox{}^{\gamma i}}
\Phi = 0 \mbox{ } ,
\label{toylin}
\ee
\be
\hat{{\tt{H}}}\Phi \equiv  
- \hbar^2 \mbox{}_{\mu^{-1}}||\pa_{\sR}||^2\Phi + \fV\Phi = \fE\Phi 
\mbox{ } .  
\label{toyquad}
\ee
There are no ordering or regularization problems here \cite{OII}.
The above parallel working, and the similarity in form  between (\ref{toylin}, \ref{toyquad}) 
and (\ref{GRQmom},\ref{WDE}) are some manifestations of the analogy between GR and RPM's.  
See also \cite{BB82, Kuchar92, B94I, RWR, OII}) for further discussion,   
and Appendix B for comparison of the realisticness of these toy models with that of 
the more habitually studied GR minisuperspace toy models \cite{DeWitt, Misner, HH, Wiltshire, eath}.   

\mbox{  }

In Secs 2-3, I consider relational `Leibniz--Mach--Barbour time' $t^{\eph(B)}$ 
as an emergent time, which is of later use in this Paper.  
The semiclassical approach \cite{DeWitt, Gerlach} leads to an a priori distinct emergent time 
\cite{LR79, Banks, HallHaw}, $t^{\sWKB}$ underlying this Paper's principal theme, the semiclassical 
approach [Sec 4--8, see also the reviews \cite{Kuchar92, Isham93, Kiefer93, Kiefer94, Kiefer99, 
Kieferbook}].  
The approach is based on an adiabatic, BO (Born--Oppenheimer)-type ansatz and approximations of types 
often associated with this (Sec 5) alongside a WKB ansatz and approximations of types 
often associated with that (Sec 6).    
In the mechanical model context, 
the BO  type ansatz for the wavefunction is\foo{Here, the $\H_{\ip}$ are  
vector-valued H-coordinates with $\ip$ taking 0 to $p$, and the $\L_{\ip}$ are 
vector-valued L-coordinates with $\ipp$ taking $p + 1$ to $n$ and 
$M$ denotes a `generic' heavy mass).  }  
\be
\Phi(\H_{\ip}, \L_{\ipp}) = \phi(\H_{\ip})|\chi(\H_{\ip}, \L_{\ipp})\rangle \mbox{ }   
\label{WKBBO} \mbox{ } ,
\ee
while the WKB ansatz is of the form
\be
\phi(\H_{\ip}) = \mbox{exp}(iM\fF(\mbox{\underline{H}}_{\ip})/\hbar) \mbox{ } .  
\label{WKB}
\ee
The formalism I consider is based on H-equations formed by $\langle\chi|$ times the 
constraints with certain approximations applied and L-equations that are ab initio 
of fluctuation type -- each constraint minus its H-equations times $|\chi\rangle$.  
I do this so as to incorporate the subtleties of geometric phase, indeed looking at whether  
the Berry \cite{Berry89, Berryrev} approach to QM extends robustly to systems which have  
subsidiary linear constraints.  
It is furthermore crucial for the quantum cosmological application that a certain `chroniferous' 
cross-term, which is neglected in BO's 
H- and L-equation treatment, is not neglected here.  
For, it is manipulation of this 
term that permits the quadratic constraint's fluctuation L-equation  to be recast (Sec 7) 
as a TDSE for the L-subsystem 
with respect to a timefunction (mostly) provided by the H-subsystem.  
One solves an approximate H-equation, then substitutes into the L-equation.  
In the cosmologically-motivated case of a single H d.o.f, 
I propose and investigate a new procedure: to invert the $t^{\sWKB}$--H 
relation so as to fully purge the H-dependence from the L-equation.  
I find that the new L-equation obtained thus is actually more general than a TDSE.  
Once this (or perhaps an approximation to it) is solved, one has enough information to evaluate the sizes 
of the omitted terms, which leads to an important application in the case of geometrodynamics in Paper II.  
Another subsequent new working is to then substitute the L-equation's solution back into a more 
accurate H-equation and thus proceed iteratively.

This Paper furthermore makes contact with a number of basic, foundational questions that have been raised 
about the semiclassical approach to the problem of time.  

\noindent B1) Does it reconcile the `paradox' between the theoretical timelessness of the universe 
and the quotidian semblance of dynamics? 
This is an issue raised e.g. in \cite{Kuchar92, Isham93, BS, B93, B94II, EOT, H99, HalliRec, HT}, and 
discussed in e.g. \cite{BV89, Kiefer94}, albeit subject to a number of further assumptions.

\noindent B2) One common set of assumptions on which semiclassical resolutions of the problem of time 
rest is the validity of the WKB procedure (which I take to mean 
ansatz plus some associated package of approximations) for the wavefunction of the universe.   
This is of crucial importance since it is via a property exclusive to the WKB ansatz 
that the trick whereby the chroniferous cross-term becomes the 
time-derivative part of a TDSE functions 
(see e.g. \cite{BS, Kuchar92, B93, Isham93, B94II, EOT} and Sec 7).    
Now, while one is accustomed to seeing WKB procedures in ordinary QM, 
N.B. that these rest on the `Copenhagen' presupposition that one's quantum system 
under study has a surrounding classical large system 
and that it evolves with respect to an external time.   
Moreover, in quantum cosmology, as the quantum system is already the whole universe, 
the notions of a surrounding classical large system and of external time cease to be appropriate  
\cite{Wheeler, Zeh86}.  
So using WKB procedures in quantum cosmology really does require novel and convincing justification, 
particularly if one is relying on it to endow a hitherto timeless theoretical framework with 
a bona fide emergent time.
Were this attainable, B1) would then go a long way toward rigorously resolving the `paradox' in the 
sense that the truly relevant procedure of inspection of L-subsystems would reveal a semblance of 
dynamics even if the universe is, overall, timeless.

Of relevance to this issue, I firstly point out (Sec 7) that a WKB regime cannot be 
expected to hold everywhere.  
Thus B1 is not a complete resolution of the problem of time, nor even of the `paradox' of the timeless 
appearence of the quantum theory  
(but there is not necessarily any need to resolve this in regions outside `quotidian experience' 
-- one cannot testify that there is a semblance of dynamics in regions in which semiclassical 
quantum cosmology does not apply.  
Secondly, I choose `ulteriorly' exactly soluble models (i.e. ones which are soluble by techniques 
outside the semiclassical approach) so as to have a framework in which extra checks are possible 
as regards whether the WKB approximation holds well in all regions of interest.  
However, I find that even addressing the question of where it holds in simple toy model cases is not 
in practice clear-cut, due to there being of the order of 20 additional, distinct quantum-cosmological 
approximations that one requires to make alongside it.  
In existing literature, these are mostly tacit or 
not touched upon by leaving vague the full extent of what the words 
`adiabatic' and `WKB' need to mean to fully cover all the simplifications required.  
Various of the below considerations also act to worsen this proliferation of approximations.  
Thus even explicit investigations in ulteriorly exactly soluble models 
only concern small pieces of configuration space characterized by all the other approximations 
being made.

\noindent B3) As regards the status of the semiclassical approach as resolving the frozen formalism 
problem, I point out that there are further senses in which fixed eigenstate universes can be frozen.  
The H- and L-subsystems could have incompatible gaps in their energy spectra,
giving a bizarre suppressed dynamics rather than a close reproduction of the L-dynamics   
(see \cite{DeWitt} and \cite{BS}).  
However, I argued in \cite{OII} that such frozennes is in practice an artefact of 
considering models with an insufficient number of degrees of freedom (d.o.f.) and a lack 
of available free particles.   
If one's model has free particles, then the precise compatibility of gaps in the energy levels 
also present does not matter, while it is L--L and L-free particle interactions that dominate 
rather than interactions between the L-subsystem and H-modes for the whole universe.  
The latter are then just tiny corrections as one would expect, rather than enforcers of high 
suppression on the L-subsystem. 
The complication that this causes is that one needs at least 3 d.o.f. in a model for 
this behaviour to be incorporable.  
This is relevant in motivating the present RPM program as it is easier to 
systematically build up our understanding with the grid of RPM's (see App. C) 
which is more densely populated and analytically tractable than the grid of minisuperspace models.

\noindent B4) Expanding on informal discussions of Barbour \cite{Barbourcom} 
and of Kiefer \cite{Kiefer93} subsequently considered by Datta \cite{Datta97}, 
I explain how different parts of the semiclassical approach yield exact and approximate 
WKB timefunctions, and that these furthermore coincide with the exact and approximate LMB timefunctions.
It is thus that the LMB approach can be considered to be a foundation underlying the semiclassical approach.
This widens the relevance of Sec 2--3's discussion of the LMB timefunction and the problems associated 
with it.

This Paper also addresses several detailed aspects of the semiclassical approach to the problem of time.  

\noindent D1) There is a need for back-reaction terms -- L-terms in the H-equation.    
(These I build in to my general working from Sec 5 onwards.)  
These are a desirable feature from the perspective of wishing to model GR, 
as a conceptually important part of GR  
(in its aspect as supplanter of absolute structure) is for
the matter to back-react on the geometry.  
There are also theory-independent reasons for back-reaction terms -- how could one subsystem provide 
time for another if the two are not coupled to each other?  
The approach I take to back-reaction terms is that these are provided by following the geometrical phase 
approach.  
As regards B2, this in turn rather increases the number of contemplable approximations.

\noindent D2)  In Sec 7 I also consider whether the combinations of approximations made in the 
literature are qualitatively and quantitatively consistent and widely applicable.  
Various of the `small terms' are interdependent in size, 
so that neglecting the one but not the other may be quantitatively dubious.
In particular, the crucial cross-term would usually be regarded as small (as BO do).  
But if terms of this nature should be kept, certain back-reaction terms should also be kept.
Thus in addition to the previous paragraph's 2 conceptual reasons for the keeping of back-reaction 
terms, there is a third, technical reason internal to the RPM's for doing so.

\noindent D3) A consequence of the true underlying many approximations character is that 
1-parameter expansions such as those used in the literature \cite{KS91} at most apply to 
small corners of the quantum cosmological configuration space.      
In the general situation considered in this Paper, one should expand in all the independent parameters!
As regards which parameters are relevant for modelling our own universe as a GR cosmology 
I begin to consider that in Paper II.

\noindent D4) There is more than one kind of geometrical phase. 
Which is relevant to quantum cosmology, absolute or relative phase?

It is also worth noting that what corrections semiclassical quantum gravity provides 
is still a developing subject in further ways -- e.g. a full physical interpretation is not yet 
available for all currently computed corrections, 
nor has the scheme producing them been checked to be robust to further 
corrections or change of gravitational theory (see however \cite{Kiefersugy}).

In Sec 8, I take my calculations further for the example of 2 linearly-coupled harmonic oscillators 
(HO's).  
This work may be seen as (leading to) an extension of the mechanical models  
in e.g. \cite{Banks, BriggsRost, Pad, PadSingh}.     
This is useful, e.g. because the treatment of inhomogeneous perturbations about 
a homogeneous cosmology involves an infinite collection of HO's by Fourier decomposition.  
The trade-off in RPM's between explicit tractability and various relevant nontriviality criteria 
relevant to problem of time strategies is laid out in Appendix C. 
The advantage of the particular examples considered is that one can solve these exactly by methods 
ulterior to the semiclassical approach to quantum cosmology; this permits further investigation of 
whether this approach's approximations and ans\"{a}tze are sensible, at least in these simple cases.  
I treat this using an iterative scheme of the type proposed in Sec 7,  
in the approximation that the potential alone picks up a time-dependent 
perturbation, treated to first order.

In Sec 9 I consider, as a useful contrast to semiclassical time, a {\sl hidden} (or {\sl internal}) 
{\sl time} approach \cite{York72, Yorktime, Kuchar81, Kuchar92, Isham93, OII}.  
The hidden time is to be found by replacing ${\cal H} = 0$ by its classical solution for a momentum 
variable, $P_{t^{\tint}} = P_{t^{\tint}}({t^{\si}}, Q^{{\mbox{\scriptsize true}}}_{\Delta}, 
P^{{\mbox{\scriptsize true}} \Delta})$.  
This could be a momentum variable obtained by performing a canonical transformation on the 
original coordinate and momentum variables, rather than one of the original momentum variables itself.    
The corresponding vposition representation quantization would then give a TDSE to supplant (\ref{WDE}).  
One candidate for $t^{\si}$ in GR is the dilational {\it York time} \cite{York72}  
$t^{\sY} = \frac{2}{3} \pi^{\alpha\beta}h_{\alpha\beta}/\sqrt{h}$.
RPM's have a dilational `Euler time' $t^{\sE} = \sum_{i = 1}^n \R_i\cdot \P^i$ \cite{OII} 
which is closely analogous to this.  
I conclude in Sec 10, including discussion of how internal and emergent timefunctions are sometimes 
aligned, sometimes approximately so, and sometimes not at all, and arguments in favour of semiclassical 
rather than internal time approaches.      
In Paper II, I transfer the conceptual knowledge gained from this study back to GR, 
considering both the general case and a specific minisuperspace example.

%==========================================================================================================================================
%==========================================================================================================================================
%=======================================================================================================
\section{Leibniz--Mach--Barbour (LMB) time} 
%=======================================================================================================
%==========================================================================================================================================
%==========================================================================================================================================

Sec 1's formulation for RPM based on the action (\ref{BBA}) or, with reparametrization invariance made 
explicit,
\be
\fS[\R_i, \di \R_i] \equiv 
\int \mbox{ }_{\mu}||\di \R||\sqrt{\fE + \fU}       
\mbox{ } , \mbox{ }
\label{Ssan}
\ee
can be interpreted as having an emergent quantity 
\be
\frac{s}{2}\sqrt{\frac{\fT}{\fE + \fU}} 
\left( = s \sqrt{\frac{\widetilde{\fT}}{\fE + \fU}} 
\right) 
\equiv \mbox{n}_{\ss}(\R_{\ip}, \dot{\R}_{\ip}, \dot{\b}) 
\label{n}
\ee 
(up to some constant time-scale $s$, thus slightly generalizing the presentation of \cite{B94I}) 
in its momentum-velocity relation (\ref{momvelrel})
and its Euler--Lagrange equation (not provided).
This quantity plays the role of a lapse, i.e. time elapsed.
Because it is dimensionally a velocity, I also denote it by $\a_{\ss}$.  
Furthermore, one can interpret the particular combination 
$\frac{1}{\mbox{\scriptsize n}_{\mbox{\tiny s}}  }\frac{\d }{\d \lambda}$ 
which occurs in the abovementioned equations as 
$\ast_{\ss} \equiv \frac{\d}{\d t_{\ss}^{\teB}}$ for $t_{\ss}^{\seB}$ the {\it LMB time}.  
Then, explicitly, integrating,  
\be
t_{\ss}^{\seB}[\R_i, \di\R_i] = t_{\ss}^{\seB}(0) + \frac{s}{2}\int\mbox{}_{\mu}||\di \R_i||/\sqrt{\fE - \fV}     
\mbox{ } .
\label{LMB}
\ee
The presence of the $t^{\seB}(0)$ term here is the familiar `freedom of choice of time-origin'.
Insofar as the lapse is emergent, LMB time (\ref{LMB}) is also an emergent notion of time.  
The role that this notion of time plays in understanding this Paper's principal theme -- the semiclassical 
approach -- will become clear in Sec 7.  
LMB time has the following characteristics.  
It is a classical time.  
It is a time provided by the system itself rather than an external time.   
It is a measure of change in the `whole' configuration\foo{I say `whole' as  
were canonical coordinates included whose kinetic terms were linear 
in their velocities, then those kinetic terms would not contribute.   
However, it is those terms which do contribute which are are those required to get Sec 4--7 to work.} 
in which respect it is somewhat like the astronomers' traditional standard of time,  
in that it pools together observations of all the significant bodies.   
Due to this, Barbour calls it `ephemeris time' (see e.g. \cite{B94I});  
however I argue against this on p. 7.  
%*page edit

\noindent\underline{Classical time lemma.} If $\mbox{n}_{\ss}$ is constant, i) conservation of energy holds.  
ii) $\lambda$ is $t^{\seB}$ up to choice of time-origin and time-scale and corresponds 
to choosing to label with a Newtonian time.  

\noindent{Proof:} i) Let constant $k = \mbox{n}_{\ss} = \frac{s}{2}\sqrt{\fT/\{\fE + \fU\}}$. 
So 
\be
\fE + \fU = \{  {s}/{2k}  \}^2\fT(\dot{\R}_i) \mbox{ } . 
\label{II}
\ee
Also, 
\be
\fT(\dot{\R}_i) = 2\sum_i \mu_i ||\dotb\R_i||^2 \mbox{ } , \mbox{ } \P^i = \frac{2k}{s}\mu_i\dotb\R_i 
\mbox{ } , 
\ee
so $\fE + \fU = \fT(\P_i)$.
ii) Rewriting the definition of $\fT$ as ${\mbox{}_{\mu}||\di \R ||^2}/{\{\d\lambda\}^2}$, 
(\ref{II}) becomes, upon solving the differential equation, 
\be
\lambda - \lambda(0) = \frac{k}{s}\int\mbox{}_{\mu}||\di \R_i||/\sqrt{\fE + \fU} \mbox{ } . 
\ee
Combine this and (\ref{LMB}) to get 
\be
\lambda - \lambda(0) = \frac{k}{s^2}
\{ t^{\seB}_{\ss} - t^{\seB}_{\ss}(0) \} \mbox{ } .  
\ee
$k = \mbox{n}_{\ss}$ also significantly simplifies the mechanical equations, which pass 
from containing $2\sqrt{\frac{\sfE + \sfU}{\sfT}}\frac{\d}{\d\lambda}$ to containing 
$\frac{\d}{\d t^{\teB}}$.  
Exercising this simplifying choice is none other than choosing to label with a Newtonian time $\Box$.

\noindent
The above simplifications also occur if a timefunction that absorbs $\mbox{n}_{\ss}$ is used: 
$\frac{\d}{\d t} = \frac{    1    }{    \mbox{\scriptsize n}_{\ss}     }\frac{\d}{\d\lambda}$
The above procedure is then the passage from the Jacobi formalism to the more habitual 
Euler--Lagrange one at the level of the mechanical equations 
(while the reverse procedure at the level of the action is the elimination of 
$\dot{t}^{\seB}$ by Routhian reduction \cite{Lanczos}).
%
%Note however that Barbour's argument for conservation of energy as a result of making this choice 
%is fallacious because it confuses $\fT(\dot\R)$ and $\fT(\P)$.  
%
Additionally, the $s = 1$ choice brings out the analogy with the GR convention most closely: 
$\mbox{n} \equiv \mbox{n}_1 = \frac{1}{2}\sqrt{\frac{\sfT}{\sfE + \sfU}} = 
\sqrt{\frac{\widetilde{\sfT}}{\sfE + \sfU}}$ is {\sl the} analogue of the GR lapse 
(c.f. Paper II).  
I use this from now on.   
$\dot{\mbox{a}} \equiv \dot{\mbox{a}}_1$.  
$\ast \equiv \ast_1$.

One of this Paper's new investigations is to see how the LMB quantity fares as regards having the 
desirable properties of timefunctions listed in Sec 1.

%--------------------------------------------------------------------------------------------------------
\noindent\underline{1) globality.} 
%--------------------------------------------------------------------------------------------------------
%
%
%
As the LMB quantity does not necessarily exist at zeros of $\fE + \fU$, it is not generally globally 
valid for a given mechanical motion.   
These are always `halting points' in the sense that $\P_{i} = 0$ there: 
$0 = \fE - \fV = \fT(\P_i)$ by conservation of energy and $\fT$ is positive-definite, 
but note that this has no bearing on what value $\fT(\dot{\R}_i, \dot{\b})$ may take at such points, 
as a consequence of the definition (\ref{n}).  
Indeed the action $\fS = \int \d\lambda\sqrt{\fT\{\fE + \fU\}}$ itself may 
well cease to make sense at such zeros, through itself becoming complex.
It may sometimes be possible to redefine the timestandard to move past such zeros, in some cases 
obtaining a fuller range of real values and in other cases as an analytic continuation into the 
complex plane.  
Complex action, momentum, time correspond to classically forbidden regions, but these can play 
a role in QM (through being penetrated by decaying wavefunctions).

%--------------------------------------------------------------------------------------------------------
\noindent\underline{2) monotonicity.} 
%--------------------------------------------------------------------------------------------------------
%
%
% 
However, if $t^{\seB}$ exists for (a given portion of) a given motion, its monotonicity is 
guaranteed: $\fE + \fU > 0$, so $\mbox{n} \geq 0$, so by (\ref{LMB}) 
$\dot{t}^{\eph(B)}  \geq 0$.
[Note this is not a $\lambda$-dependent statement by `cancellation' -- it is invariant under 
the valid reparametizations of $\lambda$ since these themselves are monotonic.]  
While existence is not compromised by sufficiently benign blow-ups in $\a$,  
i.e . those for which it remains integrable, such a blowup corresponds to the $t^{\seB}$ graph 
becoming infinite in slope.   
There may also be frozenness: at points for which the graph is horizontal, i.e. 
$\fT = 0$ or $\fE + \fU$ infinite.
Both zero and infinite slope may compromise use of $t^{\seB}$ itself to keep track 
for some ranges of mechanical motion.  
But at least in some cases, redefined timestandards may permit the following of 
motions through such points.

%--------------------------------------------------------------------------------------------------------
\noindent\underline{3) operational meaningfulness.} 
%--------------------------------------------------------------------------------------------------------
%
%
%
There are also problems with observing $t^{\seB}$ itself, or with using more readily observable 
approximations to it as detailed on page 7.   
%page* edit 

%====================================================================================================
%==========================================================================================================================================
%==========================================================================================================================================
\section{Heavy--light (H-L) split of RPM's and approximate LMB time}
%====================================================================================================
%==========================================================================================================================================
%==========================================================================================================================================

Consider situations in which the coordinates $\R_i$ $i = 1$ to $n$ can be split\foo{Note 
that this split is not in general to be considered to be aligned with the 
`subsystem'--`environment or background' split, which is rather an issue of what is under observation 
and (or) what 
is taken to be known.}   into 
heavy coordinates $\H_{\ip}$  with $\ip  = 1$     to $p$ and masses $\mu_{\ip}  =  M_{\ip}$, and 
light coordinates $\L_{\ipp}$ with $\ipp = p + 1$ to $n$ and masses $\mu_{\ipp} = m_{\ipp}$ such 
that\foo{It is to be understood that the $\epsilon$'s in this Paper are small; 
I use these instead of `$<<$' to keep a more precise account of requisite inter-relations and rankings 
among these small quantities in the approximate approaches under investigation.} 
\be
{m_{\ipp}}/{M_{\ip}} = \epsilon_{\sH\sL} \mbox{ } \mbox{ (H--L mass hierarchy) } .  
\ee
At the classical level, one of the forms for the RPM action is then
\be
\fS[\H_{\jp}, \L_{\kpp}, \dot{\H}_{\jp}, \dot{\L}_{\kpp}] = \int\d\lambda 
\sqrt{    \{\fE_{\sH} + \fU_{\sH} + \fU_{\sL} + \fJ_{\sH\sL}\}\{\fT_{\sH} +\fT_{\sL}\}    } \mbox{ } ,
\label{JHL}
\ee
for\foo{
%%%%%%%%%%%%%%%%%%%%%%%%%%%%%%%%%%%%%%%%%%%%%%%%%%%%%%%%%%%%%%%%%%%%%%%%%%%%%%%%%%%%%%%%%%%%%%%%%%%%%%%%
I use $\nM$, ${\nM}^{-1}$, $\nm$, ${\nm}^{-1}$ to denote metrics and inverses 
in direct analogy to how I previously defined $\mu$ and $\mu^{-1}$. 
I use $\fE_{\tH}$ because an energy-like separation constant $\fE_{\tL}$ will arise further on in the working.  
Then the fixed universe $\fE = \fE_{\tH} + \fE_{\tL}$.   
%%%%%%%%%%%%%%%%%%%%%%%%%%%%%%%%%%%%%%%%%%%%%%%%%%%%%%%%%%%%%%%%%%%%%%%%%%%%%%%%%%%%%%%%%%%%%%%%%%%%%%%%
}
\be
\fT_{\sH} = \mbox{ }_{\sM}||\dotb{\H}||^2
\mbox{ } , \mbox{ } 
\fT_{\sL} = \mbox{ }_{\sm}||\dotb{\L}||^2
\mbox{ } , \mbox{ } 
- \fU_{\sH} = \fV_{\sH} = \fV_{\sH}(|\H_{\ip}|, \H_{\jp}\cdot\H_{\kp}  \mbox{ alone} )
\mbox{ } , \mbox{ }
- \fU_{\sH} = \fV_{\sL} = \fV_{\sL}(|\L_{\ipp}|, \L_{\jpp}\cdot\L_{\kpp} \mbox{ alone} ) \mbox{ } ,
\ee
\be
- \fJ_{\sH\sL} = \fI_{\sH\sL} = \fI_{\sH\sL}(|\H_{\ip}|, |\L_{\ipp}|, \H_{\jp}\cdot\H_{\kp}, \L_{\jpp}\cdot\L_{\kpp}, 
\H_{\lp}\cdot\L_{\lpp} \mbox{ alone}) , 
\ee
(the `interaction potential' or `forcing term').  
Here, $| \mbox{ } |$ is the usual $\Re^3$ norm and $\cdot$ the corresponding inner product.  
The conjugate momenta are now 
\be
\nP^{\ip\alpha}_{\sH} = \frac{ M_{\ip}}{\mbox{n}}\dotb \nH_{\ip\alpha} = 
\frac{1}{2\mbox{n}}M^{\ip\jp\alpha\beta}\dotb \nH_{\jp\beta} 
\mbox{ } , \mbox{ }
\nP^{\ipp\alpha}_{\sL} = \frac{ m_{\ipp}}{\mbox{n}}\dotb \nL^{\ipp\alpha} = 
\frac{1}{2\mbox{n}}m^{\ipp\jpp\alpha\beta}\dotb \nH_{\jpp\beta} 
\mbox{ } ,
\label{HLmom}
\ee
where $\mbox{n} = \aha$.

%----------------------------------------------------------------------------------------------------
\noindent\underline{H-L split constraints.}
%----------------------------------------------------------------------------------------------------
%
%
%
The classical energy constraint now splits into 
\be
\ttH = \ttH_{\sH} + \ttH_{\sH\sL} = \fE_{\sH} \mbox{ } ,
\ee
for
\be
\ttH_{\sH\sL} \equiv \ttH_{\sL} + \fI_{\sH\sL} 
\mbox{ } , \mbox{ }
\ee
\be
\ttH_{\sH} \equiv \mbox{ }_{\sM^{-1}}||\nP_{\sH}||^2 + \fV_{\sH}
\mbox{ } , \mbox{ }
\ttH_{\sL} \equiv \mbox{ }_{\sm^{-1}}||\nP_{\sL}||^2 + \fV_{\sL} 
\mbox{ } . 
\ee
The classical ZAM constraint likewise splits into  
\be
\ttL^{\alpha} = {\ttL_{\sH}}^{\alpha} + {\ttL_{\sL}}^{\alpha} = 0 \mbox{ } , 
\ee
for
\be
{\ttL_{\sH}}^{\alpha} = {\epsilon^{\alpha\beta}}_{\gamma}\sum_{\ip = 1}^{p}\mbox{H}_{\ip\beta} \nP^{\ip\gamma}
\mbox{ } , \mbox{ } 
{\ttL_{\sL}}^{\alpha} = {\epsilon^{\alpha\beta}}_{\gamma}\sum_{\ipp = p + 1}^{n}\mbox{L}_{\ipp\beta} \nP^{\ipp\gamma}
\label{HLAM}
\ee
the H- and L-subsystems' angular momenta respectively.

%----------------------------------------------------------------------------------------------------
\noindent\underline{Approximate LMB time and its operational significance.} 
%----------------------------------------------------------------------------------------------------
%
%
%
The expression (\ref{LMB}) for emergent LMB time is now
\be
t^{\seB} - t^{\seB}(0) = 
\int\sqrt{        \frac{  _{\sM}||\di \nH||^2 +  
                          \mbox{}_{\sm}||\di \nL||^2  }
                       {    2\{\fE_{\sH} + \fU_{\sH} + \fU_{\sL} + \fJ_{\sH\sL}\}  }            } 
= t^{\seBH}
\left\{
1 + O
\left(
\epsilon_{\sH\sL}; \epsilon_{\sV}, \epsilon_{\sI}, \epsilon_{\sT} 
\right]
\right\}
\mbox{ } .
\label{mint}
\ee
for
\be
t^{\seBH} = \int\sqrt{         { _{{\sM}}||\di\nH||^2}/
                                     {  2\{\fE_{\sH} + \fU_{\sH}\}  }      }
\mbox{ } , 
\label{hint} 
\ee
\be
\left|
{\fV_{\sL}}/{\fV_{\sH}}
\right| 
= \epsilon_{\sV} 
\mbox{ }  \mbox{ (L-potential subdominance) }\mbox{ } ,
\label{epsV}
\ee
\be
\left|
{\fI_{\sH\sL}}/{\fV_{\sH}}
\right| 
= \epsilon_{\sI} 
\mbox{ } \mbox{ (interaction potential subdominance) }\mbox{ } , 
\label{epsI}
\ee 
\be
\left|
{\fT_{\sL}}/{\fT_{\sH}} 
\right|^2 = \epsilon_{\sT}  
\mbox{ } \mbox{ (L-kinetic subdominance) }\mbox{ } ,    
\label{epsT} 
\ee
all small.
I note that evaluating the integral in (\ref{hint}) provides an approximate LMB time standard.

However, I have found some reservations with LMB time or this approximation to it 
being put forward as  timefunctions.  
These are based on the following examples.  
[As these points concern the temporal 
part of relationalism, it does not matter that most of these examples are not spatially relational 
in the sense of this Paper].

In the Sun--Earth--Jupiter system, it has to be the Sun--Jupiter separation (or some very closely 
related coordinate) that is the H quantity and hence the approximate clock, rather than the more easily--observable 
Earth--Sun system.    
In this case $\fT$ and $\fV$ are dominated by their H-parts: 
using $m_{\Earth} \approx 
m_{\Jupiter}/300 \approx m_{\Sun}/300000$, the Jacobi masses go as 
$m_{\Earth}\{1- m_{\Earth}/m_{\Sun}\}$ and 
$m_{\Jupiter}\{1 - m_{\Jupiter}/m_{\Sun} + \{m_{\Jupiter} m_{\Sun}\}^2\}$;   
using  additionally near-circularity  and near-heliocentricity of the orbits  
alongside Kepler's third law to compare $\d \nR_i^2 \approx \nR_i^2\d\theta_i^2$, $i$ = Earth, Jupiter, 
the leading errors for neglecting 
the Earth are 1 part in 250 for both the kinetic term and the potential term approximations.  
But, these errors due to neglecting the Earth are sizeable compared to even vaguely modern standards 
of precision in astronomy.  
Worse, in the Sun--Jupiter--nearby star system, 
it is the Sun--nearby star separation that cannot be neglected.  
So, the H subsystem dominates the timestandard and may not be convenient to measure accurately.  
As regards Barbour calling LMB time `ephemeris time',       
while {\it ``the astronomical ephemeris expresses in numbers the actual state of the celestial sphere 
at given instants of time"} \cite{Chauvenet}, the cited source (and also e.g. \cite{Alma}) makes 
it clear that the actual such quantity that astronomers used was obtained by an elaborate iteration 
process so as to fit the positions of the defining bodies roughly within the error bounds on their 
positions, a process that neither involves reading time off an action nor is succeptible to errors 
anything like the size of those of this section as regards omitting various intuitively 
minorly-contributing bodies.  
Certainly alpha centauri doesn't enter the conventional list of input 
bodies and yet one still obtains a nicely accurate notion of time for the solar system...  
Thus, while LMB time is of ephemeris type (insofar as it is a universal time), 
is by no means {\sl the} ephemeris time notion used in astronomy, and would appear to be 
substantially inferior to it as regards setting up time standards for quasi-isolated subsystems.

Furthermore, in the second example above, the potential is dominated by the `L' Sun--Jupiter term, 
on account of $|\nL| << |\nH|$.   
Note also that approximations at the level of forces are well different 
from those at level of potential if $\R_1$ and $\R_2$ are very different in size.  
Moreover, as forces act in different directions, 
one can get force balances, unlike the strictly additive superposition of 
Newtonian gravitational potentials through these all having the same sign.    
Together, these things make approximations at the level of the forces more `intuitive' than 
at the $\fT$, $\fV$ level.

Due to this Paper's QM applications (see esp. Sec 8), I also consider a coupled pair of HO's in which 
one greatly dominates.  
Then for the various approximations to hold simultaneously, $M/m(\d\nH/\d\nL)^2 >> 1$ and 
$M\Omega^2\nH^2/m\omega^2\nL^2 >>1$ are required.   
Using that H, L both roughly undergo simple harmonic motion, the first of these conditions 
amounts to $M\Omega^2 \nH_0^2/m\omega^2\nL_0 >> 1$.  
This illustrates that some generally-distinct assumptions 
(here $\epsilon_{\sV}$ small and $\epsilon_{\sT}$ small) 
can coincide in specific examples.

%====================================================================================================
\noindent\underline{A more general critique of the approximations used so far.}
%====================================================================================================
%
%
%
If $M = M(R_{\alpha i})$, or more generally some (more GR-like) $G^{\alpha\beta ij}(\mbox{R}_{\gamma i})$  
supplants the diagonal ${2\mu_i}\delta^{\alpha\beta}\delta_{ij}$ as mass matrix  
(which often corresponds to the configuration space being curved), 
then the H--L designation ($\epsilon_{\sH\sL}$ small) may only hold patchwise.  
Another issue is that variations in `curvature', which roughly goes as 
$\d^2\nH_{\alpha\ip}/\d\nL_{\beta\ipp}^2$ over the configuration space 
allow for changes in $\d\nL_{\alpha\ipp}/\d\nH_{\beta\ip}$. 
This sharpens one part of Barbour's assertion of the importance of the configuration space geometry 
\cite{EOT}. 
Also note that it is easy to contrive for different $\fV_{\sR_i}$ to dominate in different regions 
of configuration space.  
E.g., for the Newton--Coulomb potentials, $\fV_{\sR_1}$ will dominate 
for $||\nR_1||$ sufficiently smaller than $||\nR_2||$ while $\fV_{\sR_2}$ will dominate if vice versa.  

%\vspace{4in}

%==========================================================================================================================================
%==========================================================================================================================================
%=====================================================================================================
%\macrosection{The semiclassical approach}
%=====================================================================================================
%==========================================================================================================================================
%==========================================================================================================================================

%=====================================================================================================
%=====================================================================================================
\section{Quantized H-L split RPM}
%=====================================================================================================
%=====================================================================================================

So as to provide a self-contained account of the semiclassical approach, I provide the following 
standard-type diagram, which `commutes' and is without operator ordering ambiguity \cite{OI}).  
$$
\fS[\nR_{\alpha i}, \dot{\nR}_{\alpha i},  \dot{\nb}_{\alpha}] 
\hspace{0.65in}
\stackrel{\mbox{\scriptsize H--L split}}{\longrightarrow} 
\hspace{0.5in}
\fS[\nH_{\alpha\ip}, \nL_{\alpha\ipp}, 
\dot{\nH}_{\alpha\ip}, \dot{\nL}_{\alpha\ipp}, \dot{\nb}_{\alpha}] 
\hspace{0.7in}
$$
$$
\stackrel{\mbox{\scriptsize variation, inspection}}{\mbox{\scriptsize of momenta}} 
\downarrow 
\hspace{2in} 
\downarrow
\stackrel{\mbox{\scriptsize variation, inspection}}{\mbox{\scriptsize of momenta}} \hspace{1.3in}
$$
$$
\hspace{0.4in}
\left(
\stackrel{\ttH}{\ttL^{\alpha}}
\right)
(\nR_{\alpha i}, \nP^{\alpha i}) = 
\left(
\stackrel{\fE}{0}
\right)
\hspace{0.4in}
\stackrel{\mbox{\scriptsize H--L split}}{\longrightarrow} 
\hspace{0.6in}
\left(
\stackrel{\ttH}{\ttL^{\alpha}}
\right)
(\nH_{\alpha\ip}, \nL_{\alpha\ipp}, \nP^{\alpha\ip}_{\sH}, \nP^{\alpha\ipp}_{\sL} ) = 
\left(
\stackrel{\fE_{\sH}}{0}
\right)         \hspace{0.65in}
$$
$$
\stackrel{        \mbox{\scriptsize position representation quantization}        }
         {        \mbox{\scriptsize$(\nR_{\alpha i}, \nP^{\alpha i}) \mapsto 
          (\hat{\nR}_{\alpha i}, \hat{\nP}^{\alpha i}) = 
          (\nR_{\alpha i}, {{\pa}_{\tR}}^{\alpha i})$}        } 
\downarrow 
\hspace{1.95in}
\downarrow 
\stackrel{     \mbox{\scriptsize position representation quantization 
                $(\nH_{\alpha\ip}, \nL_{\alpha\ipp}, \nP_{\tH}^{\alpha\ip}, \nP_{\tL}^{\alpha\ipp})$}     } 
         {     \mbox{\scriptsize $\mapsto 
(\hat{\nH}_{\alpha\ip}, \hat{\nL}_{\alpha\ipp}, \hat{\nP}_{\tH}^{\alpha\ip}, \hat{\nP}_{\tL}^{\alpha\ipp})   
  =    (\nH_{\alpha\ip}, \nL_{\alpha\ipp}, -i\hbar{{\pa}_{\tH}}^{\alpha \ip}, 
                                        -i\hbar{{\pa}_{\tL}}^{ \alpha\ipp}   )$}     }        
\hspace{0.1in}
$$
\be
\hspace{0.8in}
\left(
\stackrel{    \ttH    }{    \ttL^{\alpha}    }
\right)
(\nR_{\alpha i}, {\pa_{\sR}}^{\alpha i}    ) = 
\left(
\stackrel{\fE}{0}
\right)
\hspace{0.5in}
\stackrel{    \mbox{\scriptsize H--L split}    }{    \longrightarrow    } 
\hspace{0.5in}
\left(
\stackrel{    \ttH    }{    \ttL^{\alpha}    }
\right)
(\nH_{\alpha\ip}, \nL_{\alpha\ipp}, {\pa_{\sH}}^{\alpha\ip}    , 
                                    {\pa_{\sL}}^{\alpha\ipp} ) = 
\left(
\stackrel{\fE_{\sH}}{0}
\right) \mbox{ } . \hspace{0.51in}
\ee
So, by whichever path, the quantum energy constraint is 
\be
\hat{\ttH}\Phi = \hat{\ttH}_{\sH}\Phi + \hat{\ttH}_{\sH\sL}\Phi = \fE_{\sH}\Phi
\ee
for
\be
\hat{\ttH}_{\sH\sL} \equiv \hat{\ttH}_{\sL} + \fI_{\sH\sL} 
\mbox{ } , \mbox{ } 
\hat{\ttH}_{\sH} \equiv - \hbar^2\mbox{}_{\sM^{-1}}||\pa_{\sH}||^2 + \fV_{\sH}
\mbox{ } , \mbox{ }
\hat{\ttH}_{\sL} \equiv - \hbar^2\mbox{}_{\sm^{-1}}||\pa_{\sL}||^2 + 
\fV_{\sL} 
\mbox{ } . \hspace{0.05 in}
\label{full}
\ee
I supplement the above standard split with the corresponding split of the quantum ZAM constraint 
that my models require: 
\be
\hat{\uttL}\Phi = \hat{\uttL}_{\sH}\Phi + \hat{\uttL}_{\sL}\Phi = 0
\label{QAM} 
\ee
for
\be
{\hat{\ttM}_{\sH}}^{\alpha} = 
\frac{\hbar}{i}{\epsilon^{\alpha\beta}}_{\gamma}
\sum_{\ip = 1}^{p}\nH_{\beta\ip} \frac{\pa}{\pa \nH^{\gamma\ip} }
\mbox{ } , \mbox{ } 
{\hat{\ttM}_{\sL}}^{\alpha} = 
\frac{\hbar}{i}{\epsilon^{\alpha\beta}}_{\gamma}
\sum_{\ipp = p + 1}^{n}\nL_{\beta\ipp}  \frac{\pa}{\pa \nL^{\gamma\ipp} } \mbox{ } .  
\label{HLAMQ}
\ee
I next lay down the standard semiclassical approach ans\"{a}tze and approximations for the RPM, 
alongside objections to using these in the present closed universe toy model context.    
I form `less approximate' equations first, to make it clear what further approximations 
are required to go between these and more standard, more approximate forms, and also to keep 
explicit track of these smallnesses and any inter-relations between them.

%====================================================================================================
%====================================================================================================
\section{The Born--Oppenheimer (BO) procedure} 
%====================================================================================================
%====================================================================================================

By this, I mean the BO ansatz for the wavefunction, 
\be
\Phi = \phi(\H_{\jp})
|\chi_{\sfj}(\H_{\jp}, \L_{\kpp})\rangle
\mbox{ } ,
\label{BO}
\ee  
alongside a package of approximations conventionally made alongside it, 
only one of which is {\sl the} BO approximation.

The inner product used below is $\langle\chi|\chi^{\prime}\rangle = 
\int \d\L_{\kpp} \chi^*(\H_{\ip}, \L_{\kpp})\chi^{\prime}(\H_{\ip}, \L_{\kpp})$ 
for $^*$ the complex conjugate.  
$\langle{O}\rangle$ denotes the expectation $\langle\chi|\hat{O}|\chi\rangle$, 
$\overline{O}$ denotes the fluctuation $O - \langle O \rangle$, and I use 
\be
{D_{\sH}}^{\alpha\ip} = {\pa_{\sH}}^{\alpha \ip} + i{A_{\sH}}^{\alpha\ip} \mbox{ } ,
\label{Bercov}
\ee
to denote the Berry covariant derivative, and 
\be 
{D_{\sH}^*}^{\alpha\ip} = {\pa_{\sH}}^{\alpha \ip} - i{A_{\sH}}^{\alpha\ip} \mbox{ } 
\label{covstar}
\ee
for its conjugate.  
Here, ${A_{\sH}}^{\alpha\ip}$ is the Berry connection \cite{Berry84, Simon83}, 
i.e. the vector gauge potential induced by L-physics on H-physics of a nondegenerate quantum state 
that corresponds to its U(1) freedom in phase, 
\be
{A_{\sH}}^{\alpha\ip} = -i\left< \chin  \right| {\pa_{\sH}}^{\alpha\ip} \left|\chin \right>  
% = - i\int \d \nL_{\gamma \kpp} \chin^*(\H_{\jp}, \L_{\jpp})\pa_{\sH}^{\alpha\ip}\chin(\H_{\jp}, \L_{\jpp}) 
\mbox{ } .
\ee

I find that an efficient way to proceed is to establish the following identities from definitions, 
linearity and the Leibniz rule:   
\be
_{\sM^{-1}}||\pa_{\sH}||^2 \lfloor \theta\psi \rfloor = 
\psi \mbox{ }_{\sM^{-1}}||\pa_{\sH}||^2 \theta + 2_{\sM^{-1}}(\lfloor \pa_{\sH} \theta\rfloor, \pa_{\sH}\psi) +
\theta \mbox{ }_{\sM^{-1}}||\pa_{\sH}||^2 \psi 
\mbox{ } ,
\label{1}
\ee
\be
_{\sM^{-1}}||D_{\sH}||^2 \theta = 
\mbox{ }_{\sM^{-1}}||\pa_{\sH}||^2 \theta + 
2i_{\sM^{-1}}(A_{\sH}, \pa_{\sH}) \theta  +
i\theta _{\sM^{-1}}( \pa_{\sH}, A_{\sH} ) -
\mbox{ }_{\sM^{-1}}||A_{\sH}||^2\theta
\label{2}
\mbox{ } ,
\ee
\be
_{\sM^{-1}}||D_{\sH}^*||^2 \theta = 
\mbox{ }_{\sM^{-1}}||\pa_{\sH}||^2 \theta - 
2i_{\sM^{-1}}(A_{\sH}, \pa_{\sH}) \theta -
i\theta _{\sM^{-1}}( \pa_{\sH}, A_{\sH} ) -
\mbox{ }_{\sM^{-1}}||A_{\sH}||^2\theta
\mbox{ } ,
\label{3}
\ee
\be
_{\sM^{-1}}(\lfloor D_{\sH}\theta \rfloor, D^*_{\sH}\psi) = 
\mbox{}_{\sM^{-1}}(\lfloor \pa_{\sH}\theta \rfloor, \pa_{\sH}\psi) + 
\mbox{ }_{\sM^{-1}}||A_{\sH}||^2\theta\psi + 
i\mbox{}_{\sM^{-1}}(A_{\sH},\theta\pa_{\sH} \psi - \psi \pa_{\sH}\theta) 
\mbox{ } .
\label{4}  
\ee

Using additionally that $|\chi\rangle$ is normalized, and the H-derivative of this condition, the result 
\cite{Berry89}
\be
\langle \chi | _{\sM^{-1}}||\pa_{\sH}\mbox{}^2||  \lfloor \phi | \chi \rangle \rfloor = 
\mbox{ }_{\sM^{-1}}||D_{\sH}||^2\phi - e\phi
\label{5}
\ee
can be established, where the {\it generalized electric term} $e$ is the M-trace of 
{\it Berry's quantum geometric tensor} \cite{Berry89},
\be
<Q^{\alpha\beta\ip\jp}> = 
\mbox{Re}\{\langle \lfloor{\pa_{\sH}}^{\alpha \ip}\chi\rfloor| 
\{ 1 - P_{\chi} \}
\lfloor {\pa_{\sH}}^{\beta\jp} | \chi \rfloor \rangle \}\phi \mbox{ } ,  
\ee
and $P_{\chi}$ is the projector $|\chi\rangle\langle\chi|$.  
The above is Berry's \cite{Berry89} form for $e$, as opposed to the form for $e$ in e.g. \cite{BV89} 
which is related to it by the identity 
\be
\mbox{M}_{\alpha\beta\ip\jp} \langle Q^{\alpha\beta\ip\jp} \rangle = - \langle _{\sM^{-1}}||D^*_{\sH}||^2 \rangle \mbox{ } , 
\label{6}
\ee
which follows by (\ref{3}), and normalization alongside its H-derivative.  
A final useful identity is 
\be
|\chi\rangle \mbox{}_{\sM^{-1}}||D_{\sH}||^2\phi + 
2 \mbox{}_{\sM^{-1}}\left(\lfloor D_{\sH}\phi \rfloor , D^*_{\sH} | \chi\rangle \right) + 
\phi_{\sM^{-1}}||D^*_{\sH}||^2 |\chi \rangle = 
\mbox{ }_{\sM^{-1}}||\pa_{\sH}||^2\lfloor\phi|\chi\rfloor\rangle \mbox{ } .  
\label{7}
\ee
It is key for various of these identities that the array therein is to depend on $\H_{\ip}$ alone, so 
that one can take it in and out of the $\langle \mbox{ } | \mbox{ } | \mbox{ } \rangle$.

\mbox{ } 
%
%***Layout option: space suppressible.

Next, I provide for ease of reference the first few energy constraint equations of this formalism, and how they are inter-related.  
Many of the steps involve neglecting a plethora of small quantities $\epsilon$.  
To not be obstructive, I postpone most of the specification of what these are until the end of this Section.  
$$
\stackrel{\underline{\mbox{preliminary}}}{\underline{\mbox{equations}}} 
\hspace{0.05in}
\stackrel{\mbox{$-\hbar^2\mbox{}_{\sM^{-1}}||\pa_{\sH}||^2\lfloor | \chi \rangle\phi\rfloor$}  }
         {+\hat{h}|\chi\rangle\phi = \fE_{\sH}|\chi\rangle\phi} 
\hspace{0.05in}
\stackrel{\mbox{step A}}{\longrightarrow}
\hspace{0.05in}
\stackrel{\mbox{$-\hbar^2
\left\{
|\chi\rangle \mbox{}_{\sM^{-1}}   ||\pa_{\sH}||^2    \phi + 
2_{\sM^{-1}}(\lfloor\pa_{\sH}\phi\rfloor , \pa_{\sH}| \chi\rangle)  
\right.$}                                   }                                             
{+ 
\left.
\phi_{\sM^{-1}}  ||\pa_{\sH}||^2|\chi\rangle                             
\right\}
+
\hat{h}|\chi\rangle\phi = \fE_{\sH}|\chi\rangle \phi   }                                                                                                                     
\hspace{0.05in}
\stackrel{\mbox{step B}}{\longrightarrow}
\hspace{0.05in}
\stackrel{\mbox{$-\hbar^2|\chi\rangle_{\sM^{-1}}||\pa_{\sH}||^2\phi$}  }
         {+ \hat{h}|\chi\rangle = \fE_{\sH}|\chi\rangle\phi}
$$
$$
\hspace{1.4in} \mbox{step D} \downarrow \hspace{3in} \hspace{1.1in} \mbox{step C} \downarrow \hspace{1.5in} 
$$
$$
\underline{\mbox{H-equations}}
\stackrel{\mbox{$-\hbar^2\langle\chi|_{\sM^{-1}}||\pa_{\sH}||^2 \lfloor |\chi \rangle \phi \rfloor$}}
         {+ o\phi = \fE_{\sH}\phi                                                }
\hspace{1.5in}
\stackrel{\mbox{(Born--Oppenheimer}}
         {\mbox{equation \cite{BO27})}}
-\hbar^2\mbox{}_{\sM^{-1}}||\pa_{\sH}||^2\phi + o\phi = \fE_{\sH}\phi 
$$
$$
\hspace{1.6in} \mbox{step E} \downarrow \hspace{1.5in} \hspace{1.5in} \hspace{1.5in} 
\uparrow \hspace{1.5in} 
$$
$$
- \hbar^2\mbox{}_{\sM^{-1}}||D_{\sH}||^2\phi + \{e + o\}\phi = \fE_{\sH}\phi 
\stackrel{\mbox{(Berry \cite{Berry89}}}{\mbox{ H-equation)}}
\hspace{3.2in}  \uparrow \hspace{0.4in}
$$
$$
\hspace{1.6in} \mbox{step F} \downarrow \hspace{1.5in} \hspace{1.5in} \hspace{1.5in} \uparrow \hspace{1.5in} 
$$
\be
 - \hbar^2\mbox{}_{\sM^{-1}}||D_{\sH}||^2\phi + o\phi = \fE_{\sH}\phi 
\stackrel{\mbox{(Mead--Truhlar \cite{MT79} H-equation}}
{\mbox{in Berry--Simon \cite{Berry84, Simon83} geometrical form)}}
\hspace{0.4in} \stackrel{\mbox{Step G}}{\longrightarrow} \hspace{0.3in} \longrightarrow
\label{boarray} 
\mbox{ } .
\ee
%*editorial: join up arrows in bottom RH corner.  

By passage from preliminary equations (line 1) to H-equations (which covers both steps C and D), 
what is meant is 1) defining 
\be
o = \langle\chi|\hat{h}|\chi\rangle
\ee 
(which may also be regarded as the `$\H_{\ip}$ parameter dependent eigenvalue' of 
$\hat{h} \equiv \ttH_{\sH\sL} + \fV_{\sH}$, 
\be
\hat{h}(\H_{\ip}, \L_{\ipp}, \P_{\sL}^{\ipp}) | \Phi(\H_{\jp}, \L_{\jpp}) \rangle = 
o(\H_{\ip}) | \Phi(\H_{\jp}, \L_{\jpp}) \rangle \mbox{ }  ) \mbox{ } .  
\ee
2) Premultiplication by $\langle \chi |$, the acceptability of which is underlied by a 
`diagonal dominance' condition i.e. that the diagonal terms are larger than the 
$o_{\sfj\sfl}$ nondiagonal-terms, which are similarly-defined but now contain distinct 
$\langle\chi_{\sfj}|$, $|\chi_{\sfl}\rangle$):   
\be
\mbox{for } \fj \neq \fl \mbox{ } , 
\left|o_{{\sfj}{\sfl}}/o_{{\sfj}{\sfj}}  
\right|
= \epsilon_{\sBO} \mbox{ , small (BO approximation) }.
\ee
3) Making use of the normalization of $|\chi\rangle$.

Note that the diagram covers all of: BO's scheme ABC, Berry's scheme DE and the recovery of 
BO's scheme within Berry's scheme, FG.  
The merits of Berry's scheme as opposed to BO's are that it includes back-reaction -- 
important for the reasons given in the Introduction -- and moreover does so in a 
geometrically insightful way which is more precise and indeed more correct in laboratory situations 
(effects have been observed \cite{LH}, the explanation of which has been found to 
rest on geometric phase \cite{MT79, Berry84, Simon83}).   
BO's scheme ABC involves, respectively: 
expanding by (\ref{1}),  
two adiabatic neglects $\epsilon_{\sa \sw 3}$, $\epsilon_{\sa \sw 7\scr}$ small, 
and finally the above-described step C.  
Berry's move E is via identity (\ref{5}), and amounts to casting the H-equation in a geometrical form.
The context for this is an adiabatic loop in phase space, 
whence this scheme is underlied by being in a {\it classically-adiabatic} regime
(i.e. that classical H-processes are much slower than classical L-processes):   
\be
{\Omega_{\sH}}/{\omega_{\sL}} = \epsilon_{\sa} 
\ee
for `characteristic frequencies' $\Omega_{\sH}$ and $\omega_{\sL}$. 
%
%[If required, this definition can be done up more rigorously.]  
%
The suffix a indeed stands for `adiabatic', and is much used below as 
there are numerous diferent adiabatic approximations at the quantum level.  
Moves D and E can be encapsulated together as another `diagonal dominance' condition,
\be
\mbox{ for } \fl \neq \fj \mbox{ } , \mbox{ } 
\epsilon_{\sd\sBOB} = \left| \{o_{{\sfj}{\sfl}} + e_{{\sfj}{\sfl}}\}/ 
\{o_{{\sfj}{\sfj}} + e_{{\sfj}{\sfj}}\}
\right| 
\mbox{ , small .} 
\ee

Move F recovers the Mead--Truhlar equation (already known from molecular physics) via  
considering the quantum correction potential $e$ to be dominated usually by $o$ 
but just as well by $\mbox{}_{\sM^{-1}}||\pa_{\sH}||^2|\chi\rangle$: neglecting 
$\epsilon_{\sa\sw 1}$ and $\epsilon_{\sa\sw 2}$.  
Then recovering the BO equation by move G involves neglecting compensatorily $\epsilon_{\sa \sw 1}$, 
$\epsilon_{\sa \sw 2}$ (such that the whole of FG does not require these approximations to be made), and 
also neglecting $\epsilon_{\sa \sw 4}$ and $\epsilon_{\sa \sw 8}$.  
The last 2 of these are in close correspondence with the terms neglected in move B.   
However, arriving at the BO equation via Berry's route does require more work, 
reflecting that making BO's adiabatic assumptions and forming H-equations 
are non-commuting procedures.

Note that the above scheme is for H-states that are assumed to be nondegenerate, 
a yet more elaborate scheme is in general required so as to include degenerate ones.\foo{ 
%%%%%%%%%%%%%%%%%%%%%%%%%%%%%WILCZEK ZEE FOOTNOTE%%%%%%%%%%%%%%%%%%%%%%%%%%%%%%%%%%%%%%
Namely, the scheme based on the Wilczek--Zee \cite{WZ, Berryrev} connection 
\be
A_{{\sfj}}^{\alpha\ip\hat{\sfp}\hat{\sfq}} = -i
\left< {\chi}_{\sfj}\mbox{}_{\hat{\sfp}}|  {{\pa}_{\tH}}^{\alpha\ip} 
|{\chi}_{\sfj}\mbox{}_{\hat{\sfq}}\right>  
\mbox{ } ,
\ee
i.e. the non-Abelian vector gauge potential induced by L-physics on H-physics of a degenerate 
quantum state, corresponding to the now enlarged, U(D), freedom in phase.
Here, $\fj$ denotes the $\fj$th eigenstate, of degeneracy D indexed with the hatted indices. 
One then considers $\Phi = \phi(\H_{\jp})|\chi_{\sfj}\mbox{}_{\hat{\sfp}}(\H_{\jp}, \L_{\kpp})\rangle$ 
and then projects onto a distinct $\langle\chi_{\sfj}\mbox{}_{\hat{\sfq}}(\H_{\jp}, \L_{\kpp})|$.} 
%%%%%%%%%%%%%%%%%%%%%%%%%%%%%%%%%%%%%%%%%%%%%%%%%%%%%%%%%%%%%%%%%%%%%%%%%%%%%%%%%%%%%%%

Also note that Berry's H-equation contains 3 types of back-reaction terms: 
the connections in the first term, 
averaged terms included within the $o$, and 
the `electric potential' term $e$.  
In the conventional QM situation, it makes sense to talk then of 
forces associated with quantum back-reactions.  
These may be read off a form of Newton's second law 
\be
\ddot{\nH}_{\alpha\ip} = \frac{1}{M_{\ip}}
\left\{
\delta_{\ip\jp}{B_{\alpha\beta}}^{\jp\lp}(\H_{\kp})\dot{H}_{\lp}\mbox{}^{\beta} - 
\pa_{\sH\alpha\ip}
\{
e(\H_{\kp}) + o(\H_{\kp})
\}
\right\} \mbox{ } , 
\ee
which arises from coupling Hamilton's equations for the H-subsystem.  
These forces are, respectively, 
a `magnetic gauge force' based on a `magnetic field' particle-index 2-form
\be
B_{\alpha\beta}\mbox{}^{\jp\lp} = 2\hbar{\pa}_{\sH\alpha}\mbox{}^{[\jp}A_{\sH\beta}\mbox{}^{\lp]} 
\mbox{ } , 
\ee
an `electric force' and a `BO type force' (which includes 
$\pa_{\sH}\langle \chi|\fV_{\sH}| \chi \rangle$, which contains 
$-\langle \chi | \chi \rangle \pa_{\sH}\fV_{\sH} = - \pa_{\sH}\fV_{\sH}$ 
-- the classical H-force -- alongside back-reactions arising from $\fI_{\sH\sL}$  
and further quantum corrections).

\noindent\underline{L-equations.}
One can then consider 
$
\{ \mbox{preliminary equation}\} - \{\mbox{H-equation}\}|\chi\rangle  
$,
which is prima facie a fluctuation equation.  
From the top left-hand side preliminary equation in (\ref{boarray}) and the Berry version of 
the H-equation, this takes the form 
\be
\frac{1}{\hbar^2}\overline{\hat{h}} = \mbox{ }_{\sM^{-1}}||\pa_{\sH}||^2\lfloor \phi|\chi\rangle \rfloor 
-  \left\{ \mbox{ }_{\sM^{-1}}||D_{\sH}||^2\lfloor\phi\rfloor 
+  \langle _{\sM^{-1}}||D^{*}_{\sH}||^2\lfloor\phi\rfloor\rangle 
\right\}
|\chi\rangle \mbox{ } .  
\label{FLE}
\ee
An alternative form, obtained by rearrangement by (\ref{7}) is  
\be
\left\{\overline{\frac{1}{\hbar^2}\hat{h} - \mbox{ }_{\sM^{-1}}||D^*_{\sH}||^2 } \right\}|\chi\rangle = 
\frac{2}{\phi}\mbox{}_{\sM^{-1}}(\lfloor D_{\sH}\phi\rfloor, D_{\sH}^*|\chi\rangle) 
\mbox{ } .
\ee

Up to this point, one can take the working to be standard practice in QM, in which  
it is equivalent throughout to working from a TDSE input, 
while one has from the outset also a TDSE for the L-subsystem (with respect to external time).\foo{While 
%%%%%%%%%%%%%%%%%%%%%%%%%%%%%%%%%%%%%%%%%%%%%%%%%%%%%%%%%%%%%%%%%%%%%%%%%%%%%%%%%%%%%%%%%%%%%%%%%%%%
the Berry phase scheme usually starts from an external TDSE, starting from a TISE gives an analogous working.  
Further aspects in which this analogy breaks down are considered in Sec 7.}  
%%%%%%%%%%%%%%%%%%%%%%%%%%%%%%%%%%%%%%%%%%%%%%%%%%%%%%%%%%%%%%%%%%%%%%%%%%%%%%%%%%%%%%%%%%%%%%%%%%%%%
%
The quantum cosmology counterpart of path ABC but with the `chroniferous' cross-term kept 
is used in e.g. Lapchinsky--Rubakov \cite{LR79}, Banks \cite{Banks} and 
Halliwell--Hawking \cite{HallHaw}.  
It is this keeping of cross-terms that opens the way in quantum cosmology to having an emergent 
time in place of an external time, via the further WKB assumption of Sec 6 and the rearrangement 
detailed in Sec 7.  Such a working gives a novel time-dependent wave equation for the L-subsystem, with respect to an 
(approximate) time induced by the H-subsystem.  
At a less crude level, Brout and collaborators (see e.g. \cite{BV89}), and Kiefer 
(see e.g. \cite{Kieferbook}), work along path DEF, i.e. with back-reaction 
terms considered, at least at the outset.  
Doing this in no way affects the procedure by which keeping the cross-term 
and using the WKB ansatz lead to an emergent time.

For the below discussion of which approximations one may attempt to use on the L equation, 
it is useful to provide the expanded-out version of this equation: 
$$
\frac{1}{\hbar^2}
\left\{
\stackrel{\mbox{$- \hbar^2\mbox{}_{\sm^{-1}}||\pa_{\sL}||^2$}}
{+ \hbar^2 \langle\mbox{$ \mbox{}_{\sm^{-1}}||\pa_{\sL}||^2$}\rangle}
\stackrel{\mbox{$+ \fV_{\sH}$}} 
{-\langle\mbox{$ \fV_{\sH}$}\rangle}
\stackrel{\mbox{$+ \fV_{\sL}$}}
{-\langle\mbox{$ \fV_{\sL}$}\rangle} 
\stackrel{\mbox{$+ \fI_{\sH\sL}$}}
{-\langle\mbox{$ \fI_{\sH\sL}$}\rangle}
\right\} 
|\chi\rangle - 
\left\{
\stackrel{\mbox{$\mbox{ }_{\sM^{-1}}||\pa_{\sH}||^2$}} 
{-\langle\mbox{$\mbox{ }_{\sM^{-1}}||\pa_{\sH}||^2$}\rangle} 
\stackrel{\mbox{$- 2i_{\sM^{-1}}(A_{\sH}, \pa_{\sH})$}}
{+ 2i\langle\mbox{$ \mbox{}_{\sM^{-1}}(A_{\sH}, \pa_{\sH})$}\rangle}
\stackrel{\mbox{$- i_{\sM^{-1}}(\lfloor\pa_{\sH}, A_{\sH}\rfloor)$}} 
{+\langle\mbox{i$ \mbox{}_{\sM^{-1}}(\lfloor\pa_{\sH}, A_{\sH}\rfloor)$}\rangle} 
\stackrel{\mbox{$- \mbox{}_{\sM^{-1}}||A_{\sH}||^2$}}
{+\langle\mbox{$ \mbox{}_{\sM^{-1}}||A_{\sH}||^2$}\rangle}
\right\}
|\chi\rangle
$$
\be
= \frac{2}{\phi}
\left\{
\mbox{ }_{\sM^{-1}}(\lfloor \pa_{\sH}\phi\rfloor, \pa_{\sH}|\chi\rangle) 
+ \mbox{ }_{\sM^{-1}}||A_{\sH}||^2\phi|\chi\rangle 
+ i_{\sM^{-1}}(A_{\sH}, \phi\pa_{\sH}|\chi\rangle - |\chi\rangle \pa_{\sH}\phi)
\right\}
\mbox{ }.
\ee
Note that the second, sixth and seventh columns of the left-hand side 
cancel out because $\langle \mbox{ } | \mbox{ } | \mbox{ } \rangle$ is an L-integral so weightings that 
are functions of H alone can be pulled outside.  
[On the other hand, the corresponding expansion of the Berry H-equation 
merely involves applying (\ref{2}) to it, so I do not provide it.]

%========================================================================================================
\noindent\underline{Corresponding ZAM equations.}  
%========================================================================================================
%
%
%
The above scheme serves for absolute mechanics or trivially relational mechanics.  
One of this Paper's new contributions is the semiclassical treatment of the additional linear constraints 
that RPM's have (so these models serve for more detailed modelling of the situation in GR).   
I treat these linear constraints according to the natural approach which parallels  
with how the quadratic energy constraint is treated.  
The analogue of the above `cycle' is then
$$ 
\stackrel{\mbox{\underline{preliminary}}}{\mbox{\underline{ZAM equations.}}}
\stackrel{      \mbox{$\frac{\hbar}{i}
\left\{
\sum_{\ip}
\H_{\ip} \cr {\underline{\pa}_{\sH}}^{\ip}\phi + 
\right.$}      }
{      \mbox{$   
\left.
\sum_{\ipp} \L_{\ipp} \cr {\underline{\pa}_{\sL}}^{\ipp}
\right\}
\phi|\chi\rangle = 0$}      } 
\stackrel{\mbox{step A}}{\longrightarrow}
\stackrel{       \mbox{$\frac{\hbar}{i}\sum_{\ip}
\left\{
\H_{\ip} \crs \lfloor{\underline{\pa}_{\sH}}^{\ip}\phi \rfloor |\chi\rangle+\phi\H_{\ip} \crs 
{\underline{\pa}_{\sH}}^{\ip}|\chi\rangle
\right\}$}      }
{      \mbox{$ + 
\frac{\hbar}{i}\phi \sum_{\ipp} \L_{\ipp} \crs {\underline{\pa}_{\sL}}^{\ipp}|\chi\rangle = 0$}} 
\stackrel{\mbox{step B}}{\longrightarrow}
\stackrel{\mbox{$\frac{\hbar}{i}\sum_{\ip}
\H_{\ip} \crs \lfloor{\underline{\pa}_{\sH}}^{\ip}\phi \rfloor |\chi\rangle $}     }
{       \mbox{$  + 
\frac{\hbar}{i}\phi \H_{\ip} \cr {\underline{\pa}_{\sH}}^{\ip}|\chi\rangle = 0$}} 
\mbox{ } 
$$
$$
\hspace{1.6in} \mbox{Step D} \downarrow \hspace{1.5in} \hspace{1.5in} \hspace{1in} \mbox{step C } \downarrow \hspace{1.5in} 
$$
$$
\stackrel{\mbox{\underline{ZAM}}}{\mbox{\underline{H-equations.}}}
\frac{\hbar}{i}\sum_{\ip}
\left\{
\H_{\ip} \cr {\underline{\pa}_{\sH}}^{\ip}\phi + 
\langle\chi|\H_{\ip}\cr\pa_{\sH}\rangle\phi
\right\}
+ \langle \underline{\ttM}_{\sL} \rangle \phi = 0
\hspace{1.6in}
\frac{\hbar}{i}\sum_{\ip}
\H_{\ip} \cr {\underline{\pa}_{\sH}}^{\ip}\phi + 
\langle \underline{\ttM}_{\sL} \rangle \phi = 0
\mbox{ } 
$$
$$
\hspace{1.6in} \mbox{Step E} \downarrow \hspace{1.5in} \hspace{1.5in} \hspace{1.05in} \mbox{step F} \uparrow \hspace{1.5in} 
$$
\be
\frac{\hbar}{i}\sum_{\ip}
\H_{\ip} \cr {\underline{D}_{\sH}}^{\ip}\phi + 
\langle \underline{\ttM}_{\sL} \rangle \phi = 0 \mbox{ (geometric form ) }
\hspace{1.15in} \longrightarrow \hspace{1.15in} \longrightarrow 
\ee
%*editorial: join bottom RH arrows.  

By passage from preliminary equations (line 1) to H-equations (which covers both steps C and D), 
what is meant is: 

\noindent 1) define 
$
m_{\sfj\sfl} = \langle\chi_{\sfj}|\hat{\ttM}|\chi_{\sfl}\rangle
$.
2) Premultiply the preliminary equation 
by $\langle \chi |$, the acceptability of which is underlied by the `diagonal dominance' condition    
\be
\mbox{for } \fj \neq \fl \mbox{ } , 
\left|m_{{\sfj}{\sfl}}/m_{{\sfj}{\sfj}}  
\right|
= \epsilon_{\sd\sAM} \mbox{ , small }.  
\ee
3) Make use of the normalization of $|\chi\rangle$.

Path ABC is the ZAM counterpart of BO's scheme.  
It involves, respectively: expanding by (\ref{1}), the one adiabatic neglect $\epsilon_{\sa \sAM}$ 
small, and finally the above-described step C.  
Path DE is the ZAM counterpart of Berry's scheme.  
Therein, step E follows from being able to pull H outside the L-integral 
$\langle \mbox{ } | \mbox{ } | \mbox{ } \rangle$, and then by (\ref{covstar}).  
Path F parallels the recovery of the BO scheme from the Berry one.     
Note that in the ZAM cycle, this involves the same approximation as step B: 
unlike (\ref{boarray}), the above diagram `commutes'.   
This is a respect in which the linear ZAM constraint is simpler to handle that  
passes over also to the linear momentum constraint in geometrodynamics.

\noindent\underline{ZAM fluctuation L-equation.}
This is 
\be
\frac{\hbar}{i}\sum_{\ip}\H_{\ip} \cr {\underline{D}_{\sH}^{*}}^{\ip}|\chi\rangle + 
\overline{\underline{\ttM}_{\sL}}|\chi\rangle = 
\{\overline{\underline{\ttM}_{\sH}} + \overline{\underline{\ttM}_{\sL}}\}|\chi\rangle = 0
\mbox{ } ,
\label{ZAMLFluc}
\ee
and is easy to strip down, by (\ref{covstar}).  
And, as it contains no cross-terms, it is not to become a TDSE.  
What does happen, however, is that a ZAM piece features within the TDSE arising from the quadratic 
fluctuation L-equation.

%========================================================================================================
\noindent\underline{Approximations.}
%========================================================================================================
%
%
%
Another new contribution of this Paper is that I list and characterize the subsequent plethora of 
semiclassical approach approximations, many of which are made in the literature.  
Some are `tacit underliers' so that the other approximations below suffice to cover
all distinct ratios relevant to the H- and L-physics.  The `sharply peaked hierarchy' conditions,
\be
\stackrel{\mbox{max}}{\mbox{\scriptsize $\ip$, $\jp$}}
\frac{|M_{\ip} -  M_{\jp}|}{M_{\ip}} = \epsilon_{\Delta \sM} \mbox{ } , \mbox{ }  
\stackrel{\mbox{max}}{\mbox{\scriptsize $\ipp$, $\jpp$}}\frac{|m_{\ipp} -  m_{\jpp}|}{m_{\ipp}} = \epsilon_{\Delta \sm} \mbox{ }  
\mbox{ are both small ,}
\ee
are such.  
The example 
\be
\frac{m_{\ipp}  }{  M_{\ip}  } = 
\frac{    \frac{m_{\ipp} - m}{m}  m + m    }{    \frac{M_{\ip} - M}{M}M + M     } = 
\frac{  \{ \frac{m_{\ipp} - m}{m}   + 1 \}m   }{   \{\frac{M_{\ip} - M}{M} + 1\}M    } 
\mbox{ } \mbox{ } \sim \mbox{ } \mbox{ } 
\frac{m}{M}\left\{ 1 +  \frac{m_{\ipp} - m}{m} - \frac{M_{\ip} - M}{M} \right\}
\mbox{ } \mbox{ } \sim \mbox{ } \mbox{ } 
\epsilon_{\sH\sL}
\{1 + O( \epsilon_{\Delta \sM}, \epsilon_{\Delta \sm} )\}
\ee
(by binomial expansion) illustrates how using these tacit underliers allows for 
only one H--L mass ratio to feature in the other independent approximations.

The general adiabatic condition is that H-subsystem physics processes are much more 
slowly-varying than L-subsystem ones.  
There are two different `pure forms' in which this physical condition can occur at the quantum level.  
Firstly, there are quantities that are small through $|\chi\rangle$ being far less sensitive to changes 
in H-subsystem physics than to changes in L-subsystem physics, which I label `p'.  
Secondly, there are quantities that are small through $|\chi\rangle$ being far less sensitive to 
changes in L-subsystem physics than $\phi$ is sensitive to changes in H-subsystem physics, 
which I label by `$\pounds$'.  
There is also a `proto-WKB' condition that $\chi$ is more sensitive to changes in H-subsystem physics 
than $\phi$ is, which I label `w'.
Of course, this is none other than $\epsilon_{\sa\sp}/\epsilon_{\sa\pounds}$ being small.

\noindent I note that none of these in general follow from the smallness of the classical adiabatic 
parameter $\epsilon_{\sa}$, for some wavefunctions can be very steep or wiggly even for slow processes - 
like the thousandth Hermite function for the slower oscillator, say.  Nor are e.g. the various p 
quantities in general inter-related, for derivatives are unsmoothing 
while the integral has a constant of integration allowing the lower derivative to behave independently of  
the higher derivative. 
[In some conventional situations, $\epsilon_{\sH\sL}$ being small is sometimes conducive to 
some of the $\epsilon_{\sa}$ being small, but they are in general separate assumptions.] 
Inspection of the H and L equations furthermore reveals that both p and $\pounds$ 
occur in terms also containing an $\epsilon_{\sH\sL}$ type factor.  
Thus, overall, these terms in the equations are particularly small.  
There are furthermore `mixed terms' -- part p and part $\pounds$ --  which I denote by `m', 
which are to be interpreted as depending on both of the abovedescribed `pure' slownesses.    
Finally, I denote by `ZAM' those conditions which originate in that constraint, 
and `diagonal dominance' conditions in the quantum-mechanical matrices by `d'.

I set out to build these various small quantities from a minimal, primitive set of them, albeit this 
cannot be completed yet in this Section.  
Fairly primitive quantities that occur in the various equations and might be considered to be 
small or negligible are:
$$
\left| A_{\sH}^2/ \pa_{\sL}\mbox{}^2|\chi\rangle \right| = \epsilon_{\sa \sp 1}
\mbox{ } , \mbox{ } 
\left| \pa_{\sH}\mbox{}^2|\chi\rangle / \pa_{\sL}\mbox{}^2|\chi\rangle \right| = \epsilon_{\sa \sp 3}
\mbox{ } , \mbox{ }
|{\langle\pa_{\sH}\mbox{}^2\rangle|\chi\rangle}/\pa_{\sL}\mbox{}^2|\chi\rangle| = \epsilon_{\sa\sp 4} 
\mbox{ } , \mbox{ }
$$
$$
\left|
A_{\sH}^2/\langle \pa_{\sL}\mbox{}^2\rangle
\right|
= \epsilon_{\sa \sp^{\prime}1}
\mbox{ } , \mbox{ }
\left|
\mbox{}_{\sL}|| |\pa_{\sH}\chi\rangle ||^2/\langle \pa_{\sL}\mbox{}^2\rangle
\right| 
= \epsilon_{\sa \sp^{\prime}2}
\mbox{ } , \mbox{ }
|{A_{\sH}\pa_{\sH}|\chi\rangle}/\pa_{\sL}\mbox{}^2|\chi\rangle| = \epsilon_{\sa\sp 5} 
\mbox{ } , \mbox{ }
|\langle A_{\sH} \pa_{\sH}\rangle| \chi\rangle/\pa_{\sL}\mbox{}^2|\chi\rangle| = \epsilon_{\sa\sp 6}  
\mbox{ } , \mbox{ }
$$
\be
\left|
\lfloor \pa_{\sH}\phi \rfloor \pa_{\sH} |\chi\rangle/\pa_{\sL}\mbox{}^2|\chi\rangle
\right| 
= \epsilon_{\sa \sm 7\scr}
\mbox{ } , \mbox{ }
\left|
A_{\sH}\pa_{\sH}|\chi\rangle/|\pa_{\sL}\mbox{}^2\chi\rangle
\right| =
\left|
\langle \chi | \lfloor \pa_{\sH} |\chi \rfloor\rangle \pa_{\sH}\phi/\pa_{\sL}\mbox{}^2|\chi\rangle
\right| 
= \epsilon_{\sa \sm 8\scr}
\mbox{ } , \mbox{ }
\left| 
{A_{\sH}}/{\pa_{\sH}|\chi\rangle}
\right|
= \epsilon_{\sAM}
\mbox{ } , 
\ee
$$
| || A_{\sH} ||^2/{\pa_{\sH}}^2\phi | = \epsilon_{\sa \sw 1}
\mbox{ , }
| _{\sL}|| \pa_{\sH}\chi ||^2/{\pa_{\sH}}^2\phi| = \epsilon_{\sa \sw 2} 
\mbox{ , }
|{\pa_{\sH}}^2|\chi\rangle / {\pa_{\sH}}^2\phi| = \epsilon_{\sa \sw 3}
\mbox{ , }
| \langle{\pa_{\sH}}^2 \rangle|\chi\rangle/{\pa_{\sH}}^2| = \epsilon_{\sa \sw 4} 
\mbox{ } ,
$$
\be
\left|
\lfloor\pa_{\sH}\phi\rfloor\pa_{\sH}|\chi\rangle/\pa_{\sH}\mbox{}^2|\chi\rangle
\right| 
= \epsilon_{\sa \sw 7\scr} 
\mbox{ } , \mbox{ }
\left|
A_{\sH}\pa_{\sH}|\chi\rangle/\pa_{\sH}\mbox{}^2|\chi\rangle
\right| 
=  
\left|
\langle\chi|\lfloor\pa_{\sH}|\chi\rfloor\rangle\pa_{\sH}\phi/\pa_{\sH}\mbox{}^2|\chi\rangle
\right|
= 
\epsilon_{\sa \sw 8\scr} 
\mbox{ } , \mbox{ }
\left|
{A_{\sH}}/{\pa_{\sH}\phi}
\right|
= \epsilon_{\sa \sw 9\sAM}
\mbox{ } ,
\ee
\be
\left|
\langle \fV_{\sL} \rangle/ \hbar^2 \pa_{\sL}\mbox{}^2 |\chi\rangle
\right| 
= \epsilon_{\sL 1} 
\mbox{ } , \mbox{ }
\left|
\langle \fI_{\sH\sL} \rangle/ \hbar^2 \pa_{\sL}\mbox{}^2|\chi\rangle 
\right| 
= \epsilon_{\sL 2} 
\mbox{ } , \mbox{ }
\left|
\langle \pa_{\sL}\mbox{}^2 \rangle/ \pa_{\sL}\mbox{}^2 |\chi\rangle
\right| 
= \epsilon_{\sL 3} 
\mbox{ } . \mbox{ }
\ee
N.B. that not all occur concurrently: $\epsilon_{\sa\sp 3}, \epsilon_{\sa\sp 7}$ occur
along the short path of (\ref{boarray}), 
while $\epsilon_{\sa\sp4}$ and $\epsilon_{\sa\sp 8}$ occur along the long path.

As regards what happened to the abovementioned diagonal terms, 
$\epsilon_{\sd\sBO}$ is not in itself adiabatic and is kept as a primitive quantity. 
The ZAM `diagonal dominance' condition $\epsilon_{\sd\sAM}$ is also not per se adiabatic, and is kept 
as a primitive quantity.      
Moreover, for $\epsilon_{\sd\sBOB}$, 
I expand its definition (assuming the BO term $\epsilon_{\sBO}$ is largest therein):
$\epsilon_{\sd\sBOB} = \{\epsilon_{\sd\sBO} + \epsilon{\sd\sBOB^{\prime\prime}}\}
                     \{  1 + O( \epsilon_{\sd\sBOB^{\prime}})\}$
for $\epsilon_{\sd\sBOB^{\prime}} = e_{\sll\sll}/o_{\sll\sll}$ 
and $\epsilon_{\sd\sBOB^{\prime\prime}} = e_{\sll\sj}/o_{\sll\sj}$.  
It is from expanding these out from the definitions of $e$ and $o$ that 
$\epsilon_{\sp 1^{\prime}}$, $\epsilon_{\sp 2^{\prime}}$, 
$\epsilon_{\sp^{\prime}2^{\prime}}$ and $\epsilon_{\sp^{\prime}2^{\prime}}$ arise, alongside mass 
factors.
Overall, 
$\epsilon_{\sd\sBOB} = 
\epsilon_{\sd\sBO} + \epsilon_{\sH\sL} 
\{ \epsilon_{\sa \sp^{\prime}1^{\prime}} + \epsilon_{\sa\sp^{\prime}1^{\prime}} \} +
\epsilon_{\sd\sBO} \epsilon_{\sH\sL} \{ \epsilon_{\sa\sp^{\prime}1} + \epsilon_{\sa\sp^{\prime}2} \}$

\noindent 
$+ \epsilon_{\sH\sL} \{ \epsilon_{\sa\sp^{\prime}1^{\prime}} + \epsilon_{\sa\sp^{\prime}1^{\prime}} \}
O(\epsilon_{\Delta \sM}, \epsilon_{\Delta \sm)}) + ...$
though exactly what is kept in the expansions depends on the relative sizes of the 
various `small $\epsilon$ quantities'

Finally, I display various properties of the approximations in a table.  
The numbering is 1 number per numerator type. 
The last 4 columns indicate which number-letter duets are present, 
and which equation they arise in.  
The `margin column' indicates 
which terms are ignored in the previously-common 
`Born--Fock' approximation that neglects connection terms.
\begin{tabbing}
$\stackrel{\mbox{\scriptsize(powers of the connection}}{\mbox{\scriptsize in this term)}}$ 
\mbox{ } \= \underline{number \hspace{0.2in}} \mbox{ } 
\=\underline{p-adiabatic} \mbox{ } \=\underline{p$^{\prime}$-adiabatic} \mbox{ }\=  
\underline{w (proto-WKB) \hspace{0.8in}} \mbox{ } \=\underline{not adiabatic} 
\mbox{ } \= \mbox{\scriptsize for full path?}\\
\hspace{0.5in} $A_{\sH}^2$                        \>      1             \>  L \> H            \> H \>          \> no  \\
\hspace{0.5in} $A_{\sH}^2$                        \>      1$^{\prime}$  \>    \> H            \>   \>     \> yes         \\
                                                  \>      2             \>    \> H            \> H \>     \> no         \\
                                                  \>      2$^{\prime}$  \>    \> H            \>   \>     \> yes         \\  
                                                  \>      3             \>  L \>              \> H(path ABC alternative) \> \> yes   \\
                                                  \>      4             \>  L \>              \> H(path DEFG alternative) \> \> yes  \\
\hspace{0.5in} $A_{\sH}$                          \>      5             \>  L \>              \>   \>          \> no    \\
\hspace{0.5in} $A_{\sH}$                          \>      6             \>  L \>              \>   \>          \> no    \\
                                                  \>                    \>\underline{m-adiabatic}  \>              \>   \>    \>          \\  
                                                  \> 7(cross)           \>   L \>              \> H(path ABC alternative) \> \> yes\\
\hspace{0.5in} $A_{\sH}$                          \> 8(cross)           \>  L \>              \> H(path DEFG alternative) \> \> yes  \\      
\hspace{0.5in} $A_{\sH}$                          \> 9(ZAM)             \>    \>              \> H          \> H \> yes   \\             
\end{tabbing}

%====================================================================================================
%====================================================================================================
\section{The WKB procedure}
%====================================================================================================
%====================================================================================================

I take this to consist of the subsequent H-wavefunction ansatz 
\be
\phi = \mbox{exp}
\left(
iM\fF
\left(
\H_{\ip}
\right)
/\hbar
\right)
\label{wabbit}
\ee
(for $M$ a `generic heavy mass'\foo{Footnote 11 explains why I use this generic-mass form rather than 
$\sum_{\ip}M_{\ip}\fF(\H_{\ip})$.} and $\fF$ at this stage a function of an unspecified nature) 
alongside some package of conventionally-associated approximations.   
Then Berry's H-equation expanded by (\ref{2}) becomes 
\be
M^2\mbox{}_{\sM^{-1}}||\pa_{\sH}\fF||^2 - iM\hbar
\left\{
\mbox{}_{\sM^{-1}}||\pa_{\sH}||^2\fF + 2i_{\sM^{-1}}(A_{\sH}, \pa_{\sH}\fF)
\right\}
- i\hbar^2\mbox{}_{\sM^{-1}}(\pa_{\sH}, A_{\sH}) + 
\hbar^2\mbox{}_{\sM^{-1}}||A_{\sH}||^2 + e + o = \fE_{\sH} 
\mbox{ } .  
\ee
Moreover, the L-equation becomes
\be
\left\{
\frac{1}{\hbar^2}\overline{\hat{h}} + \hat{R}
\right\}
|\chi\rangle = \frac{2iM}{\hbar} 
\left\{
\mbox{}_{\sM^{-1}}(\lfloor\pa_{\sH}\fF\rfloor,\pa_{\sH})|\chi\rangle - 
i_{\sM^{-1}}(\lfloor\pa_{\sH}\fF\rfloor, A_{\sH})|\chi\rangle 
\right\} 
\mbox{ } ,
\label{preTDSE}
\ee
which is arranged so that the right-hand side exclusively and exhaustively isolates the cross-terms, 
all other types of correction terms being bundled into the left-hand side's `remainder operator' 
\be
\hat{R} = i_{\sM^{-1}}(A_{\sH}, \pa_{\sH}) - 2i\langle_{\sM^{-1}}(A_{\sH}, \pa_{\sH})\rangle - 
\mbox{}_{\sM^{-1}}||A_{\sH}||^2 - \overline{\mbox{}_{\sM^{-1}}||\pa_{\sH}||^2} \mbox{ } .  
\ee

%=======================================================================================================
\noindent\underline{Corresponding WKB ZAM equations.}  
%=======================================================================================================
%
%
%
In the case of nontrivial RPM's there is also a ZAM H-equation, which I obtain to be 
\be
0 = \sum_{\ip} \H_{\ip} \cr 
\left\{ 
M{\underline{\pa}_{\sH}}^{\ip}\fF + 
\frac{\hbar}{i}\langle\chi|\underline{\pa}_{\sH}\mbox{}^{\ip}|\chi\rangle 
\right\} 
+ \langle \underline{\ttM}_{\sL}\rangle  \mbox{ } .
\ee
As (\ref{ZAMLFluc}) does not depend on $\phi$, it is also the ZAM WKB L-equation.

%=======================================================================================================
\noindent\underline{Approximations.} 
%=======================================================================================================
%
%
%
Here I develop my new characterization and interpretation of approximations now in the WKB approximation 
setting.  Preliminarily, $\epsilon_{\Delta \sM}, \epsilon_{\Delta \sm}$ small 
are required to avoid proliferation of the other types of approximation. 
Adopting the WKB ansatz does not affect the p and L criteria,  
while the ${\pounds}$ and w criteria are modified.  
One has now not change in $\phi$ but change in $\fF$ with respect to H, alongside some power of $M/\hbar$ 
which ensure that one continues to talk about dimensionless ratios.  
I call the resulting quantities `g' and `W'.  
Note that $\epsilon_{\sg}$/$\epsilon_{\ssWKB}$ = $\epsilon_{\sp}$.   
The `WKB approximation' is that the $\fF$ is slowly varying with respect to H.  
This amounts to the following string of approximations. 
Firstly, the typical WKB assumption that 
\be
\left|\frac{\hbar\pa_{\sH}^2\fF}{M|\pa_{\sH}\fF|^2}\right| = \epsilon_{\ssWKB} \mbox{ , small , }
\ee
which is an approximation type lying outside the p, W, g classification.  
That established, while the $\pa_{\sH}\mbox{}^2\phi$ denominator of Sec 5 becomes both 
$\frac{M}{\hbar}\pa^2\mbox{}_{\sH}\fF$ and $\frac{M^2}{\hbar^2}|\pa_{\sH}\fF|^2$ in this Section, 
it is the latter which dominates and thus replaces 
$\pa_{\sH}\mbox{}^2\phi$ in passing from Sec 5's approximations to this Section's.   
Thus we obtain the following new small quantities.
$$
\left| 
\frac{\hbar}{M}   \frac{  \langle \chi| \pa_{\sH} \chi\rangle   }{    \pa_{\sH} \fF    }
\right|^2 = 
\left| 
\frac{\hbar}{M}   \frac{  A_{\sH} }{  \pa_{\sH}\fF  }
\right|^2
= \epsilon_{\sa\sW 1}
\mbox{ } , \mbox{ }
\left| 
\frac{\hbar^2}{M^2} 
\frac{    \langle \lfloor\pa_{\sH}\chi \rfloor| \pa_{\sH}\chi\rangle   }{  |\pa_{\sH}\fF|^2        }
\right|
= \epsilon_{\sa\sW 2}
\mbox{ } , \mbox{ }
\left| 
\frac{\hbar^2}{M^2} \frac{    \pa^2\mbox{}_{\sH}|\chi\rangle   }{    |\pa_{\sH}\fF|^2    } 
\right| 
= \epsilon_{\sa\sW 3}
\mbox{ } , \mbox{ }
\left| 
\frac{\hbar^2}{M^2} \frac{    \langle\chi |\pa^2\mbox{}_{\sH}|\chi\rangle   }{    |\pa_{\sH}\fF|^2    } 
\right| 
= \epsilon_{\sa\sW 4}
\mbox{ } , \mbox{ }
$$
\be
\left| 
\frac{\hbar}{M} \frac{    \pa_{\sH} |\chi \rangle \pa_{\sH}\fF    }
                     {    |\pa_{\sH}\fF|^2    }         
\right| = 
\left| 
\frac{\hbar}{M} \frac{    \pa_{\sH} |\chi \rangle     }
                     {    \pa_{\sH}\fF    }         
\right| = 
\epsilon_{\sa\sW 7\scr} 
\mbox{ } , \mbox{ }
\left| 
\frac{\hbar}{M} \frac{    \langle \chi |\pa_{\sH} |\chi \rangle \pa_{\sH}\fF    }
                     {    |\pa_{\sH}\fF|^2    }         
\right|
= \epsilon_{\sa\sW 8\scr} 
\mbox{ } , \mbox{ }
\ee
Moreover, there is now no independent $\epsilon_{\sa\sw 9} = \epsilon_{\sa\sw\sAM}$, 
as this has now become $\sqrt{\epsilon_{\sa\sW 8\scr}}$.  

Mixed terms slightly change in form from the previous Section. I denote these now by `M'.
They are
\be
\left| 
\frac{M}{\hbar} 
\frac{    \lfloor \pa_{\sH}\fF\rfloor |\pa_{\sH}|\chi\rangle     }
     {    \pa^2_{\sL}|\chi\rangle    } 
\right|
 = \epsilon_{\sa\sM 7\scr}
\mbox{ } , \mbox{ }
\left| 
\frac{M}{\hbar} 
\frac{    \lfloor\pa_{\sH}\fF \rfloor| \langle\chi \pa_{\sH}| \chi\rangle   }
     {  |\pa_{\sL}\mbox{}^2|\chi\rangle   }
\right|
=
\left| 
\frac{M}{\hbar} 
\frac{    \lfloor\pa_{\sH}\fF \rfloor A_{\sH}   }
     {  |\pa_{\sL}\mbox{}^2|\chi\rangle   }
\right|
= \epsilon_{\sa\sM 8\scr} \mbox{ } .
\ee
The handling of $o$ and $e$ is as before.

Thus, the small quantities one has to contemplate at this stage are 
$\epsilon_{\sH\sL}, \epsilon_{\sT}, \epsilon_{\sV}, \epsilon_{\sI}, \epsilon_{\Delta \sM}, 
\epsilon_{\Delta \sm}, \epsilon_{\sa}, \epsilon_{\ssWKB}$, 9 $\epsilon_{\sa\sp}$ quantities, 
2 $\epsilon_{\sa\sm}$ quantities, 
3 $\epsilon_{\sL}$ quantities, 
4 $\epsilon_{\ssWKB}$ quantities, 
$\epsilon_{\sAM}, \epsilon_{\sd\sBO}$ and $\epsilon_{\sd\sAM}$.   
The W's and M's carry their w or m precursors' H, L, connection, cross-term and full path statuses.  
The previous Section's table is then modified by these two relabellings and by the loss of 
entry aw9.  
Next note that not all of the remaining $\epsilon$'s are independent.  
This is clear from the tabulation, which reveals what excess of 
shared numerators and denominators there are. 
This affects how one can set up a full independent set of primitive quantities in terms 
of which all remaining quantities can be expressed.    
I choose to use the very cleanly adiabatic quantity
\be
|\d\H/\d\L| = \epsilon_{\sa 1} 
\ee
as a primitive quantity.  
Then $\epsilon_{\sT}$ is a derived quantity, $\epsilon_{\sT} = \epsilon_{\sH\sL}/\epsilon_{\sa 1}^2$.     
So the useful condition $\epsilon_{\sT}$ small relies on the H--L hierarchy greatly 
outstripping (at least 1 classical notion of) adiabaticity.  
I also choose to use $\epsilon_{\sa\sW 3}$ as a primitive quantity, regardless of which path is under 
consideration.  
Then there are the following dependencies.  
\be
\epsilon_{\sa\sp^{\prime}2} = \epsilon_{\sa\sW 2} \frac{\epsilon_{\sa\sW 3}}{\epsilon_{\sa\sp 3}}
\mbox{ } , \mbox{ }
\epsilon_{\sa\sW 1} = \epsilon_{\sa\sp 1} \frac{\epsilon_{\sa\sW 3}}{\epsilon_{\sa\sp 3}}
\mbox{ } , \mbox{ }
\epsilon_{\sa\sW 4} = \epsilon_{\sa\sp 4} \frac{\epsilon_{\sa\sW 3}}{\epsilon_{\sa\sp 3}}
\mbox{ } , \mbox{ }
\epsilon_{\sa\sW 7\scr} = \epsilon_{\sa\sM 7\scr} \frac{\epsilon_{\sa\sW 3}}{\epsilon_{\sa\sp 3}}
\mbox{ } , \mbox{ }
\epsilon_{\sa\sW 8\scr} = \epsilon_{\sa\sM 8\scr} \frac{\epsilon_{\sa\sW 3}}{\epsilon_{\sa\sp 3}} \mbox{ } . 
\label{interrelations}
\ee
The below-useful $\epsilon_{\mbox{\scriptsize pert}} = \fI_{\sH\sL}/\fV_{\sL}$ 
(relating to whether the HL interaction can be treated as a perturbation as regards the L-subsystem) 
is another dependent quantity, being $\epsilon_{\sI}/\epsilon_{\sV}$.

Thus I have as a full set of primitive quantities $\epsilon_{\sH\sL}, \epsilon_{\Delta \sM}, 
\epsilon_{\Delta \sm}, \epsilon_{\sa}, \epsilon_{\sa 1}, \epsilon_{\sV}, \epsilon_{\sI}, \epsilon_{\ssWKB}, 
\epsilon_{\sd\sBO},$ 8 of the 9 $\epsilon_{\sa\sp}$ (all bar $\epsilon_{\sa\sp^{\prime}2}$), 
the 3 $\epsilon_{\sL}$, 
the 2 $\epsilon_{\sa\sM\scr}, \epsilon_{\sa\sW 2}, \epsilon_{\sAM}$ and $\epsilon_{\sd\sAM}$.  
That's 24 from the quadratic constraint and 2 from the ZAM constraint, 
so 26 in total for nontrivial relational theories.   
The importance of this analysis is to point out that calculations that keep the small chroniferous cross-term 
but `randomly' throw away many other terms may be prone to inconsistencies.
%

%====================================================================================================
%====================================================================================================
\section{A suggested interpretation of the H- and L-equations}
%====================================================================================================
%====================================================================================================

\noindent\underline{Outline.}  
In the conventional QM setting, one is free to adopt a L-TDSE (with respect to external time) `from the outset' 
[in place of (\ref{preTDSE})].  
This is not possible in the context of closed-universe quantum cosmology, hence issue B1.    
The semiclassical quantum-cosmological answer to this that is used quite widely in the literature (see 
e.g. \cite{LR79, Banks, HallHaw}) is that the fluctuation L-equation (\ref{preTDSE}) can, nevertheless, 
be rearranged to obtain a TDSE with respect to an emergent time that is `provided by the H-subsystem'.  
I next underpin the usual arguments, suggest an extended procedure and survey caveats for both.

Firstly, as regards the H-equation providing a time for the L-equation, while that rather implies 
tackling the H-equation prior to the L-equation, since the H-equation in full contains functionals of 
the unknown $|\chi\rangle$ which is what the L-equation is to be solved for, 
such a prior procedure is not possible. 
Secondly, obtaining a TDSE for the L-subsytem requires formal identification of a quantity as momentum.  
This turns out to indeed be possible if (step 1) one considers the H-equation to be of 
Hamilton--Jacobi (HJ) form whence an approximation to it is solved by a $M{\fF}$ that is approximately 
Hamilton's principal function, ${\fW}$.\foo{Having made this identification, it is well-known 
(see e.g. \cite{Gold, BenFra}) that $\fW$ is in general nonseparable, so using 
$\sum_{\ip}M_{\ip}\fF(\H_{\ip})$ in (\ref{wabbit}) would have entailed building in a generally false and 
misleading feature.}   
Moreover if one considers that approximation to be the HJ equation itself, then one avoids the 
aforementioned problem through this approximate rather than full H-equation being soluble prior to the 
L-equation.    
(Step 2) Consider the solution of the approximate HJ equation to be $\fW_{0}$, while the true $\fW$ is 
$\fW_0 + \fW_{\mbox{\scriptsize correction}}$.  
Then using the approximate notion of momentum associated with $\fW_0$, 
$\P_{\sH}^{i^{\prime}} = {\pa\fW_0}/{\pa \H_{\ip}}$ one can recast the L-equation to contain 
an explicit TDSE piece built using a $t^{\WKBL}$ (Step 3).  
Next, I suggest the following extended procedure.  
(Step 4) Thus I obtain an explicit estimate $t^{\WKBH}_0$ for $t^{\WKBL}$ in terms of the 
$\H_{\ip}$ at the level of the approximate H-equation.  
(Step 5) At least in cases for which this relation is invertible, I note that the L-equation can be made 
to be free of H-dependence by using the inversion of the estimate to turn this into 
$t^{\WKBH}_0$-dependence.  
This recasts the L-equation as a time-dependent perturbation of the `ordinary L-physics' TDSE'.  
This is solved by $|\chi(t^{\WKBH}_0, \L_{\ipp})\rangle$, or, rearranging,  
$|\chi(\H_{\ip}, \L_{\ipp})\rangle$, which   
(Step $1^{\prime}$) can then be substituted into the full H-equation 
to give a more precise equation entirely in $\H_{\ip}$.
This may likewise be solved as a HJ equation (a more complicated one that is making 
some allowance for the back-reaction of the L-subsystem), on which one 
can perform the primed counterpart of the above 5-step cycle.  
[At this stage I make no claims as to whether such a scheme would require but few cycles to converge well, 
or indeed converge at all.]

%====================================================================================================
\noindent\underline{Step 1: Approximate HJ H-equation.}
%====================================================================================================
%
%
%
Regard
$\epsilon_{\sd\sBO}$, 
$\epsilon_{\sa}$, 
$\epsilon_{\sH\sL}$, 
$\epsilon_{\Delta \sM}$, 
$\epsilon_{\Delta \sm}$,
$\epsilon_{\ssWKB}$,
$\epsilon_{\sa\sp^{\prime} 1^{\prime}}$,
$\epsilon_{\sa\sp^{\prime} 2^{\prime}}$, 
$\epsilon_{\sa \sW  4}$, 
$\epsilon_{\sa \sW  8\scr}$
as small,  and also assume that the averaged counterparts of 
$\epsilon_{\sH\sL}/\epsilon_{\sa 1}^2$, 
$\epsilon_{\sV}$, 
$\epsilon_{\sI}$ are small so that
\be
o 
= \langle \hat{h} \rangle 
= \langle \ttH_{\sH\sL} + \fV_{\sH} \rangle 
= \langle \ttH_{\sH\sL} \rangle + \fV_{\sH}
= \langle -\hbar^2\mbox{}_{m}||\pa_{\sL}||^2 + \fV_{\sL} + \fI_{\sH\sL}\rangle + \fV_{\sH}
\ee 
reduces to $\fV_{\sH}$.  
Then the approximate H-equation is
\be
\mbox{}_{\sM^{-1}}||\pa_{\sH}\fW_0||^2  = \fE_{\sH}   - \fV_{\sH} \mbox{ } , 
\label{HJE}
\ee
which is a HJ equation 
(see \cite{Lanczos, GRT, Wheeler} for good accounts of the significance of these).   
The tractability in practice of this problem improves considerably if there is only one H d.o.f. 
(the separable case of the HJ equation in various H d.o.f.'s being intermediate in 
tractability).  
Formally, the solution is\foo{If this is accompanied by a nontrivial ZAM constraint, 
one can use the Lagrangian form of that to evaluate $\dot{\b}$ algebraically 
and then use that to eliminate $\dot{\b}$ from the HJ equation. 
I don't explicitly pursue this in this Paper because its GR 
counterpart is the very intractable solution of the notorious thin sandwich partial differential equation 
for the shift vector \cite{TSC}.}
\be
\fW_0(\H_{\ip}) =                     \int^{\suH_{\ip}}\mbox{}_{\sM}||\d {\H}^{\prime}_{\ip}||
                                      \sqrt{\fE_{\sH} - \fV_{\sH}({\H}^{\prime}_{\ip})}
\label{wylla}
\mbox{ } .  
\ee
If this is evaluable, one should check at this stage that $\epsilon_{\ssWKB}$ is indeed small.

%====================================================================================================
%====================================================================================================
\noindent\underline{Step 2: underlying implicit import of emergent time.}
%====================================================================================================
%====================================================================================================
%
%
%
\be
{\pa_{\sH}}^{\ip}\fW \equiv \P^{\sH\ip} = \frac{M_{\ip}}{\mbox{n}}\dotb \H^{\ip} = 
M_{\ip}
\left\{
\ast\H_{\ip}  -  \lfloor\ast\b\rfloor \cr \H_{\ip}
\right\}
\mbox{ }   
\label{HJT} \mbox{ } ,  
\ee
by the expression for momentum in HJ theory, (\ref{HLmom}) and (\ref{App1B}).

%====================================================================================================
\noindent\underline{Step 3: passing to a `TDSE' for the L-subsystem.}
%====================================================================================================
%
%
%
Next, the crucial cross-term 
$
2M\mbox{}_{{\sM}^{-1}}(  \pa_{\sH}{\fW_0},{D}^*_{\sH}|\chi\rangle) 
$
in the L-equation contains 
$$
\frac{1}{\mbox{n}}
\frac{\pa \H_{\ip}}{\pa\lambda}  \cdot \frac{\pa|\chi\rangle}{\pa \H_{\ip}}
\mbox{ } 
$$
which contains, using the chain-rule in reverse and (\ref{LMB}), 
\be
\frac{1}{\mbox{n}} \frac{\pa|\chi\rangle}{\pa\lambda} = \frac{\pa|\chi\rangle}{\pa t^{\WKBL}} \mbox{ } .
\ee

Moreover, I find the L-equation to be in full 
$$
i\hbar 2 \mbox{}_{\sM^{-1}}(\pa_{\sH}{\fW_0},D^*_{\sH}|\chi\rangle) 
= i\hbar 2 \frac{M}{\mbox{n}}  
\left\{_{\sM^{-1}}
\left(
\frac{\pa H}{\pa\lambda}, {D}_{\sH}^*|\chi
\right) 
- \left._{\sM^{-1}}
\left(
\frac{\pa b}{\pa\lambda} \cr H, {D}_{\sH}^*|\chi\rangle
\right) 
\right.
\right\}
$$
$$
= 
i\hbar\frac{1}{\mbox{n}}
\left\{\frac{\widetilde{{D}_{\sT}}^*}{D\lambda}|\chi\rangle - 
2\left.
\frac{\d \b}{\d\lambda} \cdot \H_{\ip} \cr {{D}^*_{\sH}}^{\ip}|\chi\rangle
\right. 
- \frac{\widetilde{{D}_{\sL}}^*}{D\lambda}|\chi\rangle
\right\}
$$
\be
= i\hbar
\left\{
\frac{\widetilde{{D}_{\sT}}^*}{D t^{\WKBL}}|\chi\rangle -  
\frac{\pa b}{\pa t^{\WKBL}} \cdot \overline{\ttM_{\sL}} - 
\frac{\widetilde{{D}_{\sL}}^*}{D t^{\WKBL}}|\chi\rangle
\right\}
\left\{
1 + O( \epsilon_{\sH\sL}; 
\epsilon_{\sH\sL}/\epsilon_{\sa 1}^2, \epsilon_{\sV}, \epsilon_{\sI}]
\right\}
\label{swalk} \mbox{ } .
\ee
Here, 
\be
\frac{\widetilde{D_{\sL}^*}}{D\lambda} = \frac{\pa \nL}{\pa\lambda}\pa_{\sL} - i\widetilde{A_{\sL}} 
\mbox{ } , \mbox{ }
\frac{\widetilde{D_{\sT}^*}}{D\lambda} = \frac{\pa }{\pa\lambda} - i\widetilde{A_{\sT}} 
\mbox{ }
\mbox{ for }
\mbox{ }
\widetilde{{A}_{\sL}} = -i\left<\chi \left| \frac{\pa \nL}{\pa\lambda}\pa_{\sL} \right|\chi\right> 
\mbox{ } , \mbox{ }
\widetilde{{A}_{\sT}} = -i\left<\chi \left| \frac{\pa }{\pa\lambda} \right|\chi\right> 
\mbox{ }  \mbox{ }
\ee
(these are dynamical connections as opposed to Berry ones).
In the working, use has been made of how a function of H alone can be moved inside 
$\langle \mbox{ } | \mbox{ } | \mbox{ } \rangle$,
of the scalar triple product identity, the definitions of: 
$D^*$, $A$, the overbar, angular momentum and the H-part thereof,   
and the swap of this for its L part by (\ref{QAM}).   
The functional dependence in (\ref{swalk}) arises from $\mbox{n} = \aha$ depending on L and hence 
one not being able to move this exactly through $\langle \mbox{ } | \mbox{ } | \mbox{ } \rangle$, 
which is resolved by expanding.

This gives more correction terms that involve comparing various L-derivatives, which had not been noted before: 
\be
|  \tilde{A}_{\sT} |\chi\rangle/ \pa_{\sL}\mbox{}^2 |\chi\rangle | = \epsilon_{\sL 4} \mbox{ } , \mbox{ }
|  \dot{\nL}\pa_{\sL} |\chi\rangle/ \pa_{\sL}\mbox{}^2 |\chi\rangle | = \epsilon_{\sL 5} \mbox{ } , \mbox{ }
|  \tilde{A}_{\sL} |\chi\rangle/ \pa_{\sL}\mbox{}^2 |\chi\rangle | = \epsilon_{\sL 6} \mbox{ } . \mbox{ }
\ee
The chain rule term is small if the classical adiabaticity a1 dominates over the quantum adiabatic 
ap3 term.
From this and $\epsilon_{\sT} = \epsilon_{\sH\sL}/\epsilon_{\sa 1}^2$ small,  
$\epsilon_{\sH\sL} << \epsilon_{\sa1}^2 << \epsilon_{\sa1} << \epsilon_{\sa\sp 3}$.  
i.e. mass hierarchy outstripping some kinds of adiabaticity, and adiabatic conditions varying in size.

Thus, equating (\ref{swalk}) with (\ref{preTDSE}) one obtains a `TDSE'
\be
i\hbar \frac{\pa|\chi\rangle}{\pa t^{\WKBL}}    =  
\left\{
\overline{\hat{\ttH}_{\sL} + \hat{\fI}_{\sH\sL} }  
\right\} 
|\chi\rangle +   
\frac{    \pa \b    }{    \pa t^{\WKBL}    } 
\cdot \overline{\hat{\uttL}}_{\sL} |\chi\rangle + 
\hat{ R}^{\prime}|\chi\rangle
\label{TDSEFULL}
\ee
for 
\be
\hat{{R}}^{\prime} = i\hbar\left\{\frac{D^*_{\sL}}{D t^{\WKBL}} - iA_{\sT}\right\} + \hbar^2\hat{R}
\ee
up to $\dot{\mbox{a}}$, $\langle \mbox{ } | \mbox{ } | \mbox{ } \rangle$ exchange.  
However, unfortunately this leads to two objections.  
1) further $\d/\d t^{\WKBL}$ terms in the $\hat{R}^{\prime}$ (which are small if $\epsilon_{\sW 3}, 
\epsilon_{\sW 4}$ are) so that the equation is not in general a TDSE.
2) Nor is it even satisfactory as an L-equation because the $\hat{R}$ contains H-derivatives 
(these are small if $\epsilon_{\sL 4}, \epsilon_{\sL 5}, \epsilon_{\sL 6}$ are).

It is standard that equations like (\ref{TDSEFULL}) can also be cast in a Tomonaga--Schwinger like form 
by multiplying through by $\dot{\mbox{a}}$.  
In the present case, I obtain  
\be
i\hbar \frac{\pa|\chi\rangle}{\pa \lambda}    =  
\left\{
\frac{\d \mbox{a}}{\d\lambda}
\left\{
\overline{\hat{\ttH}_{\sL} + \hat{\fI}_{\sH\sL} }    
\right\} 
+
\frac{    \d \b    }{    \d \lambda    } 
\cdot \overline{\hat{\uttL}_{\sL}}  
\right\}
  |\chi\rangle + \hat{ R}^{\prime\prime}|\chi\rangle
\label{TDSEFULL2}
\ee
for
\be
\hat{R}^{\prime\prime} = i\hbar
\left\{ 
\frac{   D^*_{\sL}   }{    D\lambda    } - 
\left< 
\chi 
\left| 
\frac{\pa}{\pa\lambda}
\right|
\chi
\right>
\right\}
+ \hbar^2\frac{   \d \mbox{a}    }{\d\lambda}\hat{R} \mbox{ } . 
\ee
I Note that this equation's $\lambda$'s can be considered to `cancel out', 
thus giving a temporally relational form.

%=======================================================================================================
\noindent\underline{Various forms of proposed approximate L-equations.}
%=======================================================================================================
%
%
%
In the L-equation, it is customary to neglect $\hat{R}$ summarily, (amounting to 
$\epsilon_{\sa\sp 1}, 
\epsilon_{\sa\sp 3},
\epsilon_{\sa\sp 4},
\epsilon_{\sa\sp 5},
\epsilon_{\sa\sp 6}$ small 
-- neglecting connection terms alongside the usual neglect of double derivatives in 
simple WKB ansatz calculations.      
Moreover, the lead chroniferous cross-term has to be regarded as non-negligible for the 
timestandard in use to emerge.  
On the other hand, $\fI_{\sH\sL}$ cannot also be dropped, as if it were, the wavefunction separates 
in H--L coordinates, but have already separated out as much H as one can in Sec 5, so $\chi$ would not 
depend on H, so a zero factor would be contained in the term which is to become 
$i\hbar\frac{\pa|\chi\rangle}{\pa t^{\WKBL}}$.  
Sometimes furthermore dropping the fluctuation terms (removing the overbars) 
is alluded to in the literature.  

The TDSE thus constructed is, modulo the H--L coupling term, `ordinary relational L physics', 
which in turn is `ordinary L physics' modulo the effect of the angular momentum correction term 
[itself absent in 1-d or if one repeats the above working in a spatially nonrelational setting].  
Thus the purported simple situation  
has `the scene set' by the H-subsystem for the L-subsystem to have dynamics, 
a dynamics which is furthermore slightly perturbed 
by the H-subsystem, while neglecting the back-reaction of the L-subsystem on the H-subsystem.  
One might even argue for the interaction term to be quantitatively negligible as regards 
the observed L-physics.

%====================================================================================================
%====================================================================================================
\noindent\underline{Step 4: explicit emergent time estimate from H-equation.}
%====================================================================================================
%====================================================================================================
%
%
%
A new proposal of mine is to extend the abovedescribed standard working as follows.  
(\ref{HJT}) in (\ref{HJE}) gives  
\be
_{\sM^{-1}}\left|\left|\frac{\pa \H_{\ip}}{\pa {t_0^{\WKBH}}} -  
\frac{\pa \b}{\pa t_0^{\WKBH}} \cr \H_{\ip} 
\right|\right|^2 = \fE_{\sH} + \fU_{\sH} \mbox{ } ,  
\ee
which can be integrated to give
\be
t_0^{\WKBH} - t_0^{\WKBH}(0) = \int^{\suH_{\ip}} \mbox{}_{\sM^{-1}}||\d_b \H^{\prime}||/
\sqrt{\fE_{\sH} + \fU_{\sH}(\H^{\prime}_{\ip})} \mbox{ } .  
\ee
This is in principle evaluable, leading to an estimate for the emergent timefunction,  
\be
t^{\WKBH}_0 = t^{\WKBH}_0(\H_{\ip})
\label{trel} \mbox{ } .  
\ee

%====================================================================================================
\noindent\underline{Step 5: inversion of estimate, giving a L-TDSE that is H-free.}
%====================================================================================================
%
%
%
Then, in the case of 1 H d.o.f. (as motivated by 
$M_{\mbox{\scriptsize Planck}} >> M_{\mbox{\scriptsize inflaton}}$ and 
scale factor $>>$ anisotropic, inhomogenous modes in GR cosmology), my proposal enables one to  
in principle invert (\ref{trel}) at least on some intervals of the mechanical motion:
\be
\nH = \nH(t_0^{\WKBH}) \mbox{ } . 
\label{tazenda}
\ee  
Thus one can formally eliminate H in favour of $t_0^{\WKBH}$ in the L-TDSE, allowing one to study it 
(or approximations to it) that are nevertheless coupled to the H-subsystem as $t_0^{\WKBH}$-dependent 
perturbations  of TDSE's. [The $\fI_{\sH\sL}(\nH_{\ip}, \L_{\ipp})$ which must be kept in the present 
context as argued above, is an obvious such perturbation upon use of 
(\ref{tazenda}).]  
%]
%
Now $\widetilde{D^*}_{\sL}|\chi\rangle / D t^{\WKBH}$ drops out 
as $t_0^{\WKBH}$ is independent of L, thus removing the abovementioned objection 1.  
The inversion can also be used to convert H-derivatives to t-derivatives, 
so one has a bona fide L-equation, thus removing objection 2 at the price of reintroducing objection 1!
However, this now in general contains first {\sl and second} time derivatives.  
Thus in some regions of configuration space, it is capable of behaving more like a 
KG equation than a TDSE, and is in all fullness more general than either of these.  
Thus the guarantee of appropriate interpretability that accompanies TDSE's is 
replaced by a difficult study of a more general wave equation.  
It is worth noting that KG-like but more complicated equations are 
prone to substantial extra impasses (see, e.g. \cite{Kuchar81}).

Explicitly, 
\be
\pa_{\sH} = \frac{\d t^{\WKBH}}{\d \nH}  \frac{\d}{\d t^{\WKBH}} = 
\sqrt{      \frac{    M    }{    2\{\fE_{\sH} + \fU(t^{\WKBH})\}    }      }\frac{\d}{\d t^{\WKBH}}
\ee
so, recursively,
$$
\pa^2\mbox{}_{\sH} = \frac{    M    }{    2\{ \fE_{\sH} + \fU(t^{\WKBH})  \}    }\frac{\d^2}{\d t^{\WKBH}\mbox{}^2} 
\mbox{ } + 
$$
\be
\frac{M}{4\{\fE_{\sH} + \fU(t^{\WKBH})\}^2}\frac{\d\fU(t^{\WKBH})}{\d t^{\WKBH}}\frac{\d}{\d t^{\WKBH}}
\mbox{ } , 
\ee
so I obtain an L-equation of the form 
$$
\{1 - P_{\chi}\}
\left\{
\left\{
i + \frac{      \hbar      }{      8\{\fE_{\sH} + \fU(t^{\WKBH})\}^2      }
\frac{\d \fU(t^{\WKBH})}{\d t^{\WKBH}}
\right\}
\hbar
\frac{    \d |\chi\rangle    }{    \d t^{\WKBH}    } 
\right.
$$
\be
\left.
+ \mbox{ }
\frac{    \hbar^2    }{    4\{\fE_{\sH} + \fU(t^{\WKBH})\}    } 
\frac{    \d^2|\chi\rangle    }{    \d t^{\WKBH}\mbox{}^2    }
+ 
\hbar^2\mbox{}_{\sm^{-1}}||\pa_{\sL}||^2    |\chi\rangle + 
\fU_{\sL}|\chi\rangle +  
\fJ_{\sH\sL}|\chi\rangle
\right\} = 0 
\mbox{ } .  
\ee

Also note that $t^{\WKBL}$ and $t^{\seB}$ are the same by the above and form (\ref{hint}, \ref{mint}) 
of (\ref{LMB}), so, collecting up the emergent time results in answer to B4, I form a new result: the  

\noindent\underline{Classical-semiclassical time lemma.}  

\noindent $t^{\seB} = t^{\WKBL}  = \left\{ t^{\WKBH} = t^{\seBH} \right\} + 
O(\epsilon_{\sH\sL}; \epsilon_{\sH\sL}/\epsilon_{\sa 1}^2, \epsilon_{\sV}, \epsilon_{\sI}]$.

\noindent
This result is why Sec 2-3 are significant for the rest of this Paper.\foo{While 
that is a nice connection to establish, it also means that worries about $t^{\seB}$ in Sec 3 carry over 
to the far more widely used $t^{\sWKB}$.} time notion $t^{\semi}$, but abbreviate it to $t^{\sem}$!   
I subsequently call this unified time notion $t^{\mbox{\scriptsize emergent(WKB--LMB)}}$ but abbreviate 
it to $t^{\mbox{\scriptsize em}}$!

%=======================================================================================================
\noindent\underline{Problems with the WKB procedure.}
%=======================================================================================================
%
%
%
Before continuing the cycle, I examine the salient deficiencies 
in the above type of `resolution of B1'.  
[N.B. this affects all the semiclassical proposals along the lines of Sec 4--6 and Steps 1--3 of Sec 7.]  
It is easy to argue out the applicability of the WKB scheme to quantum cosmology in a way which sounds, 
to many physicists, plausible and familiar and thus not requiring careful scrutiny.  
Unfortunately, the familiarity stems from the commonplace occurrence of the WKB approximation in basic 
ordinary QM, while the underlying conceptual differences between ordinary QM and quantum cosmology 
severely break that analogy (which is rather harder for the non-expert to notice, and   
was originally pointed out by Zeh \cite{Zeh86, Zeh88}).

Next I gradually lay out one way of looking at this breach.  
(Fact 1): in general solutions of HJ equations are complex.  
(Fact 2): HJ equations have 2 solutions $\fW^{\pm}$.  
(For the case in which the velocities feature solely homogeneous-quadratically in the kinetic term,  
these are $\pm$ which is `easy to omit', 
but more generally the 2 solutions are $\pm$ in the sense of being a complex conjugate pair.)
To each corresponds its own momentum.
(Fact 3): using the ansatz $\mbox{e}^{z\sfF}|\chi\rangle$ for z a complex number does not 
give bizarre equations where the above scheme gave a TDSE, because then $\fF$ itself is not 
associable with momentum.  
(Fact 4): as $\fW$ is in general complex, so are the corresponding momenta.    
This should be interpreted as in Sec 2.  
(Fact 5): there is a limitation on global validity of WKB emergent time a a problem of time resolution through 
$\fE - \fV$ having zeros.  
Often one will have oscillatory behavour on the one side and decaying behaviour on the other.  
In the present context, various careful studies have been made, allowing for more to be said.  
Not only is the WKB procedure invalid at the zeros, but  also 
near each zero, the WKB approximations are exceedingly poor. 
Thus a distinct approximation regime applies around each zero -- 
the one in the theory of connection functions.\foo{Though 
that standardly applies to ordinary differential equations, while 
the endemic nonseparability of HJ equations prevents passage to a collection of ordinary differential equations.  
Of course, that complication is avoided in the 1 H d.o.f. case that parallels many 
cosmological applications.}  
Thus one can not at all claim that a time arising from a WKB procedure is generically applicable 
over configuration space.   
%
%Zeh 86 says so in quantum cosmological context.  
%
Rather, one should expect a number of patches in configuration space where a different regime applies 
wherein emergent WKB time is not a valid answer to the problem of time.  
Additionally, if the zeros are sufficiently near to each other, 
there is no room for a WKB regime in the region between them, 
so applicability of a WKB procedure  is scarce in that region of configuration space. 
(Semi)classicality in the sense of WKB does need not occur everywhere or everywhen in a mechanical motion.

\noindent (Key fact 6): as developed by Barbour \cite{B93}, and endorsed e.g.  
by Isham \cite{Isham93}, Kucha\v{r} \cite{Kuchar92} and Joos \cite{Giu}), 
%p 2 of intro. 
the WKB ansatz is far from general from QM perspective.  
While considering just any $\mbox{e}^{i\sfF_{\tfA}}$ has the problem that just any pure such piece 
will not satisfy the deired equation, so  
$A_{1}\mbox{e}^{i\sfF_{1}} + A_{2}\mbox{e}^{i\sfF_{2}} + A_{3}\mbox{e}^{i\sfF_{3}} + ...$ does not usually make sense, 
by Fact 2, two particular such $\fF_{\sfA}$ are available: the $\fW^{\pm}$.   
Thus
\be
\{A_{+}\mbox{e}^{i\sfW_+} + A_{-}\mbox{e}^{i\sfW_-}\}|\chi\rangle
\label{breaker}
\ee
\cite{BS, B93, Zehbook} is, at the QM level, as good a guess as the WKB ansatz.  
In fact this guess is a better one, in the sense that it \cite{B93} and not the WKB ansatz \cite{Zeh88} 
is more general  -- there is 1 further d.o.f. in (\ref{breaker}).  
However, those cases of (\ref{breaker}) that are not WKB ans\"{a}tze, the chroniferous rearrangement 
ceases to work.  
[Moreover, there may be further scope for terms from `multiple saddles contributing'.]
Repeating working (\ref{swalk}--\ref{TDSEFULL}) with the generic case of (\ref{breaker}), 
one {\sl does not} obtain the emergent time term 
$i\hbar\pa/\pa t^{\mbox{\scriptsize emergent(WKB)}}$, 
but rather a cumbersome complex-valued term still containing $\fW_+$ and $\fW_-$.  
E.g. in the case of $A_+ = A_-$, $\fW \equiv \fW^- = - \fW^+$, 
one gets not a TDSE but a real equation somewhat reminiscent of a diffusion equation,
\be
\frac{\overline{\hat{h}}}{\hbar^2}|\chi\rangle = 4\frac{M}{\hbar}\mbox{tan}\left(\frac{M\fW}{\hbar}\right)
\left\{
\frac{\pa}{\pa t^{\sWKB}} - \left< \chi \left|\frac{\pa}{\pa t^{\sWKB}}\right|\chi\right>
\right\}|\chi\rangle \mbox{ }   
\ee 
(a new generalization of the result in \cite{B93}).  

That this is known to unpick the very emergence of a TDSE that is the conceptual basis for 
semiclassical quantum cosmology makes it untenable to consider the semiclassical approach by itself 
to furbish a satisfactory resolution of the problem of time (despite various claims in rather earlier literature).   
Saying that a WKB regime applies amounts to saying there is a pure wave, whose wavefronts pick out by orthogonality a 
direction which serves as timefunction.  
But this is `supposing time' rather than the sort of `bona fide emergence of time' that would be required 
to resolve the problem of time.  
In ordinary QM, one can justify the WKB ansatz as a lab set-up ``pure incoming wave".  
Or as being the product of the pre-existence of a surrounding classical large system as in \cite{Landau}.  
Or on resting on the constructive interference which underlies classicality \cite{GRT, Wheeler}, 
which amounts to applying the condition by hand to impose (semi)classicality rather than deducing 
that the world is semiclassical.  
But none of these arguments are meaningful in quantum cosmology.  
While doing so allows one to conceptualize and calculate, 
when it comes to the issue of whether one should be confident in the significance of certain 
calculations however (as is the case once contact with observations begins to appear to be possible), 
then the need to rigorously defend such a crucial unjustified assumption becomes important 
if the calculations are to be taken at all seriously as sources of prediction.

\noindent\underline{Some further justifications that have been suggested.}

\noindent 1) If the many worlds interpretation of QM is adopted,  
each piece of the typical superposition is realized in a different branch 
(see e.g. in \cite{GHS87}).  
However, one may be able to object that this is not in accord with actually experiencing a superposition.    

\noindent 2) Hopes (see e.g. Papers of Brout, Kiefer or Datta in the Bibliography, and also 
\cite{Giu}) 
have been expressed that the WKB ansatz will be independently justified by {\sl decoherence} 
\cite{Giu}.  
But these hopes come with reservations,  
e.g. \cite{Kuchar92, Isham93, CG, Giu, HT}
are between not entirely and far from optimistic about this.   

\noindent 3) Another idea is that the best one may be able to do is to impose 
a future boundary condition -- that we observe (semi)classicality here today.  
This is somewhat similar to Griffiths and Omn\`{e}s removing by hand the superposition 
states they term ``grotesque universes" \cite{Gri} due to their behaviour being 
other than we experience today.    
And to how Bojowald \cite{Bojowald} selects 
which solutions of the loop quantum gravity Hamiltonian constraint to keep. 
Though, this is disappointing in that the Copenhagen interpretation would continue to cast a shadow in 
in a part of physics in which it has no buisness to be.   
One would much prefer for quantum cosmology to be the science from which the permanently semiclassical 
behaviour of the large-scale features of the universe is a deduced feature.

A different perspective to postulating unproven suggestions is that B2 can be put to the test, 
by investigation for classes of ulteriorly exactly-soluble models.  
An extension of this would be to use the na\"{\i}ve Schr\"{o}dinger interpretation 
to provide the relative probabilities of experiencing a WKB regime within a given range of model universes.     
%

%=======================================================================================================
\noindent\underline{The many approximations problem.} 
%=======================================================================================================
%
%
%
One contribution the present Paper makes to the above debate is that it is hard to meaningfully isolate 
the testing of whether the WKB condition applies due to the plethora of other approximations made.  
It would seem that at best one 
can make a number of such and then test whether the WKB approximation holds in the 
small region of the configuration space where all those approximations are applicable.  
Thus it would in general be very drawn-out to carry out the above-suggested programs.

\noindent\underline{The many-approximations problem in connection with Detail 3.}
Some Papers \cite{KS91, Semicl, Kiefersugy} investigate quantum cosmology by expanding 
in 1 parameter.  
That however there are multiple parameters was pointed out by Padmanabhan \cite{Pad}, 
and is investigated explicitly in the present Paper.
While \cite{Pad} proceeded by considering which parameter to expand in, in the present Paper   
I point out rather that 1-parameter expansions in no matter what parameter will not in general suffice 
for beyond a corner of the quantum cosmology solution space.   
In general one would have to expand in many independent parameters.
Careful theoretical arguments may however then match certain frameworks 
with less parameters to certain relevant situations to various degrees of accuracy. 
For some consideration as to what regimes are required in GR cosmology, see Paper II.

%=======================================================================================================
\noindent\underline{Details 1 and 2 concerning back-reactions.}
%=======================================================================================================
%
%
%
The full H- and L-equations provided by my scheme are indeed intercoupled.  
However, the approximate H-equation used above contains no back-reaction.  
That's ultimately unsatisfactory for reasons given in Sec 1.  
Moreover if a cross-term is to be kept in the L-equation because of its chroniferous 
qualitative effect, it should be noted that there is a further interrelated pair of such terms, 
one in each of the full H- and L-equations.  
One may then well wish to consider keeping these terms (which is one way of ensuring that there 
is a back-reaction of the L-subsystem on the H-equation.  
[If one preliminarily asserts ``connection term neglect", this point is missed.]  
Schematically, 
\be
i\hbar <\pa/\pa t^{\sWKB}> = ||\pa_{\sH} \fW||^2 - \fE_{\sH} + \fV_{\sH} + \mbox{correction terms} 
\mbox{ } , \mbox{ }
\ee
\be
i \hbar \pa /\pa t^{\sWKB}|\chi> - i \hbar <\pa/\pa t^{\sWKB}>|\chi> = 
\widehat{\ttH}_{\mbox{\scriptsize effective}}(\mbox{L-physics})|\chi> 
\mbox{ } .
\ee

I consider two ways of handling back-reactions.

\noindent Way 1: implement by considering ab initio harder H-equations.
Manipulating this in parallel to the way the L-equation's cross-term is treated 
would seem to be a reasonable procedure.  
But that changes the qualitative type of the system:   
instead of a background H-HJ equation and a L-subsystem TDSE, one now 
has a brace of coupled integro-differential TDSE's.   
This amounts to not neglecting one of the back-reaction terms: via the connection the 
L-subsystem induces on the H-subsystem, the latter also picks up a time with respect 
to which it can be taken to run.  
This time and the L-system's 2 are, however different, as the averaged and unaveraged 
cross-terms will in general differ in size.  
Thus the notion of time emerging at such a level of detail in the semiclassical approach may indeed be 
taken to be a consequence of the entirety of the universe's contents -- this level of detail is 
sufficient for the emergent time in a WKB regime to manifest this characteristic property of its alter 
ego, LMB time.  
While the H-subsystem provides most of the timestandard like a background 
time, the L-subsystem itself has a small say in the form of the timestandard.  
Thus, if arbitrary precision is required, one has now no choice but to treat a coupled system.
The previously-suggested simple procedure of solving the H-HJ equation first is insufficient 
by itself to capture this level of detail.  
This makes sense, as $\fW$ is for conservative systems while if H interacts (weakly as that may be) with 
the L-subsystem, one expects the H-subsystem then to be more general than conservative \cite{Lanczos}.  
%`rheonomic' of p 237 Lanczos Dover Ed.  

Some previously proposed more accurate H-equation types are the habitual 
\be
\mbox{}_{\sM^{-1}}||\pa_{\sH}\fW||^2  = \fE_{\sH}   + \fU_{\sH} - \langle \ttH_{\sH\sL} \rangle 
\label{firstcorr} 
\mbox{ } ,
\ee
and Datta's
\be
\mbox{}_{\sM^{-1}}||\pa_{\sH}\fW||^2  = \fE_{\sH}   + \fU_{\sH} - e \mbox{ }  .
\label{secondcorr}
\ee
Various even fuller schemes can be assembled by retaining both of the above corrections 
and (or) retaining connection terms that correct the $\pa_{\sH}$'s.

\noindent Way 2: in loose analogy to how astronomers deal with ephemeris time at the classical level, 
one could implement the above further level of detail by an iterative scheme.

More accurate H-equations can be obtained once the L-equation has been approximately solved.
Using the approximate HJ H-equation to cancel zeroth order terms off and assuming the new double 
derivative term is negligible, I get a novel HJ equation for the H-correction:
\be
\{\fW_{1,\sH}\}^2 + 2k(\nH)\fW_{1,\sH} + K^2(\nH) = 0 \mbox{ } ,
\ee
for
\be
k(\nH) = \sqrt{2M\{\fE_{\sH} + \fU_{\sH}  \}    } - {i\hbar}\langle\chi| \pa_{\sH} |\chi\rangle    
\mbox{ } ,
\ee
\be
\frac{K^2(\nH)}{2} = 
\frac{i\hbar}{2}
\sqrt{\frac{M}{2}}        
\frac{\fU_{\sH,\sH}        }
     {        \sqrt{ \fE_{\sH} + \fU_{\sH}    }        }
-i\hbar\sqrt{        2M\{\fE_{\sH} + \fU_{\sH}\}        }\langle\chi| \pa_{\sH} |\chi\rangle
-\frac{\hbar^2}{2}\langle\chi| \pa_{\sH}\mbox{}^2 |\chi\rangle
-\hbar^2M\langle\chi| \mbox{}_{\sL}||\pa_{\sL}||^2 |\chi\rangle
+M\langle\chi| \fV_{\sL} |\chi\rangle
+M\langle\chi| \fI_{\sH\sL} |\chi\rangle    
\mbox{ } ,
\ee
so (continuing to use the + choice) 
\be
\fW_{1} = - k(\nH) + \sqrt{   k^2(\nH) - K^2(\nH)    }   \mbox{ } .   
\ee
Correspondingly, 
\be
t^{\sem}_{1} - t^{\sem}_{1}(0) = \int^{\sH} 
\frac{    \d \nH^{\prime}    }
     {    \sqrt{    2M\{\fE_{\sH} + \fU_{\sH}\}    } 
          - k(\nH^{\prime}) +  \sqrt{  k^2(\nH^{\prime}) - K^2(\nH^{\prime})  }     } \mbox{ } .  
\ee
Continuing the working requires explicit solution of the L-equation.  Then everything above 
is a known quantity so have an explicit H-equation.

Both $k$ and $K$ contain back-reaction contributions. 
The $\d|\chi\rangle/\d t^{\sem}$ term has been reformulated and absorbed into the $k$.

%=======================================================================================================
\noindent\underline{Need for complexity in order to recover some aspects of reality (B3).}
%=======================================================================================================
%
%
%
Note that in models with no back-reaction term, there is no built-in nontrivial book-keeping 
of the fact that the universe is in a fixed energy eigenstate.  
A second issue is whether the energy in the H-equation should be interpreted as the energy of the universe 
$\fE$ or as an $\fE_{\sH}$ that is only approximately equal to this.

I have thought of the following approaches to this.  
1) To simply ignore $\Delta \fE_{\sL}$ in comparison to $ \fE_{\sH}$.  
But this is conceptually dubious as fluctuations could gradually build up. 
2) Alternatively, to give a distinct $t^{\seBW}$ per $\fE_{\sH}$.  
Thus the H-problem now amounts to solving a 1-parameter family of HJ 
equations and permitting the trajectory to slip between these in response to the L-physics.  
Then L transition by $+ \Delta \fE$ shifts $\fE_{\sH}$ by $-\Delta\fE$.  
Then one has a slightly different standard with respect to 
which further transition rates are to be calculated.  
This happens for each transition in the HL model, 
but only occasionally within a HLL model with a large frequency hierarchy, so that most of the 
transitions are LL, and the interaction with the large-scale mode of the universe 
produces only a tiny correction to the motion as one would intuitively expect (see also \cite{OII}).   
Note that in absense of interaction terms, not only does the timefunction contain a zero factor but also 
the L-equation is then as frozen as the whole HL-system is, in which case the approach to B3 delineated 
in the Introduction fails.

Next, I describe the frontiers of knowledge to the above set-up.  
1) In the nonadiabatic case, keep 1 H d.o.f. and 1 L d.o.f. but now prescribe these to 
`go at same speed', so that they can nonsuppressively exchange energy.  
2) Study the more complicated but more conceptually and technically 
accurate systems in which the back-reaction is not ignored.  
Then the system's mathematics is good enough to handle the energy balance by itself.  
Adiabaticity is o.k. for a molecule in a big universe with many other molecules and 
mediating particles (photons in relativistic cosmology -- see Sec II.10).   
Note that this fits our understanding of the universe better, 
and it also leans on an existing, detailedly worked-out framework from molecular physics.

\noindent\underline{Detail 4.} 
\noindent Datta \cite{Datta97} reproaches that use of straight equation as this entails removing phase, 
which is generally inadmissible due to its acquiring a physical meaning.  
Or, geometrically speaking, this is an in general illegal simplification of fibre bundle theory, 
which masks or distorts mathematical issues corresponding to reasonably extensive pieces 
of the bundle space.
This issue is moreover quite subtle, requiring 
quite some conceptualization and nomenclature to discuss. 
Introduce $\gamma_{\mbox{\scriptsize TNS}}$, the {\sl total phase} in the case that is naturally-splittable 
by existence of an external time,
\be
\gamma_{\mbox{\scriptsize TNS}} = \gamma_{\mbox{\scriptsize geometric}} + 
\gamma_{\mbox{\scriptsize evolutionary}}
\ee
In the case in which there is no such external time however, one has, rather, a total phase 
$\gamma_{\mbox{\scriptsize T}}$
for which there is no natural such split; moreover this can be thought of as entirely geometrical, 
\be
\gamma_{\mbox{\scriptsize T}} = \gamma_{\mbox{\scriptsize geometric}^{\prime}} \mbox{ } . 
\ee
Geometric above relates to quantum phase geometry.  
The key point is that the primed and unprimed notions of geometry above are different.  
In particular, $\gamma_{\mbox{\scriptsize T}}$ is both gauge invariant (as befits its nature as a 
geometric quantum phase) and reparametrization invariant (as befits a total dynamical phase of a theory 
with no external time).  
This means overall that the primed, {\it relative quantum phase geometry} has more unphysical quantities ('overall 
gauge freedom') than the unprimed {\it absolute quantum phase geometry}.  
This means that more terms 'are gauge rather than physical' if relative quantum phase is in use.  
Being different in this way, which is used leads to differences in what is purportedly physical, 
so one should take due care to use whichever is appropriate for the situation in hand.  
Non-external time requires relative rather than absolute quantum phase geometry.  
One consequence is that gauge choice in the original sense (no inverted commas) 
also causes the zero-point energy to be shifted so that $\ttH_{\sH\sL}$ is 
renormalized to $\ttH_{\sH\sL} - <\ttH_{\sH\sL}>$.  
This then causes $<\ttH_{\sH\sL}>$ to drop out of the H-equation, hence the origin of equation 
(\ref{secondcorr}) [or more highly corrected forms]. 
The physics is then only in the fluctuating quantities, invalidating the unbarring 
of L equations also.  
Another consequence is that the argument in \cite{BV89} that a 1-d configuration space for H-subsystem 
gives automatic neglect of connection terms is nullified as regards phase effects that are relative.    
Thus the very simplest toy models can be investigated to see if relative phase effects cause 
major or minor alterations to `conventional wisdom'.  
That includes RPM's due to their implementation of temporal relationalism, 
even in the easiest cases in which nontrivial spatial relationalism is absent.

%========================================================================================================
%========================================================================================================
\section{This Paper's new procedure applied to a simple model}
%========================================================================================================
%========================================================================================================

I consider the 1-d case of a single relational H-coordinate with  
\be
\fT_{\sH} = \frac{M \dot{\nH}^2}{2} \mbox{ } , \mbox{ } 
\fV_{\sH} = \frac{M\Omega^2\nH^2}{2} \mbox{ } ,   
\ee
and a single relational L-coordinate with
\be
\fT_{\sL} = \frac{ m   \dot{\nL}^2}{2} \mbox{ } , \mbox{ } 
\fV_{\sL} = \frac{m\omega^2\nL^2}{2} \mbox{ } ,  
\ee
that are linearly-coupled:
\be
\fI_{\sH\sL} = C\nH\nL \mbox{ } .  
\ee
Then the energy constraint is
\be
\ttH \equiv \frac{\nP_{\sH}^2}{2M} + \frac{\nP_{\sL}^2}{2m} + 
\frac{M\Omega^2\nH^2}{2}      + \frac{m\omega^2\nL^2}{2} + C\nH\nL = \fE \mbox{ } ,
\ee
which, upon quantizing as described in Sec 4, gives the TISE
\be
-\frac{\hbar^2\pa_{\sH}\mbox{}^2|\Phi\rangle}{2M}  - \frac{\hbar^2\pa_{\sL}\mbox{}^2|\Phi\rangle}{2m}   + 
\frac{M\Omega^2\nH^2}{2}|\Phi\rangle + \frac{m\omega^2\nL^2}{2}|\Phi\rangle + 
C\nH\nL|\Phi\rangle = \fE|\Phi\rangle \mbox{ } .
\ee
N.B. the states are nondegenerate here, so the subtleties of footnote 8 are not required.  
%* fn number
%
Then, after applying the BO ansatz, step ADE of Sec 5, and the WKB ansatz, the H-equation is 
\be
\frac{\{\fF_{,\sH}\}^2}{2M} - \frac{i\hbar}{2M}\fF_{,\sH\sH} - 
\frac{i\hbar}{M}\fF_{,\sH}\langle\chi|\pa_{\sH}|\chi\rangle - 
\frac{\hbar^2}{2M}\langle \chi | \pa_{\sH}\mbox{}^2 | \chi\rangle -
\frac{\hbar^2}{2m} \langle \chi | \pa_{\sL}\mbox{}^2 | \chi\rangle + 
\frac{M\Omega^2\nH^2}{2} + 
\frac{m\omega^2}{2}|\langle \chi |\nL^2 | \chi\rangle + 
C\nH\langle \chi | \nL| \chi\rangle = \fE_{\sH}
\ee
and the L-equation is
\be
\{1 - P_{\chi}\}
\left\{
\frac{2i\hbar\fF_{,\sH}\pa_{\sH} + \hbar^2\pa_{\sH}\mbox{}^2}{2M} + 
\hbar^2\frac{\pa_{\sL}\mbox{}^2}{2m} - \frac{m\omega^2{\nL}^2}{2} 
- C\nH\nL
\right\}|\chi\rangle = 0  
\mbox{ } .  
\ee
Adopt the `coarsest scheme' for the H-equation:
\be
\frac{\{\fF_{0,\sH}\}^2}{2M} = \fE_{\sH} - \frac{M\Omega^2\nH^2}{2} \mbox{ } .  
\label{yoda}
\ee
This is a HJ equation, justifying $\fF_0 \longrightarrow \fW_0$ relabelling, 
and which is solved by 
\be
\fW_0^{\pm} = \pm \sqrt{2M\fE_{\sH}}\int\sqrt{1 - M\Omega^2\nH^2/2\fE_{\sH}} \mbox{ } ,
\ee
which is capable of being both real and imaginary.  
I take the + sign option.  
The oscillatory motion corresponds to $\fW_0$ real ($\fE_{\sH} > M\nH^2/2$).
Additionally, the integral can be evaluated analytically:   
\be
\fW = \sqrt{\frac{M\fE_{\sH}}{2}}\nH\sqrt{1 - M\Omega^2\nH^2/2\fE_{\sH}} + \frac{\fE_{\sH}}{\Omega}
\mbox{Arcsin}
\left(
\sqrt{\frac{M}{2\fE_{\sH}}}\Omega\nH
\right) + \mbox{ const}\mbox{ } .  
\ee
Also,
\be
M \frac{    \pa \nH    }{    \d t_0^{\sem}   } = 
\frac{M}{\mbox{n}}\dot{\nH}  = 
\nP_{\sH} = 
\fW_{0,\sH} = 
 \sqrt{2M\fE_{\sH}}\sqrt{1 - {    M\Omega^2\nH^2    }/{    2\fE_{\sH}    }    } 
\mbox{ } ,
\ee
so, 
\be
t_0^{\sem} - t_0^{\sem}(0) = 
\sqrt{        {    M    }/{    2\fE_{\sH}     }    }
\int{        \d {\nH}^{\prime}        }/{        \sqrt{      1 - {    M\Omega^2{\nH}^{\prime 2}    }/
                                                            {    2\fE_{\sH}    }      }        } 
= 
\frac{1}{\Omega}\mbox{arcsin}
\left( 
\sqrt{\frac{M}{2\fE_{\sH}}}
\Omega\nH
\right) 
\label{barde}
\mbox{ } , 
\ee  
which inverts to
\be
\nH =\sqrt{\frac{2\fE_{\sH}}{M\Omega^2}}\mbox{sin}(\Omega\{t_0^{\sem} - t_0^{\sem}(0)\})
\mbox{ } .    
\label{protopert}
\ee

A self consistency check possible at this level is, using, 
\be
\sqrt{1 - {    M\Omega^2\nH^2    }/{    2\fE_{\sH}    }    } = 
\mbox{cos}(\Omega\{t_0^{\sem} - t_0^{\sem}(0)\}) \mbox{ } ,
\label{L1}
\ee
\be
\pa_{\sH} = \frac{1}{\sqrt{2M\fE_{\sH}}\mbox{cos}(\Omega\{t^{\sem} - t^{\sem}(0)\})}
\frac{\pa}{\pa t_0^{\sem}}
\mbox{ } ,
\label{L2}
\ee
\be
\pa_{\sH}\mbox{}^2 = \frac{1}{2M\fE_{\sH}}\left\{
\frac{1}{  \mbox{cos}^2(\Omega\{t_0^{           \sem} - t_0^{\sem}(0)\})}
\frac{\pa^2}{\pa t_0^{\sem}\mbox{}^2}
+ \frac{\mbox{sin}(\Omega\{t_0^{\sem} - t_0^{\sem}(0)\})}
       {\mbox{cos}^3(\Omega\{t_0^{\sem} - t_0^{\sem}(0)\})}
\frac{\pa}{\pa t^{\sem}}
\right\}
 \mbox{ } ,
\label{L3}
\ee
\be
\epsilon_{\ssWKB} = \left|
\frac{  \mbox{sin}(\Omega\{t_0^{\sem} - t_0^{\sem}(0)\})  }
     {\mbox{cos}(\Omega\{t_0^{\sem} - t_0^{\sem}(0)\})}
\frac{\hbar\Omega}{2\fE_{\sH}}
\right| \mbox{ } ,
\ee
which is only necessarily small if $\fE_{\sH} >> \Delta \fE_{\sH}$ (or $t_0^{\sem} - t_0^{\sem}(0)$ 
is very small).  
Moreover, in the hyperbolic analogue (App. D), $t^{\sem}_0$ large has prefactor go as 
$\mbox{e}^{-2|\Omega|t_0^{\sem}}$, 
which opens up another way to attain the smallness.

%=======================================================================================================
\noindent\underline{Subsequent form of the L-equation.} 
%=======================================================================================================
%
%
%
I obtain
$$
\{1 - P_{\chi}\}
\left\{
\left\{
i + \frac{\hbar}{2M\fE_{\sH}}
\frac{\mbox{sin}(\Omega\{t_0^{\sem} - t_0^{\sem}(0)\})}{\mbox{cos}^3(\Omega\{t^{\sem} - t^{\sem}(0)\})}
\right\}
\hbar\frac{\pa |\chi\rangle}{\pa t_0^{\sem}} 
+
\frac{\hbar^2}{2M\fE_{\sH} \mbox{cos}^2(\Omega\{t_0^{\sem} - t_0^{\sem}(0)\})   }
\frac{\d^2|\chi\rangle}{\d t_0^{\sem}\mbox{}^2}
+ 
\hbar^2\frac{\pa_{\sL}\mbox{}^2|\chi\rangle}{2m} 
\right.
$$
\be
\left.
- 
\left\{
\frac{m\omega^2\nL^2}{2} + 
\frac{C}{\Omega}\sqrt{\frac{2\fE_{\sH}}{M}}\mbox{sin}(\Omega\{t_0^{\sem} - t_0^{\sem}(0)\})
\right\}|\chi\rangle
\right\} \mbox{ } ,
\label{pertmaster}
\ee
which is not always of Schr\"{o}dinger form unless certain further assumptions are made. 
Note that the first smallness is tied to the form of the consistency condition that the second derivative 
of $\fW_0$ be small being small (see preceding subsection).  
But it is the second smallness, rather, that concerns whether the equation is more KG-like.

%=======================================================================================================
\noindent\underline{Substituting back into the H-equation.}
%=======================================================================================================
%
%
%
This gives, expanding $\fW = \fW_0 + \fW_1$, using the $\fW_0$ HJ equation to cancel off some terms 
and considering $\fW_{1,\sH\sH}$ to be negligible:
\be
\{\fW_{1,\sH}\}^2 + 2k(\nH)\fW_{1,\sH} + K^2(\nH) = 0 
\ee
for 
\be
k(\nH) = \sqrt{  2M\fE_{\sH}   }\sqrt{1 - {M\Omega^2\nH^2}/{2\fE_{\sH}}} - 
{i\hbar}\langle \chi |\pa_{\sH}| \chi \rangle \mbox{ } , 
\ee
$$
K^2(\nH) = 2
\left\{
i\hbar\sqrt{\frac{M}{2\fE_{\sH}}}\frac{M\Omega^2 \nH}{2 \sqrt{1 - {M\Omega^2\nH^2}/{2\fE_{\sH}}}  }
-i\hbar \sqrt{2M\fE_{\sH}}\sqrt{1 - {M\Omega^2\nH^2}/{2\fE_{\sH}}}\langle \chi |\pa_{\sH}| \chi \rangle 
+ MC\nH\langle \chi |\nL| \chi \rangle
- \frac{\hbar^2}{2}\langle \chi |\pa_{\sH}\mbox{}^2| \chi \rangle 
\right.
$$
\be
\left.
- \frac{\hbar^2M}{2m}\langle \chi |\pa_{\sL}\mbox{}^2| \chi \rangle
+ \frac{mM}{2}\omega^2\langle \chi |L^2| \chi \rangle
\right\}
\mbox{ } .  
\ee
Then 
\be
\fW_{1} = - k(\nH) + \sqrt{   k^2(\nH) - K^2(\nH)    }    \mbox{ } . 
\ee
Correspondingly, 
\be
t^{\sem}_{1} - t^{\sem}_{1}(0) = \int^{\sH} 
\frac{    \d \nH^{\prime}    }
     {    \sqrt{   2M\fE_{\sH}  }\sqrt{  1 - {M\Omega^2\nH^{\prime 2}}/{2\fE_{\sH}}  } 
          - k(\nH^{\prime}) +  \sqrt{  k^2(\nH^{\prime}) - K^2(\nH^{\prime})  }     } \mbox{ } .  
\ee

Further progress involves solving the time-dependent perturbation of the TDSE.  
Then $|\chi\rangle$ appears as an explicit function, so k and K can be straightforwardly computed.

\noindent\underline{Proceeding with a simple perturbative treatment.}
As a simple example, suppose that only the potential perturbation is kept and that fluctuation 
terms are ignored.    
The validity of perturbation theory requires that 
\be
\epsilon_{\mbox{\scriptsize pert}} = |\fI_{\sH\sL}/\fV_{\sH}| 
\ee
itself be small.

The unperturbed equation is solved by eigenfunctions 
\be
|\fj\rangle = \frac{1}{\sqrt{2^{\sfj}\fj !}}
\left\{
\frac{m\omega}{\hbar\pi}
\right\}^{\frac{1}{4}}
\mbox{Hermite}_{\sfj}(\sqrt{m\omega/\hbar}\nL)\mbox{exp}(-m\omega \nL^2/2\hbar)
\ee
corresponding to eigenvalues $\hbar\omega(\fj + 1/2)$.  
For the perturbation, consider the above to be the situation at some $t^{\sem}_0(0)$.
Then
\be
|\fl(t^{\sem}_0)\rangle = 
\sum_{\sfj} \langle \fl(t^{\sem}_0)|\fj(t^{\sem}_0(0)\rangle \mbox{ } |\fj(t^{\sem}_0(0))\rangle
\mbox{exp}(-i\fE_{\sL}t^{\sem}_0/\hbar)
\label{spangle}
\ee
for transition (probability) amplitudes
\be
\langle \fl(t^{\sem}_0)| \fj(t^{\sem}_0(0))\rangle = 
\delta_{\sfl \sfj} - \frac{i}{\hbar}\int_{t^{\sem\prime}_0 = 
t^{\sem}_0(0)}^{t^{\sem}_0}\d t^{\sem\prime}_0 
\langle \fl(t^{\sem}_0(0))|\fI(t^{\sem\prime}_0)|\fj(t^{\sem}_0(0) )\rangle
\mbox{exp}({i\{\fE_{\sfl} - \fE_{\sfj}\}t^{\sem\prime}_0}/\hbar)
\ee
to first order in perturbation theory [i.e. with $O(\epsilon_{\spert}^2)$ corrections].  
Expressing the L in $\fI_{\sH\sL}$ in terms of the creation and annihilation operators of the 
unperturbed HO, and using their action on the eigenstates alongside the orthonormality of these, I obtain 
$$
\langle \fl(t^{\sem}_0)| \fj(t^{\sem}_0(0))\rangle = \delta_{\sfl \sfj}
- \frac{C}{2\Omega}\sqrt{\frac{\fE_{\sH}}{Mm\hbar\omega}}
\{\sqrt{\fj}\delta_{\sfl, \sfj - 1} + \sqrt{\fj + 1}\delta_{\sfl, \sfj + 1}\} \mbox{ } \times
$$
\be
\left\{
\frac{    \mbox{exp}{(i\{\{\Omega + \{\fl - \fj\}\omega\} t^{\sem}_0 - \Omega t^{\sem}_0(0)\})}
        - \mbox{exp}{(  i\{\fl - \fj\}\omega t^{\sem}_0  )}      }
     {i\{\Omega + \{\fj - \fl \}\omega\}    }           -
\frac{          \mbox{exp}(  i\{          \{    -\Omega + \{  \fl - \fj  \}\omega    \}t^{\sem}_0 
                       + \Omega t^{\sem}_0(0)          \} )           
              - \mbox{exp}(  i\{  \fl - \fj  \}\omega t^{\sem}_0 )              }
     {          i\{- \Omega + \{\fj - \fl \}\omega\}              }
\right\}
\ee
(which is a slight generalization of a computation in \cite{Landau}).  
If the classical adiabaticity approximation applies, this simplifies to 
$$
\langle \fl(t^{\sem}_0)| \fj(t^{\sem}_0(0))\rangle = \delta_{\sfl\sfj}
- C \sqrt{\frac{\fE_{\sH}}{Mm\hbar\omega}}
\{\sqrt{\nn}\delta_{\sfl, \sfj - 1} + \sqrt{\fj + 1}\delta_{\sfl, \sfj + 1}\} \mbox{ } \times
$$
\be
\left\{
\frac{     \mbox{exp}{(  i\{\fl - \fj\}\omega t^{\sem\prime}_0  }\{t_0^{\sem} - t_0^{\sem}(0)\})      }
     {     \{\fl - \fj \}\omega     }      +  
\frac{                    i
\left\{
\mbox{exp}(   i\{\fl - \fj\}\omega t^{\sem\prime}_0  ) - 
\mbox{exp}(   i\{\fl - \fj\}\omega t_0^{\sem}(0)     )     
\right\}                 }  
     {     \{\fl - \fj \}^2\omega^2    }         
\right\} 
+ O(\epsilon_{\sa}^3) + O(\epsilon_{\spert}^2)  \mbox{ } .  
\ee
Thereby, (\ref{spangle}) becomes 
$$
|\fj(t_0^{\sem})\rangle \approx
\mbox{exp}{(-i\fj\omega t_0^{\sem})}
\left\{
|\fj(t_0^{\sem}(0))\rangle - \frac{C}{\omega^2}\sqrt{\frac{\fE_{\sH}}{Mm\hbar\omega}} \mbox{ } \times
\right.
$$
\be
\left.
\left\{
\sqrt{\fj + 1}
\left\{
\stackrel{       \mbox{$\omega\{t_0^{\sem} - t_0^{\sem}(0)\} +$}        }
         {       \mbox{$i\{1 - \mbox{exp}(  i\omega  \{  t_0^{\sem}(0) - t_0^{\sem}  \}  ) \}   $}        }
\right\}
|\fj(t_0^{\sem}(0)) + 1\rangle
+
\sqrt{\fj }
\left\{
\stackrel{         \mbox{$ -\omega\{t_0^{\sem} - t_0^{\sem}(0)\} + $}        }
{         \mbox{$i 
\left\{
1 - \mbox{exp}(-i\omega \{  t_0^{\sem}(0) - t_0^{\sem} \}  )
\right\}$}      }
\right\}
|\fj(t_0^{\sem}(0)) - 1\rangle
\right\}
\right\} \mbox{ } .  
\ee
Analysis of the coefficient of the correction term reveals it to be 
$\frac{\epsilon_{\mbox{\tiny pert}}\epsilon_{\sa}}{\sqrt{\epsilon_{\tT}}}
\sqrt{\frac{\Delta \sfE_{\sL}}{\sfE_{\sH}}}$ which will be small unless, e.g.,  
$\epsilon_{\sT}$ is vastly smaller than $\epsilon_{\spert}$ or $\epsilon_{\sa}$.

Now, the action of $\nL$, $\nL^2$, $\pa_{\sL}\mbox{}^2$ on this is straightforwardly computible  
using creation--anihilation operators.  
One may then compute the action of $\pa_{\sH}$, $\pa_{\sH}\mbox{}^2$ or, passing from H to $t^{\sem}_0$, 
the action of the first and second derivatives with respect to the emergent time.  
Thus, one can specifically check which of the epsilons are indeed small in this regime.  
And also, one can build the next level H-equation.  
These calculations are too lengthy to enclose here, so I merely sketch some notes about them.  
They show that considering just the above perturbation to first order is reasonable if 
$\epsilon_{\spert}^2$ is very small and 
$\Delta \fE_{\sL}, \Delta \fE_{\sH} << \fT_{\sL}, \fT_{\sH}, \fE_{\sH}, \fV_{\sH}$ (i.e. 
that the gaps in the energy levels are much smaller than the energies of each system involved).  
Whether the double time derivative  and the correction to the the prefactor of the 
single time derivative are negligible is interesting due to the qualitative changes 
to the system if these are kept.  
See Paper II. 
See also Appendix E for a useful checking method that is available for the above model for detailed 
investigations of such schemes.  
%
%Presenting the above calculation could be a further paper.  

%==========================================================================================================================================
%==========================================================================================================================================
%=====================================================================================================
\section{Internal time approach}
%=====================================================================================================
%==========================================================================================================================================
%==========================================================================================================================================

I provide this Section for contrast of its timefunctions and their properties with those 
of the semiclassical approach.

%==========================================================================================================================================
\noindent\underline{Scale and Euler times.}
%==========================================================================================================================================
%
%
%
One might consider using (a function of) scale as a timefunction, e.g. 
$t^{\mbox{\scriptsize scale}} = \frac{1}{2}\mbox{log}J$ for $J$ the moment of inertia.
However, the sign of the derivative of such can sometimes reverse, so it fails the monotonicity 
criterion for a timefunction.  
Now, switching the coordinate and momentum statuses of the conjugate quantities $\sqrt{J}$ and 
$E \equiv \sum_i \P^i\cdot\R_i$ by a canonical transformation has the advantage that the 
dilational object $E$ has monotonicity guaranteed, at least in certain substantial sectors \cite{OII}, 
by the Lagrange--Jacobi relation, 
\be
\dot{E}  = 2\fT + l\fV = 2\fE + \{l - 2\}\fV
\label{LJR}
\ee  
for systems whose potentials are homogeneous of degree $- l$.
One such sector, for which scale can `bounce', is 

\noindent $S_1 = \{ \fE > 0, {\cal V} \leq 0, l < 2\}$, where l is minus the 
degree of the power of the homogeneous potential ${\cal V}$.  
Here, $E = t^{\sE}$ serves better as a time.  
This is a close parallel of the advantages in certain sectors of York's dilational time 
over Misner's scale time (see II.9).  
Though note that there are other sectors in which both scale time and Euler time 
can serve as monotone time functions, e.g. $S_2 = \{ \fE > 0, {\cal V} \geq 0, l < 2\}$.

%==========================================================================================================================================
\noindent\underline{Examples of use of Euler time.}
%==========================================================================================================================================
%
%
%
Simple specific examples for which the Euler time is suitably monotonic are (c.f. Appendix D):  
1) two upside-down HO's, with the magnitude of the coupling constant $|C|$ not exceeding $\sqrt{mM}wW$. 
2) The free-free subcase of that ($C = w = W = 0$).  
I will also show that case 3) of a free--HO system has monotonic Euler internal time locally in 
internal time: on certain finite intervals.

%==========================================================================================================================================
\noindent\underline{Shape--scale coordinates for homogeneous quadratic potential HL models.}
%==========================================================================================================================================
%
%
%
These examples employ a version of the 1-d, 3-particle technique of \cite{OII}. 
Namely, the version with masses kept explicit, for ease of comparison with preceding Sections.   
This technique is based on the changing to the configuration space `polar coordinates' 
\be
\sqrt{{M_1}/{f}}\mR_1 = \mbox{e}^{\sigma}\mbox{sinS}  \mbox{ } , \mbox{ }
\sqrt{{M_2}/{f}}\mR_2 = \mbox{e}^{\sigma}\mbox{cosS}  \mbox{ } ,  
\ee
which invert to
\be
\sigma = \mbox{ln}(J)/2 = \{1/2\}\mbox{ln}
\left(
\{{M_1\mR_1^2 + M_2\mR_2^2}\}/{f}
\right)
\mbox{ } , \mbox{ }
\mS = \mbox{arctan}
\left(
{\sqrt{M_1}\mR_1}/{\sqrt{M_2}\mR_2}
\right) 
\mbox{ } ,
\ee
for $J$ the moment of inertia and $f$ a convenient fiducial quantity serving 
to straighten out the physical units.
In terms of the new coordinates, 
the action $\fS_{\sJ} = \int \fL \d\lambda = \int \sqrt{\{\fE - \fV\}\fT}\d\lambda$'s 
kinetic term $\fT$ is the conformally flat expression 
\be
\fT = f \mbox{e}^{2\sigma}\{\dot{\sigma}^2 + \dot{S}^2\}
\mbox{ } .
\ee
The conjugate momenta are then 
\be
\mP_{\sigma}   \equiv \frac{\pa \fL}{\pa\dot{\sigma}} = f\mbox{e}^{2\sigma}\dot{\sigma}/\mbox{n} 
\mbox{ } , \mbox{ }
\mP_{\sS}^i \equiv \frac{\pa \fL}{\pa\dot{\mS}} = f\mbox{e}^{2\sigma}\dot{\mS}/\mbox{n}
\mbox{ } .
\ee
Then, working via the original coordinates, I observe that
\be
\mP_{\sigma} = \sum_{i = 1}^2 
\left\{
\dot{\R}_i
\right\} \cdot \R_i/\mbox{n}
= \sum_{i = 1}^2\P_i\cdot \R_i \equiv t^{\sE} \mbox{ } .
\ee

%==========================================================================================================================================
\noindent\underline{The internal time method.}
%==========================================================================================================================================
%
%
%
Following the method outlined on p. 4,
%*page 
one requires a canonical transformation such that 

\noindent $t^{\sE}$ is now a coordinate and 
$-\sigma$ is its conjugate momentum \cite{OII}.   
$\ttH = \fE$ is to be replaced by the explicit expression 
$\mP_{t^{\tE}} \equiv -\sigma = -\sigma(t^{\sE}, \mS; {\mP}_{\sS}) \equiv 
\ttH^{\mbox{\scriptsize true}}(t^{\sE}, \mS; {\mP}_{\sS})$ obtained by solving $\ttH = \fE$ at the classical level 
for $\sigma$.  
This new expression being linear in one of the new momenta, quantizing it gives a TDSE 
in the new position representation: 
\be
i\hbar\frac{    \pa\Phi    }{    \pa t^{\sE}    } = \widehat{H}^{\mbox{\scriptsize true}}
\left(
t^{\sE}, \mS, \hat{{\mP}}_{\sS} 
\right) 
{\Phi} \mbox{ } .
\ee

What remains to be done for the examples at hand is to solve for $\sigma$ the equation 
\be
\fE \equiv {\cal H}
\left(
t^{\sE}, \mS, -\sigma, \mP_{\sS}
\right) 
= \fT + \fV = \frac{ f\mbox{e}^{2\sigma} }{2}
\left\{
\left\{
\frac{    {\mP}_{\sigma}    }{     f\mbox{e}^{2\sigma}    }
\right\}^2 + 
\left\{
\frac{    {\mP}_{\sS}    }{     f\mbox{e}^{2\sigma}    }
\right\}^2
\right\} + \fV
= \frac{  \mbox{e}^{-2\sigma}  }{  2  }
\left\{
{t^{\sE}}^2 + \mP_{\sS}^2
\right\} + \fV(\sigma, \mS) 
\mbox{ } .  
\ee
This is explicitly possible in the homogeneous quadratic potential cases in hand.

%==========================================================================================================================================
\noindent\underline{Explicit examples of Euler internal TDSE's.}
%==========================================================================================================================================
%
%
%
Another working original to this Paper is that all the cases under consideration are covered by 
\be
\fV = f A \mbox{e}^{2\sigma}/2 \mbox{ } ,
\ee
for $A = A(\mS)$ a subcase-specific shape function.   
For examples 1 and 3 [$A(\mS)$ not everywhere zero], the equation to solve is then 
\be
\frac{  \mbox{e}^{-2\sigma}  }{  2f  }
\left\{
{t^{\sE}}^2 + \mP_{\sS}^2
\right\} 
- \frac{fA}{2}\mbox{e}^{2\sigma} = \fE 
\ee
which gives
\be
\mbox{e}^{2\sigma} = 
\left\{    
- \fE \pm \sqrt{  \fE^2 + A\{{t^{\sE}}^2 + \mP_{\sS}^2\}      }
\right\}
/{    fA    } \mbox{ } .  
\ee  
Thus the Euler internal TDSE is 
\be
i\hbar\frac{\pa \Phi}{\pa t^{\sE}} = 
\left\{
\mbox{ln}
\left(
\sqrt{        \fE^2 + 
A
\left\{
{t^{\sE}}^2 - \hbar^2\frac{\pa^2}{\pa \mS^2}
\right\}  - \fE        }
\right) 
- \mbox{ln}(fA)  
\right\}  
\Phi \mbox{ } . 
\ee

\mbox{ }

Example 2 is simpler.  
Now, the classical equation to solve is
\be
\mbox{e}^{2\sigma} = \{{{t^{\sE}}^2 + P_{\sS}^2}\}/{2f\fE} \mbox{ } , 
\ee
which leads to the Euler internal TDSE
\be
i\hbar\frac{\pa \Phi}{\pa t^{\sE}} = \frac{1}{2}
\left\{
\mbox{ln}(2f\fE) - \mbox{ln}
\left(
{t^{\sE}}^2 - \hbar^2\frac{\pa^2}{\pa \mS^2}
\right)
\Phi 
\right\} \mbox{ } .
\ee

%----------------------------------------------------------------------------------------------------
\noindent\underline{Are approximate WKB-LMB and Euler times aligned?} 
%----------------------------------------------------------------------------------------------------
%
%
%
I find that these are related by
\be
t^{\sE} = \mL\mP_{\sL} + \mH\mP_{\sH} 
= \frac{1}{\mbox{n}}
\left\{ 
M\mH\frac{\pa\mH}{\pa\lambda} +m\frac{\pa\mL}{\pa\lambda}
\right\}
= \frac{1}{2}\frac{\pa\{M \mH^2 + m \mL^2\}}{\pa t^{\mbox{\scriptsize em:L}}   } \mbox{ } .  
\ee
An approximation to this that I use often below is  
\be
t^{\sE} = 
\frac{M}{2}\frac{\pa\mH^2}{\pa t^{\mbox{\scriptsize em:L}}}
\left\{
1 + \frac{O(\epsilon_{\sH\sL})\sqrt{O(\epsilon_{\sV})O(\epsilon_{\sa})}}{O(\epsilon_{\sa 1})}
\right\} 
\mbox{ } .
\label{quirrel}
\ee

In the case of the H-subsystem being an HO, this reads 
\be
t^{\sE} = \sqrt{2\fE_{\sH} M}\mH\sqrt{1 - {M\Omega^2\mH^2}/{2\fE_{\sH}}  }
\left\{
1 + \frac{O(\epsilon_{\sH\sL})\sqrt{O(\epsilon_{\sV})O(\epsilon_{\sa})}}{O(\epsilon_{\sa 1})}
\right\}  
\mbox{ } .
\ee
Thus, from (\ref{bollah}), Euler internal and WKB--LMB emergent times are approximately aligned 
in all of these examples for small $t^{\mbox{\scriptsize em:L}}$ or $\nH$:
\be
t^{\sE} 
\approx 
\frac{\fE}{\Omega}\mbox{sin}(2\Omega \{t^{\mbox{\scriptsize em:H}} - t^{\mbox{\scriptsize em:H}}(0)\}) 
\approx 
2\fE\{t^{\mbox{\scriptsize em:H}} - t^{\mbox{\scriptsize em:H}}(0)\}   \mbox{ } ,
\label{HOOK}
\ee
but more generally they are not aligned.  
From the less approximate form above, can also see that 
$t^{\sE}$ is only monotonic for 
$t^{\mbox{\scriptsize em:H}} - t^{\mbox{\scriptsize em:H}}(0) \in [-\pi/4\Omega, \pi/4\Omega]$ 
(or some other such half-period).  
In this particular case there is not any trouble in defining a different inversion piecewise 
for further pieces which at least match up at the $(n + \frac{1}{2})\pi$ points.

In the case of the H-subsystem being an upside-down HO, (\ref{quirrel}) reads 
\be
t^{\sE} = \sqrt{2\fE_{\sH} M}\mH\sqrt{1 + {MW^2\mH^2}/{2\fE_{\sH}}  }
\left\{
1 + \frac{O(\epsilon_{\sH\sL})\sqrt{O(\epsilon_{\sV})O(\epsilon_{\sa})}}{O(\epsilon_{\sa 1})}
\right\}  \mbox{ } .
\ee
Thus, from (\ref{HUDOH}), Euler internal and WKB--LMB emergent times are often approximately aligned  
(taken to mean up to choice of origen and scale) for small 
$t^{\mbox{\scriptsize em:L}}$ or $\nH$:
\be
t^{\sE} 
\approx 
\frac{\fE}{W}\mbox{sinh}(2W\{t^{\mbox{\scriptsize em:H}} - t^{\mbox{\scriptsize em:H}}(0)\})
\approx
{2\fE}{W}\{t^{\mbox{\scriptsize em:H}} - t^{\mbox{\scriptsize em:H}}(0)\} \mbox{ } ,
\label{HOupsie}
\ee
but more generally are not aligned.

Finally, for H free, the approximation in (\ref{quirrel}) is not valid, 
but one can use the even more general 
approximation 
$$
t^{\sE} = \frac{1}{2}\frac{\pa M\nH^2}{\pa t^{\mbox{\scriptsize em:L}}}
\left\{
1 + \frac{O(\epsilon_{\sH\sL})O(\nL/\nH)}{O(\epsilon_{\sa 1})}
\right\}
$$
\be
=
\sqrt{2\fE_{\sH} M}\mH
\left\{
1 + \frac{O(\epsilon_{\sH\sL})O(\nL/\nH)}{O(\epsilon_{\sa 1})}
\right\} 
\approx
2\fE_{\sH}\{t^{\mbox{\scriptsize em:H}} - t^{\mbox{\scriptsize em:H}}(0)\}
\left\{
1 + \frac{O(\epsilon_{\sH\sL})O(\nL/\nH)}{O(\epsilon_{\sa 1})}
\right\}
\ee 
so the two time standards are directly aligned in a sizeable portion of the configuration space.

Thus, in these examples, $t^{\emL}$ is more widely useable.  
It also  
exists for scale-invariant models \cite{B03}, characterized by the constraint $E = 0$ which implies 
that the Euler quantity is frozen and thus unavailable as a timefunction.   
Although some portions of Newtonian mechanics have guaranteed global monotonicity for $t^{\sE}$, 
solutions outside this portion may still possess intervals on which $t^{\sE}$ is monotonic.

Also note that the internal time examples given above work as well for non-interacting HL-systems.  
Though this is not now a conceptual necessity, as H does not now impose a timefunction on L, but 
rather both contribute to a joint timefunction.
However, while everything in the universe contributes in this species by species respect, 
note the lack of role within for the potential -- makes it look more artificial or imposed, 
as the details of species from the potential plays no (direct) role in the construction 
of the timefunction.

Criticisms of internal approaches themselves are: 
1) the above equations are subject to a tangle of ordering ambiguities and well-definedness issues 
(a point also made by Blyth and Isham \cite{IB75} in minisuperspace examples).    
2) From my work above, I also note that the internal approach's equations do not look anything like the equations encountered in the various conventional 
approaches to HO's. 
These criticisms are outside the more usual `practical impasse to exact classical solutions' objections 
that one finds in the internal time literature, and thus may merit further investigation.  
It may be easier to investigate these issues though careful choice of RPM models and quantization 
procedures than through considering the parallel situation in the minisuperspace examples of \cite{IB75}. 
As a possible first step, I've done some work with approximations on the above TDSE's to obtain 
slightly more familiar equations, but I'll present that elsewhere \cite{Ian}.  
%
%Roots and sums.  Spectral theorem result for roots.  [to be developed elsewhere].

%=====================================================================================================
%=====================================================================================================
%=====================================================================================================
\section{Conclusion}  
%=====================================================================================================
%=====================================================================================================
%=====================================================================================================

\noindent\underline{The Semiclassical approach to the problem of time in quantum gravity.}
This Paper considers this in the relational particle model (RPM) arena.  
The semiclassical approach hinges upon the WKB ansatz in order for this notion of time to emerge, and 
upon the retention and subsequent rearrangement of a cross-term that is, in conventional QM, 
usually regarded as small and discarded.   
By these means, heavy (H) background physics provides a timestandard for local, light (L) 
physics subsystems to run with respect to.  
This procedure is widely said in the literature to yield a time-dependent Schr\"{o}dinger equation.

I have provided a geometric phase formalism for the semiclassical approach to particle mechanics, 
which I have extended to the case of RPM's, which are useful because they include linear constraints 
in close analogy to the situation in GR.  
For these models, I have shown that the emergent WKB and Leibniz--Mach--Barbour (LMB) notions of time can be 
identified.  
I provided a general framework in which to view the various approximations made in the literature.  
This will permit to judge under which circumstances expanding in only 1 parameter is justifiable, 
while setting the scene for more generally applicable multiparameter expansions.  
I showed that in the cosmologically-relevant setting with 1 H d.o.f., in general the L-subsystem 
is {\sl not} governed by a time-dependent Schr\"{o}dinger equation, but rather by a more general 
time-dependent wave equation.  
I proposed and investigated an iterative procedure for solving this and examined a simple regime for 
2 linearly coupled HO's in more detail in this respect.  
The L-subsystem back-reacts on the H-subsystem and so itself 
contributes (a bit) to the timestandard.  
The H-subsystem plays a perturbative role as well as a chroniferous one in the L-physics.
This in turn opens the possibility that such an emergence of time for the L-subsystem may on occasion be 
betrayed by noticeable deviation from the na\"{\i}ve L-physics due to this perturbation term, although 
the semiclassical approach works just as well if this perturbation's effects are unobservably small.  
Thus, indications are that, unlike claimed in e.g. \cite{BS, EOT}, quantum cosmology causes small 
corrections rather than gross distortion as regards local physics, provided that the overall model for 
the universe is of sufficient complexity.

This Paper has also contributed to the discussion of a number of basic and detailed issues within 
the semiclassical approach. 
If the universe is overall timeless as implied by the Wheeler--DeWitt equation, 
there is an underlying requirement to explain the local semblance of dynamics   
The applicability of the WKB ansatz is truly important in the sense that if this does not apply, one no 
longer gets anything like a time-dependent Schr\"{o}dinger equation for the L-physics.     
Whether this ansatz is widely applicable is also considered for this Paper's toy models in Appendix E.
In situations in which this ansatz is applicable, then its means of working through 
retaining and manipulating a small term has further qualitative, and perhaps quantitative, 
repercussions through the equations containing further similar small terms which are nevertheless 
habitually neglected in the literature.    
I have also pointed out other relations in the sizes of various small terms, which restrict how 
freely one can neglect some but not other small terms.  
This will be important in providing detailed, consistent semiclassical calculations, 
as required to make cosmological predictions with more confidence and accuracy. 
Investigation of approximations is in general important for the semiclassical approach to the 
problem of time in quantum gravity, as this is only meant to be an approximate resolution, so we 
would like to work out in detail in which regimes this does and does not work out.  
See Paper II for more on this issue.

%-------------------------------------------------------------------------------------------------------
\noindent\underline{Comparison with hidden time approach and timefunction.}
%-------------------------------------------------------------------------------------------------------
%
%
Monotonicity and non-frozenness considerations indicate that emergent semiclassical 
WKB--LMB time can be more widely applicable than hidden dilational Euler time. 
The free H-model has globally monotonic Euler time and good Euler--WKB--LMB time alignment, the 
upside-down HO H-model has globally monotonic Euler time but only good Euler--WKB--LMB time alignment 
for small times, and the HO H-model only has Euler time locally and only good Euler--WKB--LMB time 
alignment for small times.  
By these examples, and by scale-invariant models having Euler time rendered useless by frozenness,  
emergent WKB--LMB time looks to be a more widely applicable notion.  
Note also that Euler time is not so attuned to the contents of the Universe as WKB--LMB time is.  
For, it itself has a energy- and potential-independent overall form (its time derivative does 
however depend on energy and potential of the universe).

%-------------------------------------------------------------------------------------------------------
\noindent\underline{Further Work.}
%-------------------------------------------------------------------------------------------------------
%
%
This Paper's main 2 coupled HO example can be further explored in the following ways.  
1) Find a perturbation formalism appropriate for the averaged and kinetic-term perturbations 
that also show up in (\ref{pertmaster}).   
2) For the perturbation explored in the present Paper, or the more full one of 1), 
explicitly form and solve the second iteration's H-equation.  
3) Consider at which stage higher order perturbation theory becomes necessary.  
4) Do such iterative scheme converge?

There is also the issue of whether to apply higher-order WKB techniques \cite{KS91} (which is relevant  
when the associated small quantity $\epsilon_{\ssWKB}$ is insufficiently small for 
${\epsilon_{\ssWKB}}^2$ to be entirely negligible). 
Moreover, WKB techniques are but one of a family of techniques in semiclassical QM \cite{Recent}.  
Does the chronifer role and the difficulty in justifying the semiclassical regime pervade all of this?

More detailed models and which features of the recovery of the semblance of dynamics 
and of GR they are necessary for are listed in Appendix C.  
See Appendix B and Paper II for parallel minisuperspace investigations.  
I hope to build this up toward a formalism for the more complicated case of inhomogeneous perturbations 
about minisuperspace so as to be able to detailedly justify and model the quantum origin of microwave 
background fluctuations and of galaxies (extending e.g. Halliwell and Hawking's work 
\cite{HallHaw}).

\mbox{ }

\mbox{ }

\mbox{ }

%====================================================================================================
%====================================================================================================
\noindent{\bf{\large Acknowledgments}}
%====================================================================================================
%====================================================================================================

\mbox{ }

\noindent I thank: Professor Don Page, for discussions on time, histories, records and the 
semiclassical approach and for prompting me to consider homogeneous quadratic potential specific examples.  
Dr. Julian Barbour for interesting me in relational particle models in the first place.  
Professor Gary Gibbons for highlighting my awareness to geometrical issues.  
Professor Niall \'{O} Murchadha for discussions.   
Dr Julian Barbour and an anonymous referee for comments on earlier versions of the manuscript.  
Dr Fay Dowker, Professor Malcolm MacCallum, Lord Wilson of Tillyorn, Claire Bordenave 
and Eve Jacques for being supportive.  
The Killam foundation, for funding me at the University of Alberta in 2005 for the first part of this 
work, and Peterhouse for funding me since.  

\mbox{ }

%====================================================================================================
%====================================================================================================
\noindent{\Large{\bf Appendix A: Group-corrected operators}}
%====================================================================================================
%====================================================================================================

\mbox{ }

\noindent This generalizes my arbitrary (group)-frame method \cite{Lan} which in turn is 
a derivation of Barbour--Bertotti best matching \cite{BB82}.  
For an operator $\Box$, canonical coordinates $\{q_{\Gamma}\} = \fQ$ and auxiliary quantities 
$g_{\Delta}$ which correspond to a basis for the infinitesimal motions of a group $G$ 
acting on $\fQ$, define the $G$-corrected operator $\Box_{g_{\Delta}}$ acting on $\fQ$ 
by 
\be
\Box_{g_{\Delta}}q_{\Gamma} = \Box(q_{\Gamma}) - 
\stackrel{        \mbox{\Huge $\longrightarrow$}        }
         {        G(\Box(g_{\Delta}))        }               
q_{\Gamma} 
\mbox{ } , 
\ee
where $\stackrel{\mbox{\Large$\longrightarrow$}    }{    G(g_{\Delta})    }q_{\Gamma}$ is the $G$-action on 
$\fQ$ under the infinitesimal motion $g_{\Delta}$.

\noindent Example 1: for the rotations acting on the Jacobi coordinates $\R_i$, 
\be
\dotb \R_{i} \equiv \dot{\R}_i - \dot{\b} \cr \R_i
\mbox{ } \mbox{ (} \mbox{ } \Circ{} \mbox{ is an enlarged dot so that $\b$ can be `hung' on it )} 
\mbox{ } , 
\label{App1}
\ee
\be
\de \R_{i} \equiv \Ast\R_{i} - \lfloor\Ast{\b}\rfloor 
\cr \R_{i} 
\mbox{ } \mbox{ for } \mbox{ }  \Ast \equiv
                                         {\pa}/{\pa t^{\seBW}} 
\mbox{ } ,
\label{App1B}
\ee
\be
\di \R_{i} \equiv \d{\R}_{i} - \d\b \cr \R_{i} \mbox{ } .
\label{App2}
\ee
While the notation used is manifestly 3-d, it nevertheless encompasses also 
dimension 1 (no correction) and dimension 2 (the case of a single correction $\nb$, 
under the proviso that $\nb_3 \equiv \nb$ \cite{OI}).

\noindent Example 2: for the 3-diffeomorphisms acting on the 3-metrics $h_{\alpha\beta}$, 
\be
\DotB h_{\alpha\beta} \equiv \dot{h}_{\alpha\beta} - \pounds_{\dot{\suB}}h_{\alpha\beta} \mbox{ } ,
\label{App3}
\ee
\be
\De h_{\alpha\beta} \equiv \Ast{h}_{\alpha\beta} - 
\pounds_{\sAst\suB}h_{\alpha\beta}   
\mbox{ } \mbox{ for } \mbox{ }  \Ast \equiv
                                         {\pa}/{\pa T^{\seBW}} 
\mbox{ } ,
\label{App3B}
\ee
\be
\Di h_{\alpha\beta} \equiv \d{h}_{\alpha\beta} - \pounds_{\d{\suB}}h_{\alpha\beta}  
\label{App4}  
\mbox{ } ,
\ee
for $\pounds_{\sB}$ the Lie derivative with respect to the vector field $\mbox{B}_{\alpha}$.  

%\vspace{2in}

\mbox{ }

%==========================================================================================================================================
%==========================================================================================================================================
%====================================================================================================
\noindent{\bf{\Large Appendix B: GR features present in RPM's versus those present in Minisuperspace}}
%====================================================================================================
%==========================================================================================================================================
%==========================================================================================================================================

\mbox{ }

\noindent Like GR, RPM's have linear constraints (nontrivial for $d \geq 2$).  While MSS does not have nontrivial 
linear constraints (at least in absense of matter).    
But RPM's have positive-definite kinetic term, unlike GR's which is indefinite (at least for anything more 
than homogeneous, isotropic and matter-free models which have a 1-d configuration space for which 
the issue of definite or indefinite is not even defined), 
and the simplest RPM's configuration spaces aren't curved.  
Minisuperspace models have more specific potentials than RPM's.  
And these are quite complicated specific potentials, while for RPM's the freedom of choice 
permits one to consider simple and well-behaved choices like free or HO models.  

\mbox{ }  

%==========================================================================================================================================
%==========================================================================================================================================
%====================================================================================================
\noindent{\bf{\Large Appendix C: Where the examples in Papers I, II stand within the RPM framework: a guide to some nontriviality criteria}}
%====================================================================================================
%==========================================================================================================================================
%==========================================================================================================================================

\mbox{ }  

\noindent 
The specific examples in this Paper are $d = 1$, $N = 2$ RPM's.    
This Appendix serves to place these in context of further RPM's and of other treatments.

%-----------------------------------------------------------------------------------------------------
\noindent\underline{Spatial dimension.}
%-----------------------------------------------------------------------------------------------------
%
%
%
\noindent In $d = 1$ there are only translations, which are much easier to quotient out than 
rotations. 
$d = 1$ gives configuration spaces that are only (pieces of) flat space.
While this is useful for carrying out technical and conceptual checks for various approximations to RPM's, 
one would then wish to move onto the now decidedly distinctive $d > 2$ RPM's.  
See \cite{2d} for $d$ = 2.  
Classically, many genuinely $d = 3$ issues would require at least 4 particles (as $d=3$, $N=3$ 
can be considered in a plane and is thus equivalent in many ways to $d = 2$, $N=3$).

%-----------------------------------------------------------------------------------------------------
\noindent\underline{Configuration space dimension.}
%-----------------------------------------------------------------------------------------------------
%
%
Using that the dimension of the absolute configuration space 
\{$\fQ_{\sq}(N, d)\}$ is $Nd$, of the translation-irrelevant configuration space  
\{$\fQ_{\sR}(N, d)\}$ is $nd \equiv (N - 1)d$,     
and of the translation- and rotation-irrelevant configuration space  
\{$\fQ_{\bar{\sR}}(N, d)$\} is $Nd - d(d + 1)/2$, 
and also considering their scale-invariant counterparts\foo{These bear resemblance to 
maximally-sliced GR.} \cite{B03, OII} of one dimension less,
one can use the following elementary geometrical considerations.  
Denote configuration space dimension by $k$ and dimension of subconfiguration space $S$ by $k(S)$.  
1) The $k = 1$ configuration spaces are always geometrically 
trivial in the sense that they are intrinsically flat: the line, or pieces thereof 
(perhaps identified to make a circle) and (or) decorated with special 
points. 
2) The $k = 2$ ones are in general curved but are always conformally flat.  
Useful preliminaries toward problem of time studies for these are in \cite{GM, OII, 2d}.  
3) The $k \geq 3$ ones are no longer in general conformally flat.  
Less is known at this level of nontriviality; \cite{GM, OII} do not venture that far, while  
the $d = 3$, $N > 3$ geometry is harder \cite{Gergely} and less explored than for $N = 3$.
See however \cite{Montgomery, Kendall}.

2), 3) could serve to investigate themes for which this Paper's specific examples are too simple.  
One such theme is Barbour's speculation that the configuration space geometry plays an important role 
in the selection of records that encode a semblance of history.  
For 2), conformal flatness means that nonflatness in $\fT$ can be transferred into a redefined $\fV$ 
(at least piecewise, as conformal factors cannot admit zeros).  
Moreover, as 3) are no longer generally conformally flat, the distinction between $\fT$ and $\fV$ effects 
becomes fully nontrivial.

As regards semiclassical models that work reasonably well, 
note that we need $k \geq 2$ to have a H--L split.  
Moreover, addressing some aspects of nonfrozen L physics requires 
$k \geq 3$ via requiring a H--L split and $k(L) \geq 2$.  
Some back-reaction effects require $k(H) \geq 2$ or $Q_H = S^1$ [though consideration of 
relative phase is another way out].  
The following table identifies the RPM's with the smallest nontrivial values of $k$.

%========================================================================================================
\noindent\underline{Configuration space dimension for various available RPM's.}
%========================================================================================================
%
%
%
\noindent
\hspace{0.2in}
\begin{tabbing}
\underline{usual RPM} \hspace{0.2in} \= \underline{N = 3} \hspace{0.1in} \mbox{ } 
\=\underline{N = 4} \hspace{0.1in} \=   \underline{N = 5}   \mbox{ }\=  
\hspace{0.5in} \underline{scale-free RPM} \hspace{0.2in} \mbox{ } 
\= \underline{N = 4} \hspace{0.1in}
\= \underline{N = 5} \hspace{0.1in}
\= \underline{N = 6} \\
\underline{d = 1}  \>  2  \>  3  \>  4  \> \hspace{0.5in} \underline{d = 1}   \>  2   \>  3  \> 4  \\
\underline{d = 2}  \>  2  \>  4  \>     \> \hspace{0.5in} \underline{d = 2}   \>  3   \>    \>   \\
\underline{d = 3}  \>  2  \>     \>     \> \hspace{0.5in} \underline{d = 3}   \>  4   \>    \>   \\
\end{tabbing}

%==========================================================================================================================================
%==========================================================================================================================================
%=====================================================================================================
\noindent{\bf {\Large Appendix D: Upside down and free counterparts.}}
%=====================================================================================================
%==========================================================================================================================================
%==========================================================================================================================================

\mbox{ }

\noindent The following examples are useful in various places in this Paper.  
For the upside-down HO H-subsystem, replace $\Omega$ by $iW$ in (\ref{yoda}). 
Then, one can similarly deduce that 
\be
\fW_0 = \sqrt{2M\fE_{\sH}}\int^{\sH} d \nH^{\prime}\sqrt{1 + {MW^2\nH^{\prime 2}}/{2\fE_{\sH}}} 
\ee
and 
\be
t^{\sem}_0 - t^{\sem}_{0}(0) = 
\frac{1}{W}\mbox{arsinh}
\left( 
\sqrt{{M}/{2\fE_{\sH}}}
W\nH
\right) 
\mbox{ } ,
\label{bollah}  
\ee
which inverts to 
\be
\nH =\sqrt{    {2\fE_{\sH}}/{MW^2}    }\mbox{sinh}(W\{t^{\sem}_0 - t^{\sem}_{0}(0)\})
\mbox{ } .   
\label{HUDOH}
\ee
For the free H-subsystem, set $\Omega = 0$ in (\ref{yoda}).  
Then 
\be
t^{\sem}_0 - t^{\sem}_{0}(0) = \sqrt{{M}/{2\fE_{\sH}}} \nH 
\mbox{ } ,
\label{garde}
\ee
which inverts to 
\be
\nH = \sqrt{{2\fE_{\sH}}/{M}}\{t^{\sem}_0 - t^{\sem}_0(0)\} \mbox{ } .
\label{HF}
\ee

%=====================================================================================================
\noindent{\Large{\bf Appendix E: Exact treatment of HO's by coordinate transformation}}
%=====================================================================================================

\mbox{ }

\noindent It is intuitively obvious that an H--L system's H-part will roughly go like the H-system in isolation.  
From this, the pure H-part of the ground state for an HO will go like
\be
\mbox{exp}\left(-{M\Omega \nH^2}/{2\hbar}\right) \mbox{ } ,
\ee
which agrees with $\mbox{e}^{i\sfW_0^-/\hbar}$ for H large ($\mbox{H} > 2E/M\Omega^2$). 
The pure H-part of the free particle wavefunction is 
\be
\mbox{exp}(\sqrt{2M\fE}\nH)
\ee
which agrees with $\mbox{e}^{i\sfW_0^-/\hbar}$ for all H.  
And, the pure H-part of the ground state for an upside down HO will go like
\be
\mbox{exp}\left(\pm i{M W \nH^2}/{2\hbar}\right) \mbox{ } ,
\ee
which agrees with $\mbox{e}^{i\sfW_0^{\pm}/\hbar}$ for H large ($\mbox{H} > 2E/M\Omega^2$). 
Note also that the last 2 match well for classically allowed, oscillatory WKB regimes.

The advantage of exact ulterior solubility, moreover, is in being able to 
form arbitrarily precise expansions for the pure H and HL parts of the wavefunction.  
That the 2 linearly-coupled HO's example is ulteriorly exactly soluble: 
under the rotation and mass rescaling
\be
\left(
\stackrel{\mbox{$\nH$}}{\mbox{$\nL$}}
\right) =
\left(
\stackrel{    \mbox{$\frac{1}{\sqrt{M}} \mbox{ } 0$}    }{    0 \mbox{ } \frac{1}{\sqrt{m}}     }
\right) 
\left(
\stackrel{    \mbox{cos$\alpha \mbox{ }$ sin$\alpha$}    }{    -\mbox{sin}\alpha \mbox{ } \mbox{cos}\alpha}
\right) 
\left(
\stackrel{\mbox{$\nX_-$}}{\mbox{$\nX_+$}}
\right) 
\label{roma} \mbox{ } ,
\ee
with 
\be
\mbox{tan}2\alpha = {\gamma}/\{\omega^2 - \Omega^2\} \mbox{ } , \mbox{ } \gamma = {2c}/{\sqrt{Mm}}  
\mbox{ } ,
\ee
this becomes the Lagrangian 
\be
\mbox{\sffamily L} = \frac{\dot{\nX}_-^2 + \dot{\nX}_+^2}{2} - \frac{k_-\nX_-^2 + k_+\nX_+^2}{2} 
\mbox{ } ,
\ee
where 
\be
2k^2_{\pm} \equiv \Omega^2 + \omega^2 \pm \beta 
\mbox{ } , \mbox{ } 
\beta \equiv \sqrt{\gamma^2 + (\omega^2 - \Omega^2)^2} 
\mbox{ } ,
\ee
i.e. the Lagrangian for 2 uncoupled HO's.  
The quantization in $\nX_{\pm}$ variables is then immediately separable, 
yielding 
\be
\Phi = \sum_{m = 0}^{\infty}\sum_{n = 0}^{\infty} 
\frac{1}{\sqrt{2^{\sfj}\fj !}}
\mbox{Hermite}_{\sfj}(\sqrt{k_-}\nX_-)\mbox{e}^{-\sqrt{k_-}\sX_-^2/2} 
\frac{1}{\sqrt{2^{\sfl}\fl !}}
\mbox{Hermite}_{\sfl}(\sqrt{k_+}\nX_+)\mbox{e}^{-\sqrt{k_+}\sX_+^2/2}
\ee
subject to $\fE_{\sfl} + \fE_{\sfj} = \fE_{\su}$.  
Then e.g., by inversion of (\ref{roma}), the pure-H prefactor of $\Phi_{00}$ 
is 
\be
\mbox{exp}\left(-\frac{H^2}{2\hbar}
\left\{ 
k_+m + k_-M + \{k_+m - k_-M \}\{\Omega^2 - \omega^2\}/\beta
\right\}
\right) 
\mbox{ } . 
\ee
Then a large number of uses of binomial expansions via the definitions of 
$\gamma$, $\beta$ and $k_{\pm}$ reveal that this goes as
\be
\mbox{exp}
\left(
-\frac{M\Omega \nH^2}{2\hbar}
\left\{
1 - \frac{        \epsilon_{\mbox{\scriptsize pert}}^2\epsilon_{\sH\sL}        }{        2        } 
\left\{
1 + \frac{\epsilon_{\sH\sL}}{\epsilon_{\sa 1}^2}
\right\} 
+ 
\frac{        \epsilon_{\sH\sL}^2\epsilon_{\mbox{\scriptsize pert}}^2   }{        2\epsilon{\sa}      } 
+ ...
\right\}
\right) \mbox{ } .  
\ee
By such means, ulteriorly exactly soluble models can be used to probe 
the plethora of possible approximations schemes that the semiclassical approach gives rise to.  

\mbox{ }

%=====================================================BIBLIOGRAPHY==========================================================================

\end{document}